\documentclass[preprint]{aastex}

\usepackage[breaklinks]{hyperref}
\usepackage{amsmath}
\usepackage{url}
\usepackage{color}

\newcommand{\gray}{$\gamma$-ray}
\newcommand{\grays}{$\gamma$-rays}

\newcommand{\hi}{H{\sc i}}
\newcommand{\htwo}{H$_{2}$}
\newcommand{\hii}{H{\sc ii}}

\newcommand{\Xco}{$X_{\rm CO}$}

\newcommand{\BeTen}{$^{10}$Be}
\newcommand{\BeNine}{$^{9}$Be}
\renewcommand{\deg}{^{\circ}}

\newcommand{\fermi}{{\it Fermi}{}}

\shorttitle{The spectrum of isotropic diffuse gamma-ray emission between 100 MeV and 820 GeV.}
\shortauthors{The \textit{Fermi}-LAT Collaboration}

\begin{document}


\title{The spectrum of isotropic diffuse gamma-ray emission between 100 MeV and 820 GeV}

\author{
M.~Ackermann\altaffilmark{1,2}, 
M.~Ajello\altaffilmark{3}, 
A.~Albert\altaffilmark{4}, 
W.~B.~Atwood\altaffilmark{5}, 
L.~Baldini\altaffilmark{6}, 
J.~Ballet\altaffilmark{7}, 
G.~Barbiellini\altaffilmark{8,9}, 
D.~Bastieri\altaffilmark{10,11}, 
K.~Bechtol\altaffilmark{12,13}, 
R.~Bellazzini\altaffilmark{6}
% ...
}
\altaffiltext{1}{Deutsches Elektronen Synchrotron DESY, D-15738 Zeuthen, Germany}
\altaffiltext{2}{email: markus.ackermann@desy.de}
\altaffiltext{3}{Department of Physics and Astronomy, Clemson University, Kinard Lab of Physics, Clemson, SC 29634-0978, USA}
\altaffiltext{4}{W. W. Hansen Experimental Physics Laboratory, Kavli Institute for Particle Astrophysics and Cosmology, Department of Physics and SLAC National Accelerator Laboratory, Stanford University, Stanford, CA 94305, USA}
\altaffiltext{5}{Santa Cruz Institute for Particle Physics, Department of Physics and Department of Astronomy and Astrophysics, University of California at Santa Cruz, Santa Cruz, CA 95064, USA}
\altaffiltext{6}{Istituto Nazionale di Fisica Nucleare, Sezione di Pisa, I-56127 Pisa, Italy}
\altaffiltext{7}{Laboratoire AIM, CEA-IRFU/CNRS/Universit\'e Paris Diderot, Service d'Astrophysique, CEA Saclay, 91191 Gif sur Yvette, France}
\altaffiltext{8}{Istituto Nazionale di Fisica Nucleare, Sezione di Trieste, I-34127 Trieste, Italy}
\altaffiltext{9}{Dipartimento di Fisica, Universit\`a di Trieste, I-34127 Trieste, Italy}
\altaffiltext{10}{Istituto Nazionale di Fisica Nucleare, Sezione di Padova, I-35131 Padova, Italy}
\altaffiltext{11}{Dipartimento di Fisica e Astronomia ``G. Galilei'', Universit\`a di Padova, I-35131 Padova, Italy}
\altaffiltext{12}{Kavli Institute for Cosmological Physics, University of Chicago, Chicago, IL 60637, USA}
\altaffiltext{13}{email: bechtol@kicp.uchicago.edu}
% ...








\begin{abstract}
The \gray{} sky can be decomposed into individually
detected sources, diffuse emission attributed to the
interactions of Galactic cosmic rays with gas and radiation fields,
and a residual all-sky emission component commonly called the
isotropic diffuse \gray{} background (IGRB). The IGRB comprises all
extragalactic emissions too faint or too
diffuse to be resolved in a given survey, as well as any residual Galactic
foregrounds that are approximately isotropic. 
The first IGRB measurement with the Large Area Telescope (LAT) on board the \textit{Fermi Gamma-ray Space
Telescope} (\textit{Fermi}) used 10 months of sky-survey data and
considered an energy range between 200~MeV and 100~GeV.
Improvements in event selection and characterization of
cosmic-ray backgrounds, better understanding of the diffuse Galactic
emission, and a longer data accumulation of 50 months, allow for a
refinement and extension of the IGRB measurement with the LAT, now covering the
energy range from 100~MeV to 820~GeV. The IGRB spectrum shows a significant
high-energy cutoff feature, and can be well described over nearly four
decades in energy by a power law with exponential cutoff having a spectral index of $2.32\pm0.02$ and a break
energy of $(279\pm52)$~GeV using our baseline diffuse Galactic emission
model. The total intensity attributed to the IGRB is $(7.2\pm0.6)
\times 10^{-6}$~cm$^{-2}$~s$^{-1}$~sr$^{-1}$ above 100~MeV,
with an additional $+15$\%/$-30$\% systematic uncertainty due to the
Galactic diffuse foregrounds.
\end{abstract}


\keywords{gamma rays: diffuse background, diffuse radiation}

\section{Introduction}\label{sec:introduction}

The universe is filled with electromagnetic radiation, which can be
characterized by a cosmological energy density and spectrum.
This extragalactic background light (EBL) is energetically
dominated by thermal relic radiation from the last scattering surface observed
as the cosmic microwave background. Different physical processes
characterize the EBL in each waveband---starlight in the
optical, thermal dust emission in the infrared, and emission from active
galactic nuclei (AGN) in X rays. The extragalactic \gray{}
background (EGB) provides a non-thermal perspective on the cosmos, which
is also explored through the cosmic radio background as well as
extragalactic cosmic rays (CRs) and neutrinos.

The EGB represents a superposition of all \gray{} sources, both
individual and diffuse, from the edge of the Milky Way to the edge of
the observable Universe, and is thus expected to encode diverse
phenomena (see \citet{Dermer:2007} for a comprehensive review).
Guaranteed contributions arise from established
extragalactic \gray{} source classes including AGN, star-forming galaxies,
and \gray{} bursts. The beamed emission from blazars is
sufficiently bright that statistically large samples of individual
sources have now been detected to cosmological distances \citep{Ackermann2011:2LAC}. Accordingly, the
flux distribution of blazars even below the detection threshold for
individual sources can in principle be estimated from a relatively firm empirical
basis through an extrapolation of the observed flux distribution
\citep[e.g.,][]{Abdo2010:HighLatitude,Ajello2012:FSRQ,Ajello2014:BLLac}, although a
consensus has not yet been reached \citep[e.g.,][]{Singal:2012}.
For other populations, such as star-forming galaxies
\citep{Pavlidou:2002,Thompson:2007,Fields:2010,Makiya:2011,Stecker:2011,Ackermann2012:Star-forming}
and AGN with jets oriented obliquely to our line of sight \citep{Inoue:2011,DiMauro:2014}, the cumulative 
intensity is almost entirely unresolved by current instruments; calculations of the flux
distribution incorporating physical models and/or
multiwavelength scaling relations must be invoked to estimate
their EGB contributions. There are additional theoretically well-motivated
extragalactic source classes too faint to have been individually
detected thus far, including galaxy clusters and their associated large scale
structure formation shocks \citep{Colafrancesco:1998,Loeb:2000}. 

At energies $\gtrsim100$~GeV, the interaction length for $\gamma$~rays with photons of
the UV/optical/IR EBL becomes much shorter than a Hubble length,
thus defining an effective \gray{} horizon. The electromagnetic cascades
initiated by both very-high energy \grays{} \citep{Coppi:1997} and
ultra-high energy CRs \citep{Berezinskii:1975} interacting
with the EBL create truly diffuse EGB contributions. 

Finally, more exotic processes such as
dark matter annihilation/decay may be present, though as yet
unrecognized, in the EGB \citep{Bergstrom:2001,Ullio:2002,Taylor:2003}.

From an observational standpoint, there are two main challenges in measuring the
EGB. One is to model the diffuse Galactic emission (DGE) created by
CR interactions with interstellar gas (ISG) and interstellar radiation
fields (ISRF), 
which is comparable to the EGB intensity at energies $\gtrsim1$ GeV even at the Galactic poles,
and therefore represents a strong foreground to the EGB measurement.
The second challenge is separating cosmic $\gamma$~rays 
from CR induced backgrounds at the detector level. For instruments in low Earth orbit, the CR intensity can exceed that of the
EGB signal by a factor of up to $\sim10^6$. In addition there is a sizable flux of secondary particles that are produced by
interactions of CR in Earth's atmosphere.
 

The existence of all-sky \gray{} emission
was first realized experimentally using 621 candidate $\gamma$~rays collected by the
\textit{OSO-3} satellite \citep{Clark:1968,Kraushaar:1972}, while \cite{Fichtel:1975,Fichtel:1978} reported the first
spectral measurement of an isotropic diffuse background using the
\textit{SAS-2} satellite. Analyses using more sensitive instruments capable of
detecting individual extragalactic sources began reporting the
residual all-sky average intensity after subtracting
individual sources and DGE templates \citep[e.g.,][using
  EGRET]{Sreekumar:1998,Strong:2004}. The remaining emission component is found
to be approximately isotropic on large angular scales and is commonly
called the isotropic diffuse \gray{} background (IGRB). The sum
of the IGRB and individually resolved extragalactic sources represents an
upper limit to the total EGB intensity, since residual unresolved
Galactic emissions may be present in the IGRB.
For example, CR interactions with gas \citep{Feldmann:2013} or radiation
fields \citep{Keshet:2004} in the extended halo of the Milky Way, unresolved
Galactic sources such as millisecond pulsars \citep{Faucher-Giguere:2010}, as well as CR
interactions with solar system debris \citep{Moskalenko:2009} and the solar
radiation field \citep{Moskalenko:solarIC,Orlando:2007,Orlando:solarIC}
have been considered as sources of approximately isotropic emission on
large angular scales. 

The intensity attributed to the IGRB is \textit{observation-dependent} because more sensitive
instruments with deeper exposures can extract fainter extragalactic
sources, whereas the total EGB intensity (assuming complete
subtraction of all Galactic emissions) is the fundamental quantity.

The Large Area Telescope (LAT) on board the \textit{Fermi Gamma-ray Space
Telescope} (\textit{Fermi}) is the first instrument with sufficient
collecting area and CR-background rejection power to measure the
IGRB at energies $>100$~GeV. Since launch into low-Earth
orbit on 11 June 2008, the LAT has operated primarily in
a sky-survey mode that combined with a large field of view (2.4~sr)
and good spatial resolution ($\sim1\deg$ at 1~GeV) has
enabled the most detailed studies of the DGE to date. The LAT is a
pair-conversion telescope consisting of a precision tracker and imaging calorimeter, which
are used together to reconstruct \gray{} directions and energies, and a
surrounding segmented anti-coincidence detector (ACD) to
identify charged particles entering the instrument. \cite{Atwood:2009} provide a description of
the \textit{Fermi} mission and LAT detector as well as details of the
on-orbit calibration. Ground data processing, event selection, and
instrument response functions are provided in
\cite{Abdo2009:On-orbit}, \cite{Ackermann2012:EnergyScale}, and
\cite{Ackermann2012:Performance}. 

A first measurement of the IGRB spectrum
between 200~MeV and 100~GeV based on 10 months of LAT data
was published in \cite{Abdo2010:EGB}. In this paper, we present a
refinement and extension of that analysis based on 50 months of
sky-survey observations. Multiple improvements in event
classification, Galactic foreground and CR background models,
and analysis techniques have been implemented. 
Together with increased statistics, these updates allow for an extension of the LAT IGRB
measurement by over a decade in energy, now covering the range from 100~MeV to 820~GeV.

\clearpage

\section{Data samples}

50 months of LAT data recorded between 5 August 2008 and 6 October
2012 are used for this analysis, corresponding to a
total observation time of 1239 days\footnote{LAT data recording is disabled for $\approx13$\% of the total on-orbit
  time, during passages through the South Atlantic Anomaly (SAA), a region with extremely high charged particle backgrounds. 
  Only observation periods that passed data quality monitoring and where the angle between the LAT z-axis and the Zenith 
  was below 52$\deg$ are used for this analysis. Note that the actual livetime is $\sim$ 8\% smaller than the 
  1239 days observation time quoted here due to instrumental deadtime associated with event latching and readout.}. 
  The events have been reprocessed with an updated instrument calibration which improves the agreement 
  between data and simulation of the energy reconstruction quality, the point spread function (PSF), and certain classification 
variables, and thereby reduces systematic uncertainties
\citep{Bregeon2013:P7REP}\footnote{The reprocessed data are available from the
\emph{Fermi Science Support Center} (http://fermi.gsfc.nasa.gov/ssc/)
together with a list of caveats regarding their usage.}.

The LAT IGRB analysis poses especially stringent requirements on the
\gray{} purity of the event selection since both the signal and
CR-background spatial distributions are quasi-isotropic. The residual CR background contamination must be
reduced to a relatively small fraction of the total isotropic
intensity in order to measure the IGRB with acceptable systematic
uncertainty because the (not perfectly known) CR background is directly
subtracted from the total isotropic intensity in the final step of evaluating the IGRB.

The predefined event classes publicly available from the \emph{Fermi Science Support Center}
including {\tt P7ULTRACLEAN} have insufficient CR background rejection
performance for the IGRB analysis energies below $E<400$~MeV
and energies above $E>100$~GeV. Therefore, we developed two
dedicated event samples for the IGRB analysis with distinct selection
criteria at low and high energies. The IGRB intensity measurements
reported in Section \ref{sec:results} use the
`low-energy' sample for the energy range 100~MeV -- 13~GeV, and
the `high-energy' sample for the energy range 13~GeV --
820~GeV\footnote{Each sample includes events from the full
LAT energy range. The labels `low-energy' and `high-energy' refer to
the energy regime for which the event classifications have been
optimized. The energy overlap between samples allows for consistency
checking between the low-energy and high-energy analyses (described in
Section \ref{sec:analysis}), a feature we used to verify that the specific choice
of crossover energy between 5~GeV and 50~GeV does not affect the
accuracy of our quoted results for the IGRB intensity.}. 
Multiple event classifications are necessary in order to
obtain the best possible compromise between statistics and low
CR backgrounds across the full LAT energy range since the
compositions and interactions of CR backgrounds in the low- and high-energy
regimes are rather different.
The modifications to the baseline {\tt P7ULTRACLEAN} classification for
the two event samples used in this work are described below, and summarized in Table \ref{tab:EventSelection}.

\subsection{Event selection}
\label{subsec:selection}


The low-energy sample is a strict subset of events classified as
photons according to the {\tt P7ULTRACLEAN} event class definition \citep{Ackermann2012:Performance}.
To reduce the residual background of secondary electrons, positrons, and
protons produced by CR interactions in the Earth's atmosphere, which are
the primary concern in the low-energy IGRB analysis, the following
additional criteria are imposed:

\begin{description}
\item[Tracker veto I:]{Part of the tracker is used as an additional veto
  to complement the ACD. Specifically, we require the reconstructed \gray{} trajectory 
  to cross at least two layers of active silicon strip detector area
  without producing hits in these detectors. This selection criterion
  significantly increases the efficiency of vetoing charged
  particles entering the LAT.}
  
\item[Tracker veto II:]{We discard events for which the reconstructed pair-conversion vertex 
  lies in the three $x$-$y$ double-layers of the tracker closest to
  the calorimeter. Comparisons of low-background and high-background
  on-orbit data sets as well as comparisons of \gray{} and CR-background Monte Carlo
  simulations have shown that these events suffer a higher background contamination fraction.}

\item[Deposited charge veto:]{$\gamma$~rays convert into an electron-positron pair while most of the
background events involve a single charged particle. Therefore, we
require the charge deposited in the first tracker layer following
the interaction vertex to be $> 1.5$ times the value expected for a
minimum ionizing particle, which is typically indicative that two particles crossed the layer rather than one by itself.}

\item[Incidence angle veto:]{Events arriving from directions
  $>72\deg$ off the LAT boresight are rejected because there is
  increased CR background leakage for such highly inclined events.}
\end{description}

The new event class for the low-energy sample is denoted as
{\tt P7REP\_IGRB\_LO} in the remainder of this manuscript to distinguish it from
the publicly available standard event classes. 
The sky-averaged exposure of the {\tt P7REP\_IGRB\_LO} selection is 66\% of the 
exposure of the corresponding {\tt P7ULTRACLEAN} selection for survey mode
observations (see Figure \ref{fig:exposure}), when compared at 
the energy of maximum exposure. The estimated residual
CR background rate is reduced by a factor of $\sim3$ around 200~MeV, where the background
rate is highest.

As a final step to define the low-energy sample, events from measured directions $>90\deg$ off the Earth zenith are
rejected to limit contamination by photons from the Earth limb
\citep{Abdo2009:Albedo}. 

For the high-energy sample, we use a relaxed event selection compared to {\tt P7REP\_IGRB\_LO}.
At energies above $13$~GeV, CR primaries in the form of protons and heavier nuclei dominate the
background flux. The rejection power for CR nuclei 
is sufficient for this analysis if one requires only the condition
described above as `Tracker veto I' in addition to the standard {\tt P7ULTRACLEAN} event class definitions.
We implement the `Incidence angle veto' as for the low-energy event class.

The standard {\tt P7ULTRACLEAN} event classification rejects events
for which the positions of the primary interaction vertex and
the reconstructed shower maximum are separated by $>12$ radiation
lengths as measured along the shower axis.
This selection criterion was introduced to reject CR
events with bad shower reconstructions that would sometimes result
in large apparent depths for the shower maxima---a strategy that
works well for energies $\lesssim500$~GeV, but removes a significant 
fraction of $\gamma$~rays $\gtrsim500$~GeV. Therefore this selection
criterion is removed in the event selection for the high-energy
sample. The very moderate increase in residual CR background arising 
from this removal is overcompensated by the `Tracker veto I' 
condition that was introduced for this event class.

The distinct classification scheme for the high-energy sample is
denoted as {\tt P7REP\_IGRB\_HI} in contrast to the {\tt P7REP\_IGRB\_LO}
classification used for the low-energy sample.
At high energies, the reconstructed arrival directions of CR-induced
atmospheric $\gamma$~rays are confined to angles very close to the Earth limb ($113\deg$ from the Earth
zenith) due to the reduced width of the PSF ($\sim0.1\deg$ above
10~GeV). Therefore the zenith angle veto condition described above for
the low-energy sample is modified to reject only photons
from directions $>105\deg$ off the Earth zenith.

The {\tt P7REP\_IGRB\_HI} selection has a peak exposure of about 85\%
of the peak exposure of the standard {\tt P7ULTRACLEAN} selection, and
surpasses the {\tt P7ULTRACLEAN} selection in acceptance $\gtrsim700$~GeV.

Both new event selections were cross-validated against the standard P7ULTRACLEAN event selection
by performing the analysis described below also on the P7ULTRACLEAN dataset.
We obtain consistent results in the energy range in which we can perform the analysis 
even in the presence of the higher CR background of the P7ULTRACLEAN selection (400~MeV -- 100~GeV).

\clearpage

\begin{deluxetable}{lcc}
\tabletypesize{\scriptsize}
\tablecaption{Event selection criteria for low-energy and high-energy
  samples, including modifications with respect to {\tt P7ULTRACLEAN}}
\tablewidth{0pt}
\tablehead{
\colhead{} & \colhead{Low-energy}  &
\colhead{High-energy }
}
\startdata
Data sample / event selection & {\tt P7REP\_IGRB\_LO} & {\tt P7REP\_IGRB\_HI} \\ 
\hline
Add tracker veto I & Y & Y \\
Add tracker veto II & Y & N \\
Add deposited charge veto & Y & N \\
Remove calorimeter shower maximum veto & N & Y \\
Incidence angle veto & $>72\deg$ & $>72\deg$ \\ 
Zenith angle veto & $>90\deg$ & $>105\deg$ \\
\enddata
\tablecomments{See Section \ref{subsec:selection} for detailed descriptions of the event
  selection criteria. The low-energy and high-energy event samples are used to derive the IGRB intensity in the
100~MeV--13~GeV and 13~GeV--820~GeV energy ranges,
respectively.}
\label{tab:EventSelection}
\end{deluxetable}

\clearpage

\subsection{Instrument response functions}

New sets of dedicated instrument response functions (IRFs) were created
for the low-energy ({\tt P7REP\_IGRB\_LO}) and the high-energy ({\tt P7REP\_IGRB\_HI}) event classes 
via Monte Carlo simulation of $\gamma$~rays.
The energy range of the new IRFs is 17.8~MeV to 1.78~TeV.

\begin{figure}
\epsscale{0.7}
\plotone{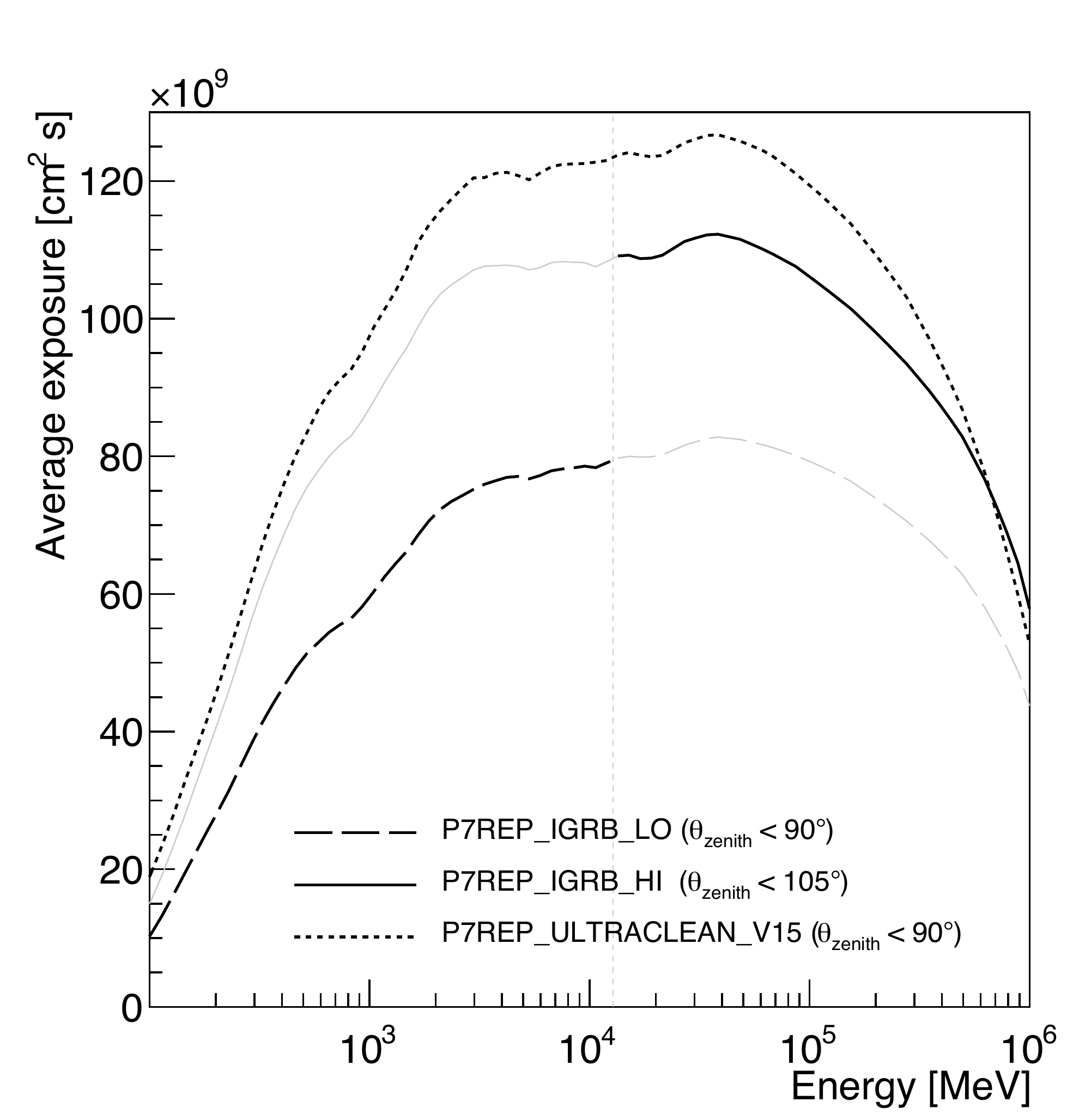}
\caption{Comparison of the sky-averaged exposure for the {\tt P7REP\_IGRB\_LO}, {\tt P7REP\_IGRB\_HI}, 
and {\tt P7ULTRACLEAN} event selections. Thick lines indicate the respective energy ranges for which the
 {\tt P7REP\_IGRB\_LO} and {\tt P7REP\_IGRB\_HI} event classes are used in this analysis.\label{fig:exposure}}
\end{figure}

Figure \ref{fig:exposure} shows the sky-averaged
exposures obtained for the low-energy and the high-energy samples using the 
corresponding {\tt P7REP\_IGRB\_LO} and the {\tt P7REP\_IGRB\_HI} IRFs,
respectively. The exposure that would be obtained for the same observation
period, but using the standard {\tt P7\_ULTRACLEAN} event sample with IRFs
{\tt P7REP\_ULTRACLEAN\_V15} is plotted for comparison. 
For the {\tt P7REP\_IGRB\_HI} selection, there is an overall drop in exposure due to using part of
the silicon tracker to veto charged particles. 
However, at the highest energies, especially $>300$~GeV, this loss
is increasingly counteracted by the removed shower maximum constraint 
in the {\tt P7REP\_IGRB\_HI} class (see Section \ref{subsec:selection}). The {\tt
  P7REP\_IGRB\_LO} selection has a significantly lower average exposure
than {\tt P7REP\_IGRB\_HI}. This loss of exposure is acceptable at low
energies where this event class is used since the IGRB measurement 
is not limited by statistics below tens of GeV.

In-flight PSF corrections available for the IRFs corresponding to
standard event classes have not been applied to the {\tt P7REP\_IGRB\_LO} and {\tt P7REP\_IGRB\_HI} IRFs.
The corrections were motivated by small differences observed in the 
PSF of the original ({\tt P7}) on-orbit and simulated LAT data at energies
$\gtrsim1$~GeV \citep{Ackermann2012:Performance,Ackermann2013:PSF}. We verified directly 
that such small corrections, mitigated in the 
reprocessed data \citep{Bregeon2013:P7REP}, do not significantly affect this analysis. 
This is expected since it is performed on a spatial grid of about 0.9 deg, 
considerably larger than the typical high-energy PSF.

\subsection{Residual cosmic-ray background}
\label{subsec:particle_background}

Charged and neutral CRs misclassified as $\gamma$ rays by
the multivariate event classification algorithms mimic an isotropic
flux that is indistinguishable from the IGRB. In addition, genuine $\gamma$ rays from the Earth's
atmosphere that have directional reconstruction errors sufficient to
bypass the zenith angle veto become a source of apparent extraterrestrial emission over
the full sky. In this work, the term `CR background' includes CR-induced $\gamma$~rays from the atmosphere.

Our estimation of the residual CR background event rate is based on Monte
Carlo simulations of the relevant particle species in the near-Earth environment, namely CR nuclei and
electrons, as well as their atmospheric secondaries.
We simulate both CR backgrounds and signal $\gamma$~rays
and extract the distributions for reconstructed event properties with
the greatest discrimination power at low and high energies respectively: 
the multivariate event classifier output and the transverse shower size. 
The distributions for simulated background and signal are compared to the distributions for the flight
data to quantify the level and associated uncertainty of the CR background.
A detailed description of this method can be found in \cite{Ackermann2012:Performance}. 

To account for atmospheric $\gamma$~rays surviving the zenith angle veto, an updated
phenomenological model for the Earth emission based on LAT
observations is included in the Monte Carlo
simulation. Atmospheric $\gamma$~rays can bypass the
zenith veto either by being reconstructed in the extreme tail of the
PSF, or by entering from the back side of the instrument and
being reconstructed as though coming from the front. Although such
catastrophic mis-reconstructions are rare, the Earth emission is sufficiently
bright that the expected event rate is non-negligible at energies
$\lesssim1$~GeV \citep{Bechtol:2012}. For the stringent zenith angle selections
used in this work, the residual contamination of atmospheric
$\gamma$~rays is expected to be comprised primarily of back-entering
events. The reconstructed directional distribution of back-entering atmospheric
$\gamma$~rays in particular is approximately isotropic. 

\begin{figure}
\epsscale{1.1}
\plottwo{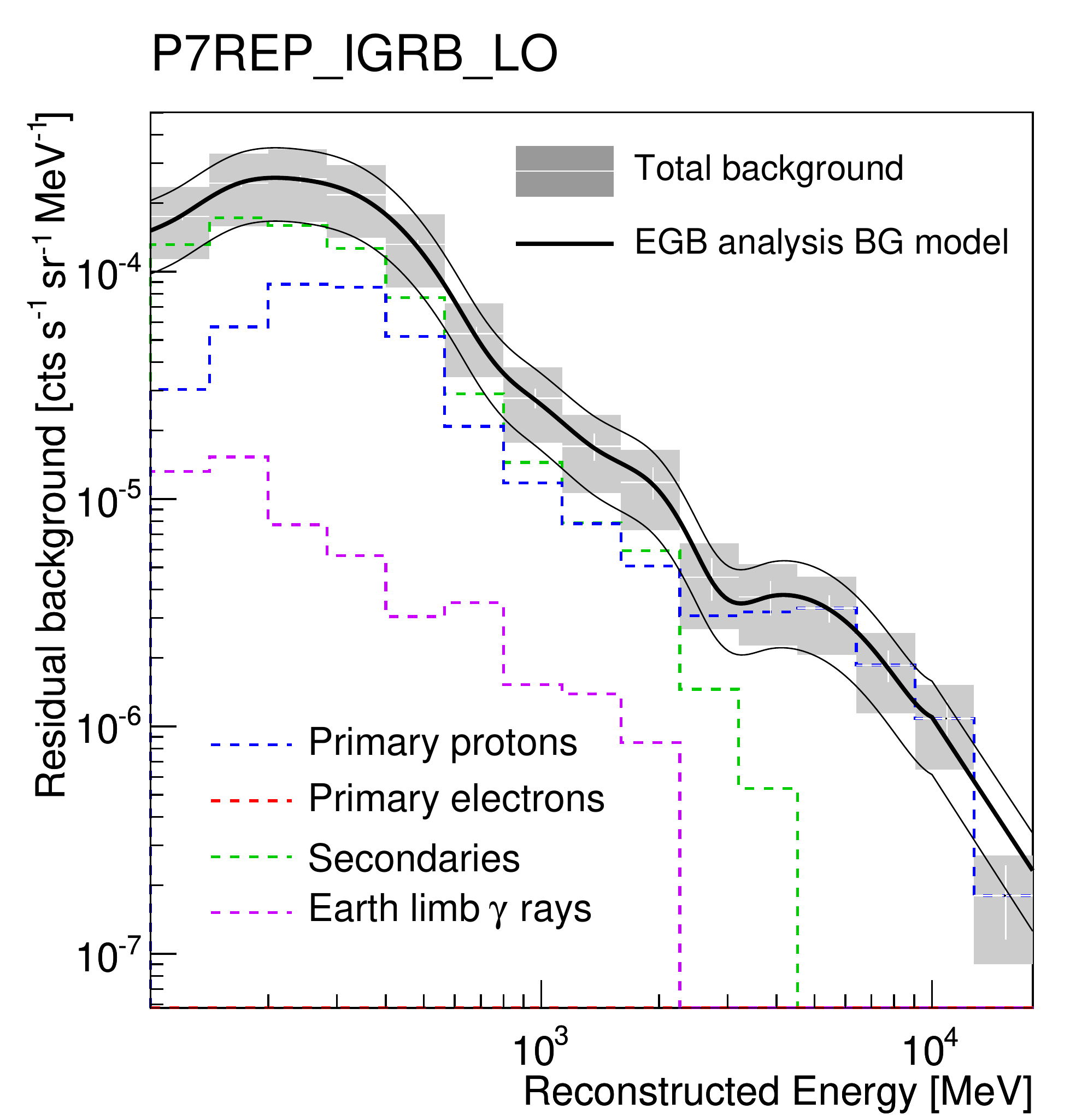}{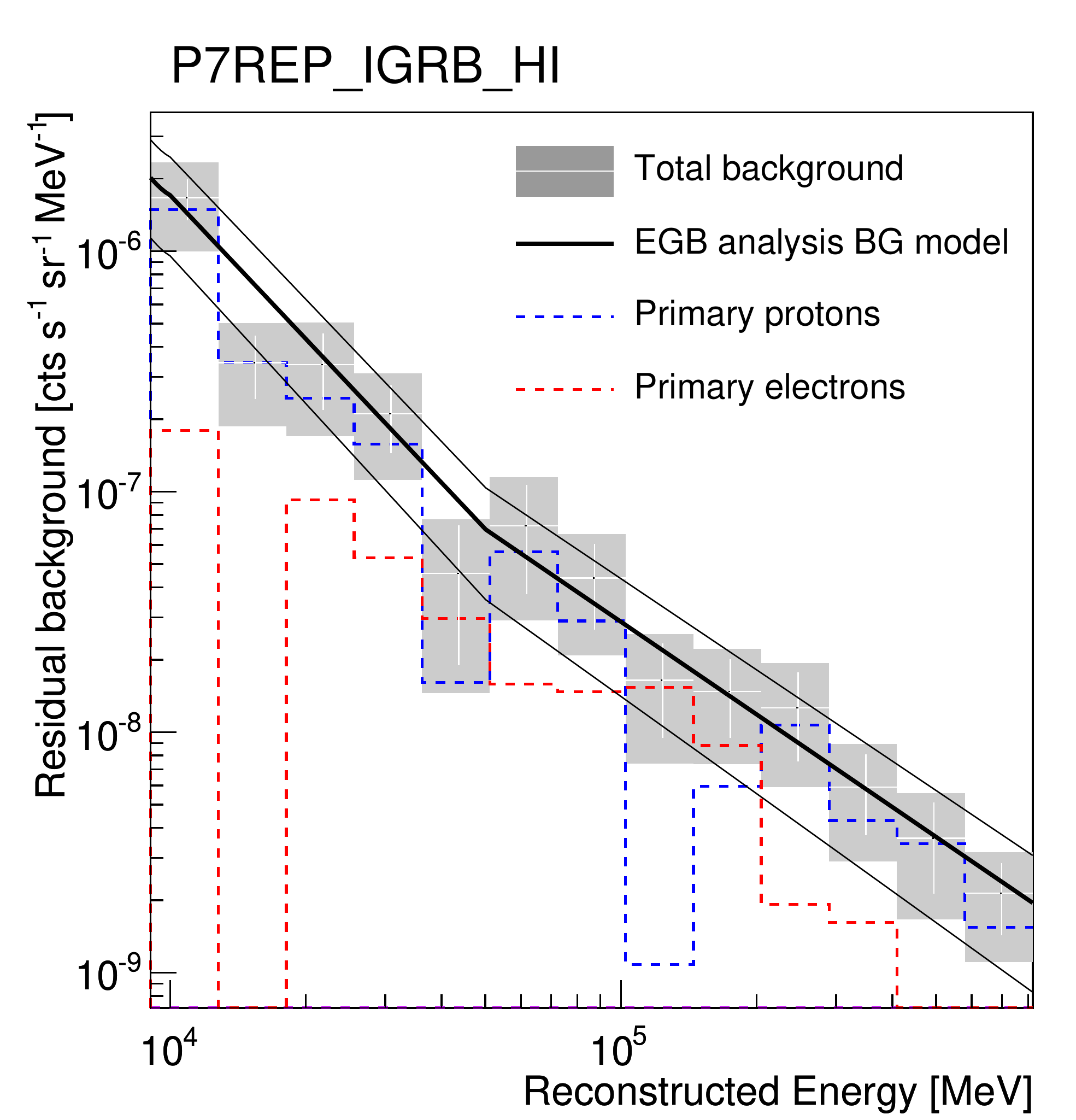}
\caption{Comparison of the rates of residual CR background events
  encountered in the {\tt P7REP\_IGRB\_LO} ({\it left}) and {\tt P7REP\_IGRB\_HI} ({\it right}) event selections.
The individual contributions from primary protons, electrons, secondary CR, and $\gamma$~rays produced
in the atmosphere are shown based on the respective Monte Carlo predictions. The white crosses and 
gray boxes denote the total CR background contamination level including (gray boxes)
 and not including (white crosses) the systematic uncertainties. A  
band consisting of three black lines displays the model which is used for the 
level (thick line) and the uncertainty (thin lines) of the CR background in 
the IGRB analysis. \label{fig:backgroundRates}}
\end{figure}

Figure \ref{fig:backgroundRates} shows the residual CR background
rates as a function of \textit{reconstructed} energy for the {\tt
  P7REP\_IGRB\_LO} and the {\tt P7REP\_IGRB\_HI} classes. Note that the event energy is reconstructed
under the hypothesis of a front-entering $\gamma$~ray and in general does not represent the actual
energy for hadrons. At high energies, primary protons and electrons both contribute
significantly to the CR background. Although protons are far more abundant than
electrons in the environment of the LAT, there is also greater
rejection power against protons since analysis of the shower shape in
the calorimeter can be used to tag and remove protons in addition to the
veto power obtained from the ACD. All contributions shown in Figure
\ref{fig:backgroundRates} have been adjusted from the raw Monte Carlo
predictions based on event property comparisons between the simulated and flight data, as described above.
The CR background contamination after re-scaling agrees with the raw predictions from Monte Carlo 
simulation to within 35\%, depending on energy, and this maximum
discrepancy is used as a measure of the systematic uncertainty in the CR background
contamination. 

The full uncertainty in residual CR background rates, shown 
in Figure \ref{fig:backgroundRates}, has been calculated by adding
systematic and statistical uncertainties in
quadrature. For the {\tt P7REP\_IGRB\_HI} event class at energies
above 10~GeV (relevant for the high-energy sample), the statistical
uncertainties are large due to the limited size of the simulated 
residual background sample\footnote{The existing background rate
estimates were derived from several million CPU hours of CR
simulation. Significant gains in precision might be achieved in the future when more computing power becomes available.}. 
Therefore, instead of using bin-by-bin estimates for 
the CR background rates in the high-energy IGRB analysis, the rates are obtained
from a fit to the (re-scaled) simulated background rates between 10~GeV
and 820~GeV using a simple broken power law function in energy with a
break at 50~GeV. 
A spline interpolation of the background rates is used as the CR 
background model below 10~GeV.

\clearpage

\section{Derivation of the IGRB spectrum}
\label{sec:analysis}

The spectrum of the IGRB is derived in a two-step procedure. First, the
spectrum of the isotropic component is determined as part of a multi-component maximum likelihood fit of
template maps to the observed LAT counts using the tools and 
method described in \citet{Ackermann:2009}. 
This isotropic component is attributed to the sum of the IGRB and
misclassified CR backgrounds (Section \ref{subsec:particle_background}). 
Second, the CR background contribution is subtracted from the
isotropic component to obtain the IGRB spectrum. 

In the multi-component maximum likelihood fit, we create template maps
containing the number of LAT counts expected during the observation period
for various contributions to the \gray{} sky. Each template map is
based on a model or measurements of the respective contribution. 
The \gray{} emission observed by the LAT is modeled using five template maps in addition to the isotropic emission
and 403 point sources that are fitted individually.
 Two template maps describe the DGE. One map is used for the solar disk and 
inverse Compton (IC) emission associated with the solar radiation field. One map describes the
local diffuse emission from Loop~I and the Local Loop. A last map is used to describe contributions from 
established point sources that are not individually fitted. 

The normalization of each template is
fitted individually for each energy bin in the energy range between
100~MeV and 13~GeV using the low-energy event sample. 
This fit is performed by maximizing the likelihood to obtain the
observed number of counts in each pixel given our model. We denote
this fit result as the `low-energy fit'.
Above 13~GeV, the normalizations of the Galactic foreground templates are kept fixed over the full high-energy range,
i.e. we use the spectral shape that is provided in the templates 
to model the foreground in addition to the spatial information. 
Only the normalizations of the isotropic template and the point
sources are fitted for each individual energy bin above 13~GeV. 
To determine the fixed template normalization factors, we first fit
the normalization of each Galactic foreground template in the six
energy bins between 6.4~GeV and 51~GeV using the same procedure as for
the low-energy fit (the number of events above 51~GeV is too low to
robustly fit all foregrounds individually in each
energy bin). 
For each respective foreground template, the
average normalization factor from these six
energy bins is applied to all of the energy bins between 13~GeV and
820~GeV. The bin-by-bin fitting and average-value scaling
procedures yield consistent spectral forms in the 6.4~GeV to 51~GeV
range (cf. Figures \ref{fig:modelGalacticA}, \ref{fig:modelGalacticB}, and \ref{fig:modelGalacticC}
Appendix \ref{sec:appendixA}), providing confidence to
the extrapolation above 51~GeV based on the spectral model shape of
each foreground component.
Moreover, the shapes of the high-energy spectra of the dominant Galactic foreground contributions (i.e.,
the interactions of CRs with ISG and ISRF) can be derived quite robustly based on the well measured 
local CR electron and nucleon spectra. Modeling uncertainties are therefore small.
    
After the Galactic foreground template normalizations are fixed to
these average values, a multi-component maximum likelihood fit is
performed using the high-energy event sample in the energy range between 13~GeV and
820~GeV, which we denote as the `high-energy fit'.

The template maps are
binned on a HEALPix\footnote{\url{http://healpix.jpl.nasa.gov/healpixSoftwareGetHealpix.shtml}} grid of order 6 \citep[$\approx0.9\deg$ pixel size;][]{Gorski:2005} in Galactic
longitude and latitude, and in 26 energy bins 
between 100~MeV and 820~GeV.  
Galactic diffuse emission dominates the intensity of the \gray{}
sky at low Galactic latitudes. This emission originates from 
the interactions of CRs with ISG and the ISRF.
In the former case, $\gamma$ rays are produced through the generation and decay of neutral
pions, or through non-thermal bremsstrahlung; in the latter case,
$\gamma$ rays are 
produced by the IC process. We consider the \gray{} emission due to interactions with ISG
and the ISRF separately in this analysis. The spatial distribution of $\gamma$ rays from
interactions with ISG is well correlated to the column density of ISG
along a given line of sight, whereas $\gamma$ rays created through
interactions with the ISRF form a comparatively smooth emission component peaked
at the Galactic center. We use the GALPROP\footnote{\url{http://galprop.stanford.edu}} code
\citep{Strong:1998,Vladimirov:2011} to obtain templates for these two components. A
detailed discussion of the modeling of the Galactic diffuse 
emission will be presented in Section \ref{sec:foreground} and Appendix \ref{sec:appendixA}.
We refer to the two template maps as the `\hi{}~+~\hii{}' template and the `Inverse Compton' template 
below.

The second prominent contribution to the \gray{} sky is from the individually resolved LAT \gray{} sources. 
We include the 403 sources from the second LAT source catalog
\citep[2FGL;][]{Nolan2012:2FGL} from above and below the Galactic 
plane ($|b|>2\deg$) that are detected with a test statistic (TS) larger than 200 in that catalog as individual 
templates. The 1215 sources with TS values less than 200 are
added to a common template using the spectral information
found in the 2FGL catalog. Additionally, we add a template for a source
at the position of CRATES J231012$-$051421 \citep{Healey:2007} 
after a localized excess in the residual map was found in a first
iteration of this analysis at a position consistent with the CRATES
source. Due to the difference in the observation time between this work
and the 2FGL catalog (50~months vs. 24~months) and the intrinsic 
variability of many high-latitude LAT sources, both extra sources and 
changes in their time averaged spectra can be expected.
However, no systematic search for additional sources too faint 
to be identified on the residual map was performed on the data samples used in this analysis. 
We note that due to the considerably more stringent event selection used in this work in 
comparison to the 2FGL catalog, the effective gain in exposure is much smaller than what 
is indicated by the difference in observation length. The threshold in \gray{} flux for detecting 
sources in this sample would only be marginally lower than the flux threshold for the source sample 
in the 2FGL catalog.

A third contribution to the \gray{} sky that is included as a component in the likelihood fit 
is the \gray{} emission related to the Sun. $\gamma$ rays are produced by
CR collisions with the outer atmosphere of the Sun and by IC
scattering of CR electrons off the solar radiation field. The solar disk and
extended IC emission has been measured using the LAT \citep{Abdo2011:Sun}. We
use the Solar System Tools \citep{Johannesson2013:SolarSystemTools} from the LAT \emph{ScienceTools} (v9r30p0) to
create a template for the time-averaged solar emission in our observation period
which is based on this measurement. This template is denoted as `Solar Disk + IC' throughout this work.
To avoid bias at high energies where the solar spectrum is not well known by
measurements, we do not use the $\pm 3\deg$ region around the ecliptic
plane in the fit for energies above 13~GeV (in the energy
range where the high-energy sample is used). The \gray{} emission from the Moon \citep{Abdo2012:Moon} has been 
neglected in this work since the Moon does not feature an extended IC contribution that could bias a measurement
of the IGRB. 
 
Structures are seen in the diffuse \gray{} emission that are
correlated to Loop~I \citep[seen most prominently in the region of the North Polar Spur;][]{Casandjian:2009}. 
Loop~I is also bright in the 408 MHz radio continuum survey of 
\cite{Haslam:1982} indicating a local overdensity of high-energy electrons or stronger magnetic fields
in that region. A detailed investigation
of the spectrum and spatial distribution of the \gray{} emission from this region has not yet been performed, but
is outside the scope of this paper. We therefore use a simple geometrical model \citep{Wolleben:2007} for 
the synchrotron emission from Loop~I and the Local Loop to generate a template for the \gray{} emission from these 
structures (referred to as the `Loop~I~/~Local Loop'
template). 

Systematic uncertainties associated with the foreground templates mentioned above and other foreground modeling choices, e.g., 
the optional inclusion of an additional template for the \textit{Fermi} Bubbles, are discussed in Section \ref{sec:systematics}.

Certain regions in the vicinity of the Galactic plane have been masked
and the corresponding pixels have not been used in the likelihood
fit. The shape of the mask has been chosen to reduce systematic
uncertainties connected to the Galactic diffuse foreground emission by
excluding regions in which the column density of the ISG is not
dominated by the atomic and ionized hydrogen in the vicinity of our
solar system. Details of the definition of the mask are listed in Appendix \ref{sec:appendixC}.
Figure \ref{fig:DataPlusMask} shows the integrated LAT counts above
100~MeV that are used for this analysis and the excluded regions.

\begin{figure}
\epsscale{0.9}
\plotone{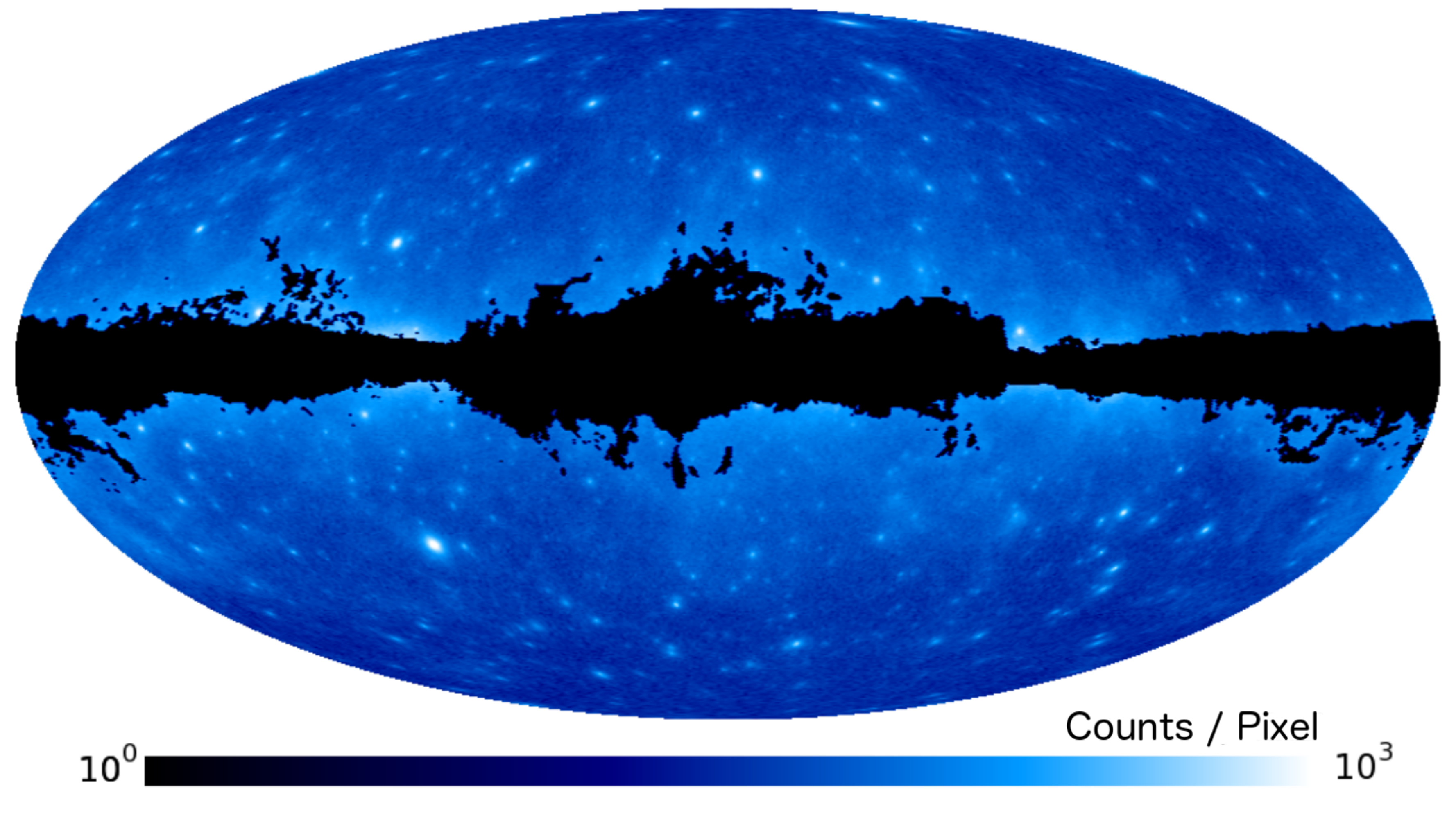}
\caption{Map of counts observed by the \textit{Fermi} LAT above
  100~MeV using a Mollweide projection in Galactic coordinates with a pixel
  scale of $\approx0.9\deg$. The color scale is logarithmic. Overlaid is the
  mask used in this analysis to exclude regions from the template fitting
  procedure (see Appendix \ref{sec:appendixC} for details).
\label{fig:DataPlusMask}}
\end{figure}

\clearpage

\section{Foreground diffuse Galactic emission models}
\label{sec:foreground}

At high Galactic latitudes ($|b| \gtrsim 10\deg$) the ISG is dominated by
atomic gas clouds within a few hundred parsecs, a range in which we do not 
expect a significant change in the density of the CRs that interact with these
clouds \citep{Ackermann2013:LocalCRGradient}. Therefore the spectrum and spatial distribution
of the $\gamma$~rays arising from the CR interactions with atomic gas are 
relatively well constrained by measurements of the gas cloud distributions 
and direct measurements of local CR spectra. The proximity
of ISG seen at high Galactic latitudes permits the use of a single
model template as opposed to multiple templates describing the ISG at
various distances, as is typical for other DGE studies.
Only a small fraction of the ISG-related \gray{} emission arises 
from ionized hydrogen gas, which has a larger scale height than the
atomic gas and a less well known distribution on the sky.

The spectrum and intensity distribution of the IC-related \gray{} emission 
on the sky can only be predicted from a global modeling of 
CR propagation and interaction in the Galaxy.
It is highly dependent on the injection and propagation of electrons in 
the Galactic plane and into the Galactic halo. It further depends on the 
spatially varying spectrum of the ISRF, which is more uncertain than the distribution 
of the ISG.

\subsection{Reference foreground models}
\label{sec:foregroundReference}

An extensive study comparing LAT data to GALPROP-based predictions of the DGE  
has been published in \cite{Ackermann2012:Diffuse}.
However, that study was restricted to a selected set of propagation
and CR injection scenarios. In particular, it was assumed that the diffusion coefficient is constant throughout the
Galaxy and the source population that injects the electrons
is the same as the source population that injects the nuclei. These
choices are well motivated by Occam's razor, but we demonstrate in Appendix \ref{sec:appendixA} 
that template maps for the IC emission that are derived from DGE models of this type
lead to inconsistencies when used in our multi-component likelihood fit.
For example, the spectrum of the IC emission predicted by the model
does not match the spectrum obtained in the fit to the LAT data.
Such a mismatch is critical for the IGRB study if it originates from an inaccurate model 
of the IC intensity distribution on the sky. In this case, the isotropic template could partially compensate for the
inaccuracies of the IC intensity distribution and thereby lead to a biased IGRB measurement.

 
We therefore extend the study of foreground models to include two reference models for
propagation and injection scenarios with more degrees of freedom than 
those considered in \citet{Ackermann2012:Diffuse}. One allows
for distinct populations injecting CR electrons and
nuclei. The other allows a variation of the diffusion
coefficient with radial distance from the Galactic center and height
above the Galactic plane. These two reference models are described in more
detail in Appendix \ref{sec:appendixA}. We denote them as foreground
models `B' and `C' respectively, to distinguish them
from foreground model `A' that is derived from the class
of DGE models studied in \citet{Ackermann2012:Diffuse}. The principal features of and differences between the three 
reference foreground models are summarized in Table \ref{tab:foregroundSummary}.

\begin{deluxetable}{lc}
\tabletypesize{\scriptsize}
\tablecaption{Comparison of Benchmark Galactic Foreground Models}
\tablewidth{0pt}
\tablehead{
\colhead{Foreground} & \colhead{Main features and differences with respect to other DGE models}
}
\startdata
Model A & Sources of CR nuclei and electrons trace pulsar distribution; \\
 & constant CR diffusion coefficient and re-acceleration strength through Galaxy \\
\hline
Model B & Additional electron-only source population near Galactic center, \\
 & these electrons are responsible for majority of IC emission; \\
 & local source of soft CR electrons needed to explain CR electron spectrum at Earth below 20 GV \\
\hline
Model C & Sources of CR nuclei and electrons more centrally peaked than pulsar distribution; \\
 & CR diffusion coefficient and re-acceleration strength vary with Galactocentric radius and height \\
\enddata
\tablecomments{See Appendix \ref{sec:appendixA} for a more detailed description of these three reference DGE models.}
\label{tab:foregroundSummary}
\end{deluxetable}

Our main concern in this work is to investigate whether
the fitted spectral features of the IGRB depend on the specific type of foreground
model chosen. It is not the aim or scope of this work to
perform a quantitative study of whether one of the alternative foreground model
classes matches the LAT data better than another.
For simplicity, we use model A as a baseline for
the purpose of quoting certain results and testing variations of
DGE model parameters. However, we do not view model A as
canonical or preferred over the other models.

All three foreground models assume diffusive CR transport with re-acceleration in the interstellar medium. 
The diffusion coefficient has a power-law dependence on rigidity with index $\delta=0.33$, as 
expected from a Kolmogorov spectrum of magnetic turbulences. 
CR propagation and injection parameters within each model are chosen to obtain good agreement between the predicted local
spectra of interstellar CRs and actual CR measurements after solar modulation effects have been taken into account.
In particular, we require good agreement with the measured proton and helium spectra, the electron spectrum,
and the B/C and \BeTen/\BeNine{} ratios\footnote{One notable exception is the electron spectrum for foreground model B that is 
tuned to reproduce the observed IC emission spectrum rather than the measured CR electron spectrum. 
A detailed description of the various injection and propagation parameters for all three 
foreground models can be found in Appendix \ref{sec:appendixA}.}.

The distribution of \htwo{} and \hi{} gas in the Galaxy is modeled
based on the microwave survey of \citet{Dame:2001} and the radio
survey of \citet{Kalberla:2005}, respectively (see also Appendix \ref{sec:appendixC}). Details regarding the gas distribution
modeling can be found in \citet{Ackermann2012:Diffuse}. An analytic model is used for the distribution 
of ionized hydrogen in the Galaxy \citep{Gaensler:2008}. We use the ISRF model introduced in \cite{Porter:2008}, which
is available within GALPROP. 
We take into account the anisotropy of the ISRF by calculating for
192 uniformly distributed lines-of-sight the ratio between the
predicted IC emission from a full anisotropic calculation and the
prediction assuming that the ISRF is isotropic. This set of ratios is
then interpolated and applied as a multiplicative correction to all the generated IC templates.

It has been shown that interstellar dust can trace gas that is not seen
in \hi{} or CO surveys \citep{Grenier:2005}. 
Therefore we use the E($B-V$) visual reddening map provided by
\citet{Schlegel:1998}, a tracer of the interstellar dust column density, to
estimate the total ISG column density along a line-of-sight. We use the procedure described in
\citet{Ackermann2012:Diffuse} to obtain a conversion factor between the magnitude of reddening in the E($B-V$)
map and the \hi{} gas column density (denoted as \hi-to-dust ratio
below). The procedure uses a linear regression between the E($B-V$)
map and the \hi{} and CO surveys to obtain the \hi-to-dust
ratio in regions of the sky where E($B-V$)$<5$~mag. The fit depends slightly on the
spin temperature $T_S$ that one assumes to correct for the opacity of the 21~cm-line in the \hi{} surveys.
We find a \hi-to-dust ratio of
$7.9\times10^{21}$~cm$^{-2}$~mag$^{-1}$ 
assuming the widely used value of $T_S$=$125$~K for the spin temperature \citep[e.g.,][]{Kulkarni:1988}, 
and use this \hi-to-dust ratio in our analysis. Note that this value
is different from the value used in \citet{Ackermann2012:Diffuse} 
where the higher spin temperature of $T_S=150$~K was used for deriving
this conversion factor. The lower spin temperature used here
leads to smaller residuals between the E($B-V$) and the \hi{} and CO survey
derived gas column densities at high Galactic latitudes that are relevant for the 
IGRB analysis.

\subsection{Additional foreground models used for systematics studies}
\label{sec:foregroundSystematics}

In addition to our three reference foreground models, we consider
further variations of foreground models to assess the systematic uncertainties 
for the derivation of the IGRB that are related to the modeling of the DGE. 
Specifically, we study variations 
of the size of the CR halo between 4~kpc and 10~kpc,
a variation of the distribution of CR sources in Galactocentric radius,
a model where CRs are not re-accelerated in the Galaxy,
models with an extra foreground template for the \textit{Fermi} Bubbles,
a higher radiation field in the Galactic Bulge,
a lower random Galactic magnetic field than in the default models,
and a variation of the \hi-to-dust ratio by 10\%.

Foreground model A serves as the baseline model for these
variations. To assess the stability of the IGRB measurement with respect to assumptions regarding the 
halo size and the CR source distribution we use three models discussed in
\citet{Ackermann2012:Diffuse} chosen to cover the extreme 
values for the CR halo size (4~kpc vs. 10~kpc) and radial source
distribution (traced by pulsars vs. traced by SNRs) in the range of 
models studied there. 

\citet{Strong:2011} suggest that CR propagation models where CRs are not re-accelerated in the 
Galaxy (so called plain diffusion models) describe the synchrotron emission observed from the Galaxy better
than present models with re-acceleration. We therefore include such a plain diffusion model in our investigations.

\citet{Dobler:2010} and \citet{Su:2010} have noted the existence of large-scale structures of residual diffuse
\gray{} emission above and below the Galactic center region that became well known as the `\textit{Fermi} Bubbles'.
Although well established as significant features, the Bubbles were not included in the reference foreground models.
The \textit{Fermi} Bubbles have been studied exclusively as a residual
after subtracting a model of the DGE and isotropic diffuse emission. No
template for their shape has yet been derived from independent observations. 
Using a template derived from a strictly empirical excess of \gray{} emission
might lead to a bias in the other components of the fit, including the isotropic template.
Since neglecting the emission from the \textit{Fermi} Bubbles
might bias the fit as well, we tested the effects of including template maps for the \textit{Fermi} Bubbles in
the multi-component fit. Two models for the intensity distribution of 
the \textit{Fermi} Bubbles on the sky were tested. The first is a simple geometric template used for 
the investigation of systematic uncertainties in studies of the \gray{} emission from SNRs \citep{DePalma:2013}. 
The second is a template derived from the residual \gray{} emission \citep{Ackermann2014:Lobes}.


Our knowledge of the ISRF in the inner Galaxy is limited. We therefore repeat the IGRB fit against an ISRF model 
that assumes a factor ten higher stellar luminosity in the Galactic
Bulge. Such a model is still compatible with constraints derived
from observations of the ISRF in the solar neighborhood. 

We also test the impact of the assumed random magnetic field strength
by generating a foreground model with a lower random magnetic field
strength of 3~$\mu$G in the solar neighborhood, to compare with the reference models A and B
which use a value of 7.5~$\mu$G for the random magnetic field there 
(see in this context Appendix \ref{sec:foregroundA}). 

The difference in the \hi-to-dust ratio between the spin temperature value of
$T_{S}=150$~K adopted in \citet{Ackermann2012:Diffuse} and the
widely used spin temperature value of $T_{S}=125$~K that was used in
this work is of order 10~\%. We test foreground models with
variations of the \hi-to-dust ratios by $\pm$10~\% to determine the
impact of this parameter on the IGRB measurement.

The impact of all described variations in modeling the foreground DGE
on the spectrum of the IGRB is discussed in Section \ref{sec:systematics},
together with a model-independent study to assess the impact of un-modeled residuals in the foreground emission.

\clearpage

\section{Results}\label{sec:results}

\subsection{IGRB spectrum}

The likelihood fitting technique introduced in Section \ref{sec:analysis} is used to derive the 
spectrum of the isotropic emission for the three different DGE foreground 
models described in Section \ref{sec:foreground}. The residual
particle background contamination is subtracted from the 
isotropic component to obtain the spectrum of the IGRB.
 
\begin{figure}
\epsscale{1.1}
\plottwo{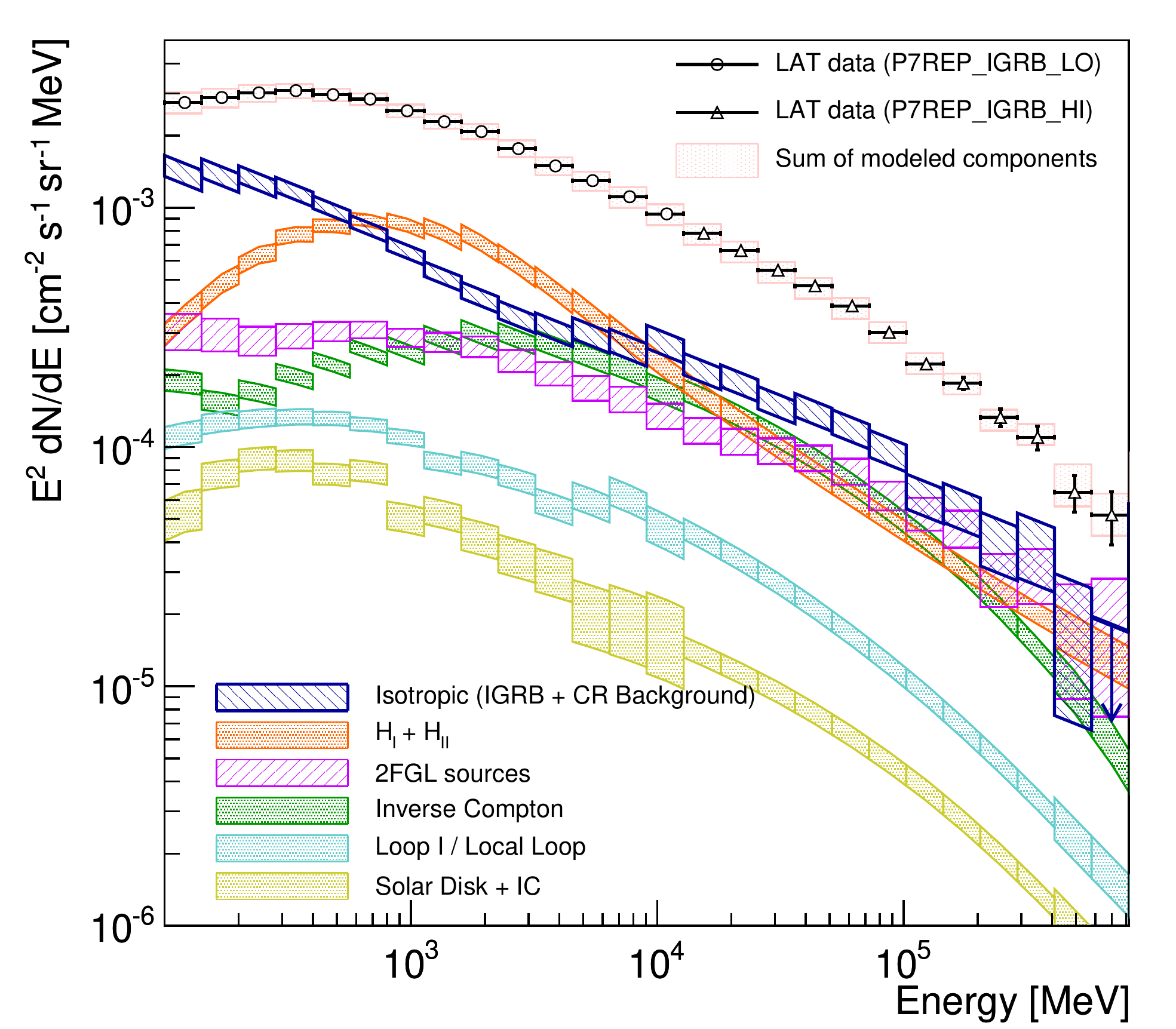}{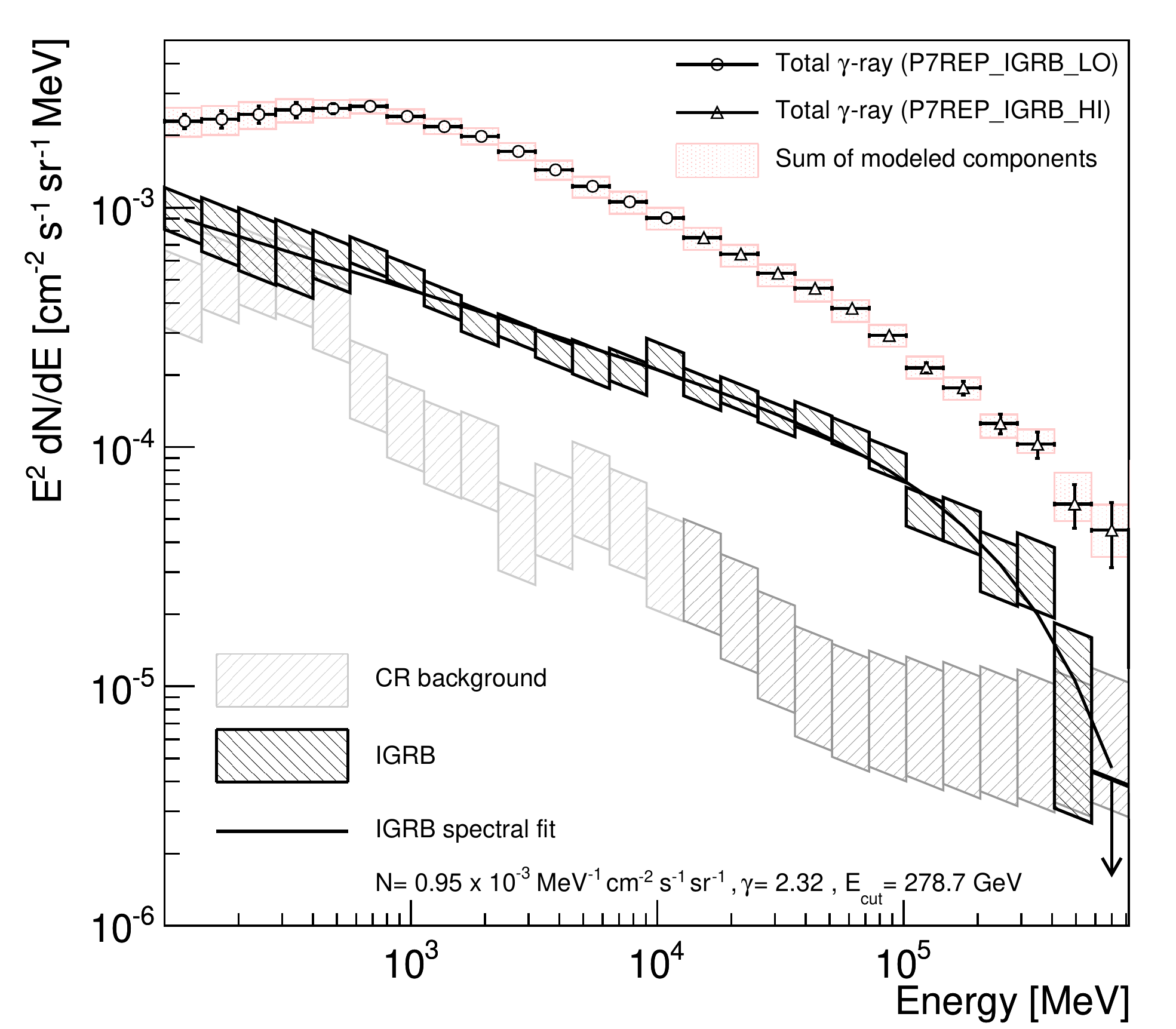}
\caption{Results of the IGRB fit for foreground model A. Average intensities for Galactic 
latitudes $|b|>20\deg$ are shown.
{\it Left:} Intensities attributed to the different foreground templates, the isotropic emission,
and the individually resolved sources in the multi-component likelihood fit. The isotropic emission and the 
individually resolved sources are fitted separately in each energy bin. All other components  
are fitted individually in each energy bin below 13~GeV and included with fixed normalizations above this energy.
The total intensity obtained from the IGRB fit is compared to the total intensity observed by the LAT.
Error bars include statistical errors as well as systematic errors from the uncertainty in the LAT effective area
parametrization.  See text for details.
{\it Right:} IGRB and CR background contributions to the isotropic emission. The line indicates the best-fit
IGRB spectrum with a power-law plus exponential cutoff spectral model. Spectral parameters are given in the 
legend. The total intensities obtained from the fit and measured by the LAT are shown with
the CR background contributions subtracted in this graph.}
\label{fig:fitModelA}
\end{figure}

\begin{figure}
\epsscale{1.1}
\plottwo{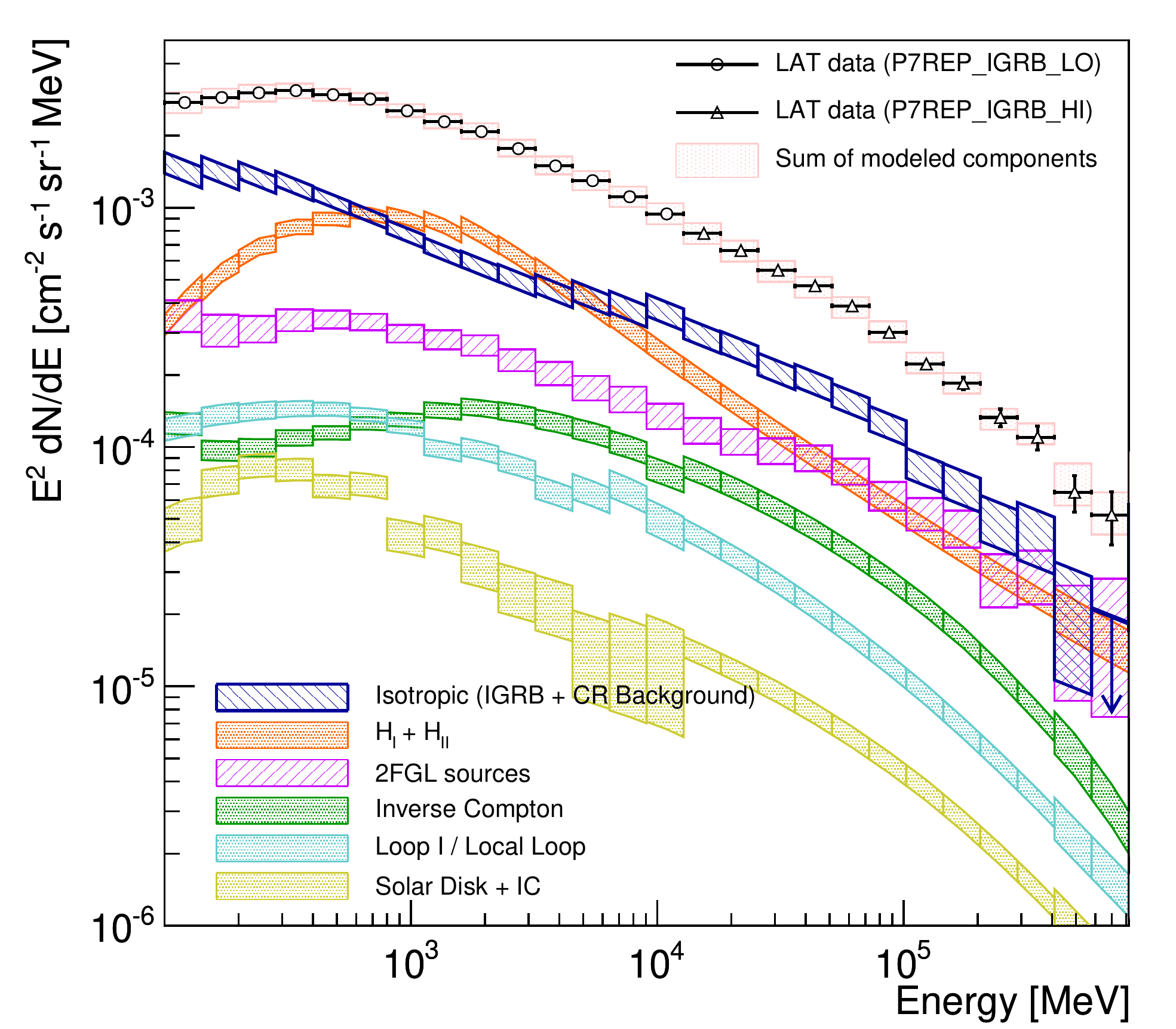}{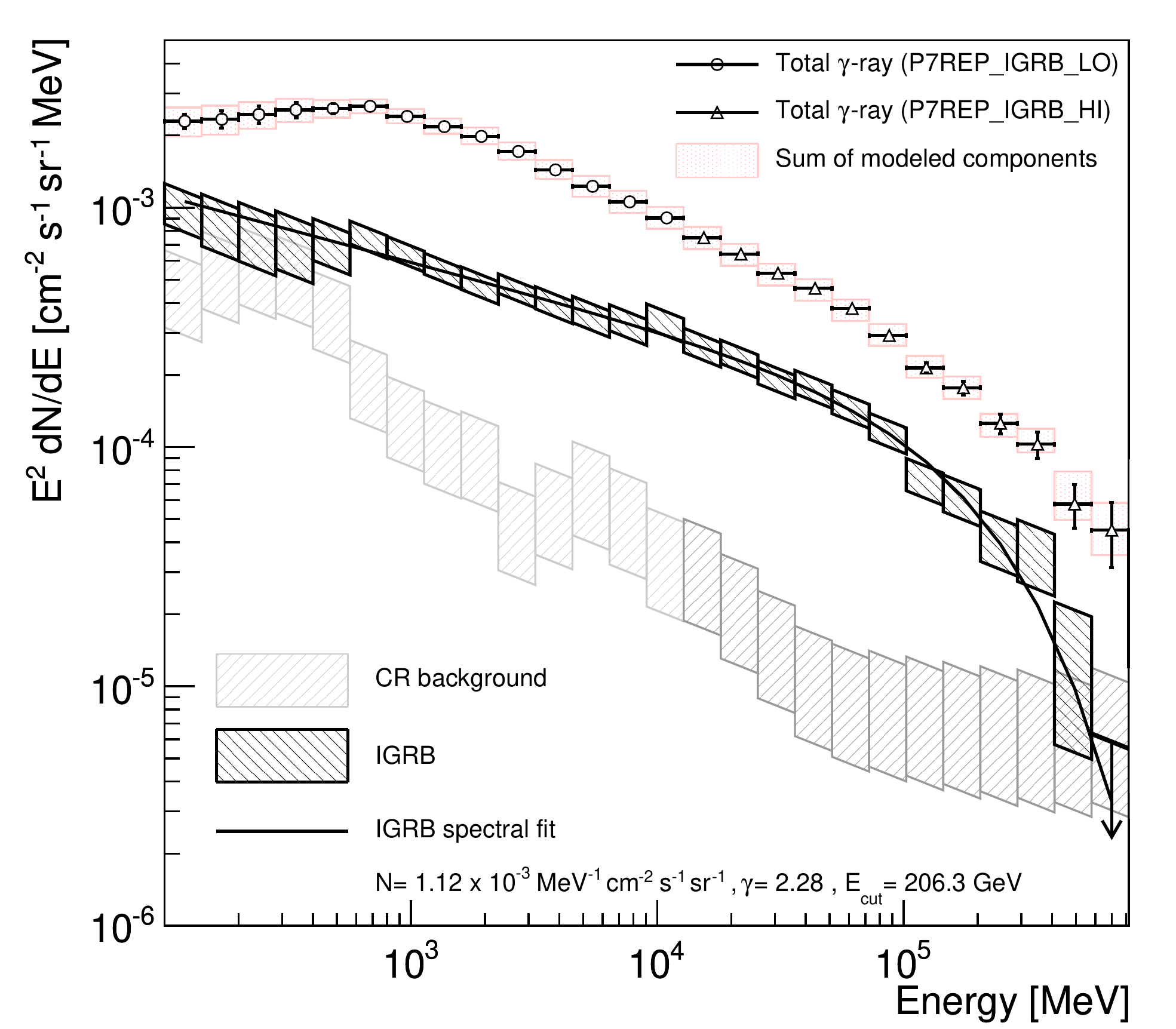}
\caption{Results of the IGRB fit for foreground model B. See Figure \ref{fig:fitModelA} for legend.}
\label{fig:fitModelB}
\end{figure}

\begin{figure}
\epsscale{1.1}
\plottwo{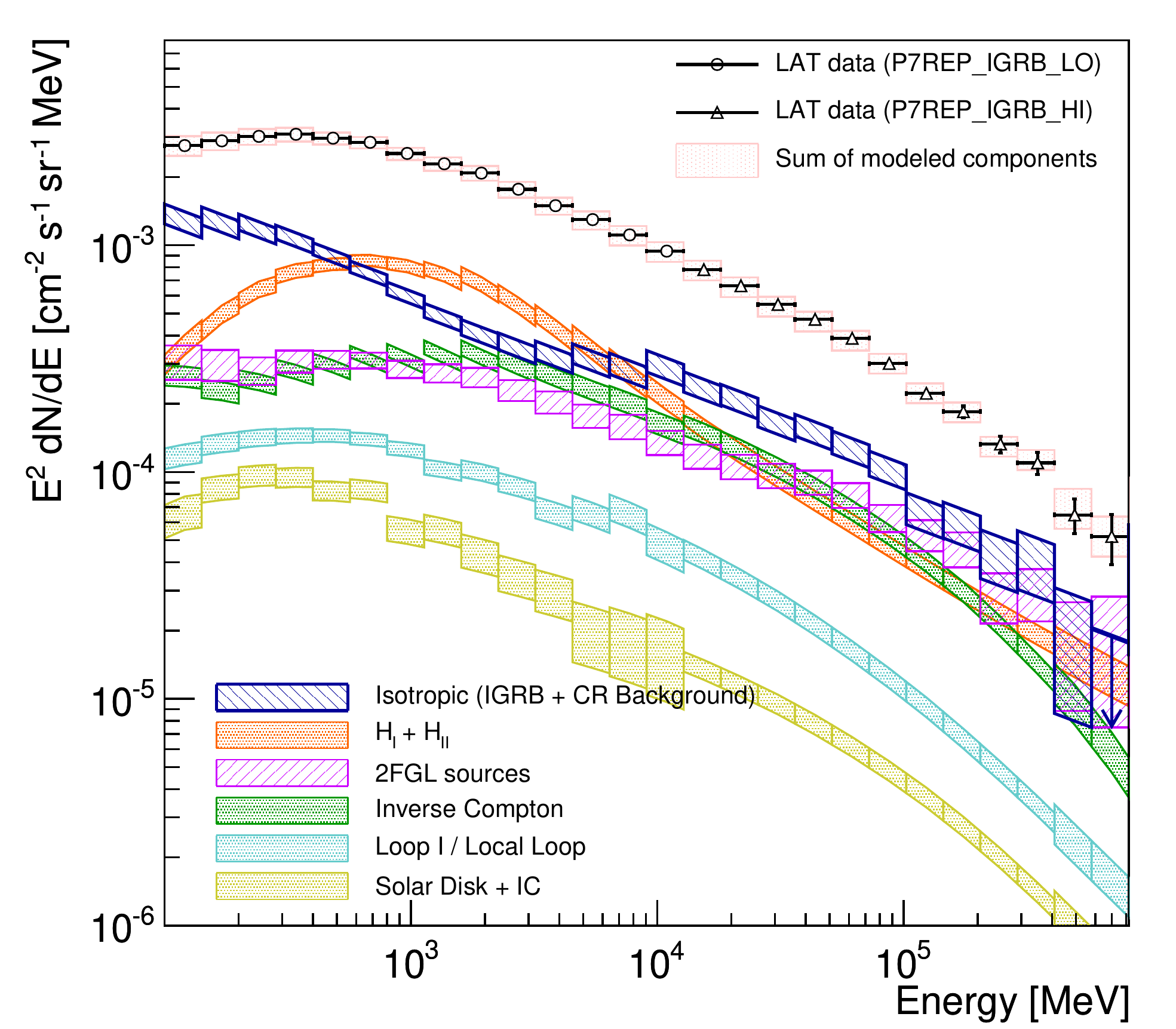}{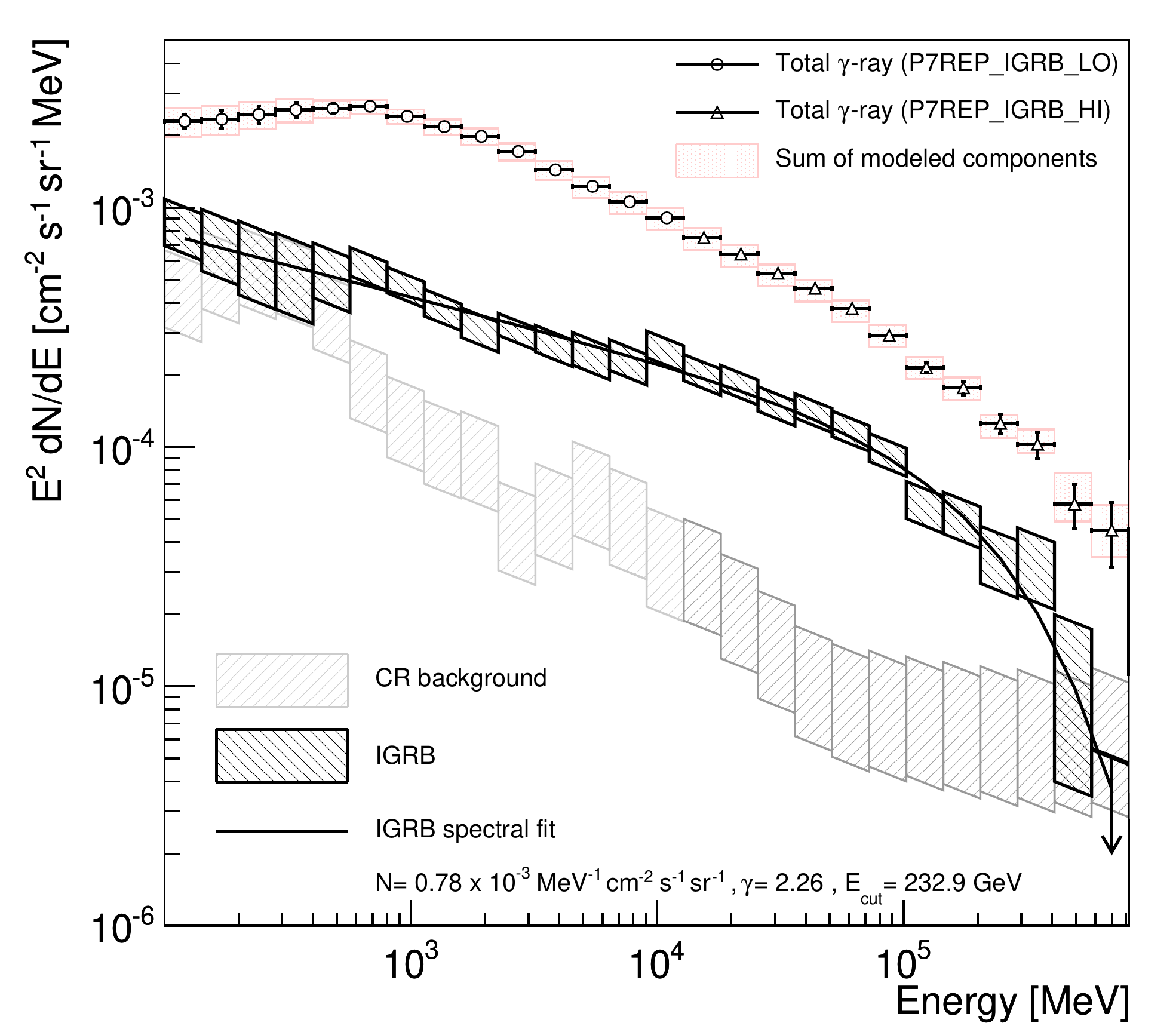}
\caption{Results of the IGRB fit for foreground model C. See Figure \ref{fig:fitModelA} for legend.}
\label{fig:fitModelC}
\end{figure}

Figures \ref{fig:fitModelA}, \ref{fig:fitModelB}, and \ref{fig:fitModelC}
show the results of the fits using foreground models A, B, and C, respectively. 
Each figure displays the average high-latitude ($|b|>20\deg$) intensities
attributed to the isotropic emission (IGRB plus CR background), 
the individual sources, two DGE components, the solar emission, and the local foreground templates. 
The sum of these intensities is compared to the average \gray{} intensity observed by 
the LAT. A separate graph shows the contributions of the IGRB and of the residual CR contamination
to the isotropic emission.
Since the isotropic and IGRB intensities in the highest energy band are compatible with zero 
within the $1\sigma$~uncertainty range we quote upper limits in that energy band\footnote{Statistical 
uncertainties on the isotropic emission have been calculated using the MINOS algorithm of the
MINUIT minimization package \citep{James:1975}. The position of the upper limit 
corresponds to the upper bound of the $1\sigma$ uncertainty interval.}.
The error bars displayed for the individual components in the three figures include the statistical
uncertainty and the systematic uncertainty of the effective area parametrization \citep{Ackermann2012:Performance}
added in quadrature. The error bars for the IGRB component
additionally contain the systematic uncertainty due to subtracting a not perfectly known CR background
contamination, also added in quadrature. 

The IGRB intensities corresponding to foreground models A, B, and C
are compared in Figure \ref{fig:igrb_spectrum_comparison}.
Numerical values for the IGRB intensities per energy band when using foreground model A are presented in
Table \ref{tab:IGRBResults}. Intensities for other 
foreground models can be found in the electronic supplementary material to this article. The IGRB
intensity shows a clear cutoff at high energies, independent of the foreground model. 
A $\chi^2$ regression of the IGRB spectrum using a power law with
exponential cutoff (PLE) spectral model of the form 

\begin{equation}
\frac{dN}{dE} = I_{100} \; \left ( \frac{E}{100\,\mathrm{MeV}} \right )^{-\gamma} \,
\exp \left ( \frac{-E}{E_{\mathrm{cut}}} \right )
\label{equ:expcutoff} 
\end{equation}

\noindent results in low $\chi^2$ values for all three foreground models
and can therefore be considered suitable to characterize the IGRB spectrum. 
We further try to fit the IGRB intensities with a single power law (PL) and a 
smoothly connected broken power law (BPL).
The fit parameters for the PLE model, as well as the $\chi^2$ values for all
fitted spectral hypotheses are summarized in Table \ref{tab:FitResults}. 
The $\chi^2$ values for the BPL and the PLE spectral models are similar enough 
that the two hypotheses are indistinguishable in the energy range observed. We prefer to quote
fitted parameter values for the PLE model given its lower
number (three vs. four) of free parameters.  The PL model is disfavored 
independent of the foreground model based on the high $\chi^2$ values of 
$88$~(foreground model A), $151$~(foreground model B), and $106$~(foreground model C)
for 23~degrees of freedom. Note that the $\chi^2$ value cannot be
easily interpreted in terms of a significance for the agreement between spectral model and data because the error
bars of the IGRB spectrum are systematics dominated over most of the energy 
range, and therefore correlations between bins are expected. These correlations are also 
responsible for the rather small $\chi^2$ values (when compared to the number of degrees of freedom) for the
PLE and BPL spectral models.

The large difference in $\chi^2$ between the PL and the PLE models even when neglecting bin-to-bin correlations 
can still be interpreted as a robust evidence against a simple power-law spectrum. For the benchmark models this $\chi^2$  
difference is larger than 61 at just one added degree of freedom in the model. We also calculated the $\chi^2$  
differences between the PL and PLE models for the additional foreground models used in the investigation 
of the foreground related systematics that are summarized in Table 4. The $\chi^2$ difference between the PL and PLE 
models is 45 or larger in all of these foreground variations.

Finally we calculated the $\chi^2$ difference between PL and PLE models if we add the foreground model related systematic 
uncertainties to the instrument related uncertainties. Since we do not know the correlations between the bins introduced 
by the systematic errors we adopt a worst-case scenario here assuming that the dominant fraction of the systematic error 
is fully correlated at low energies and anti-correlated between low and high energies (E~$>$300~GeV). Such a hypothetical 
anti-correlation in the systematics might artificially enhance indications of a cutoff. Even in this worst-case scenario 
we still find $\chi^2$ differences exceeding 25 between the PL and PLE scenarios for our 3 benchmark models.

Residual maps of the relative deviations in intensity between model and data in different regions of the sky can be found
in Appendix \ref{sec:appendixB}. None of the models considered is a perfect description of 
the data and large-scale residuals at the 25\% level appear in many parts of the sky. 
No Galactic foreground model is unambiguously preferred 
over the others based on the level and distribution of the residual \gray{} emission. Prominent residuals 
include the \textit{Fermi} Bubbles, but
other un-modeled residuals appear in different parts of the sky.
Also, there seems to be less intensity observed in the south polar
region than its northern counterpart, a feature that cannot be modeled
with the classes of foreground models considered here which exhibit a
symmetrical CR density about the Galactic plane by construction. 
 
In the next section, we present a study of the general impact of such un-modeled residuals on the 
IGRB spectrum. The study is not restricted to the \textit{Fermi} Bubbles, but
applies to all features in the residual maps.

\subsection{Systematic uncertainties from foreground modeling}
\label{sec:systematics}

The fitting procedure is applied to the variants of the foreground models introduced in
Section \ref{sec:foregroundSystematics} in the same way as for the benchmark models A,~B, and C.
Table \ref{tab:sysModelVar} summarizes the spectral parameter and $\chi^2$
values obtained for the IGRB when using each of the foreground model variants.
For most of the variant models, the general shape of the IGRB
spectrum is not affected; a simple power law with exponential cutoff remains a valid parametrization. 
The single exception is the 
effect from changing the source distribution. When using the distribution of
SNRs as a tracer of the CR source density, a second apparent spectral feature
manifests as a dip in the IGRB spectrum at a few GeV. This dip in the IGRB 
compensates the higher intensity
of the IC emission with respect to other foreground models.
A good fit of the shape of the IGRB spectrum can be obtained only by 
describing it as a sum of two components. We choose two components
each having a power-law plus exponential cutoff spectrum to describe 
the resulting IGRB spectrum for this foreground model version.

A second investigation of systematic uncertainties is aimed at the residuals visible in the maps 
shown in Appendix \ref{sec:appendixB}.
We study the variations of differences between LAT data and our model within four very 
high-latitude ($|b|>60\deg$) overlapping regions in which the IGRB is responsible for a substantial fraction of 
the total \gray{} emission. We use the Galactic
north pole, the Galactic south pole, the $|b|>60\deg$ region 
facing the inner Galaxy, and the $|b|>60\deg$ region facing the outer Galaxy. For each of these regions, we 
calculate the spectral residual and the renormalization factor for the
IGRB needed to obtain the best overall agreement between
model and data in the corresponding region. The resulting adjustment
factors are between 0.7 and 0.95 depending on the region and foreground model. Note that all adjustment factors 
are $<1$, i.e., the observed \gray{} intensity at $|b|>60\deg$ is lower than that predicted by the models. 
This overestimation of the \gray{} intensity could arise
either from an overestimation of the Galactic foreground at high latitudes or from 
an overestimation of the isotropic intensity, and therefore represents a 
systematic uncertainty for our analysis.

\begin{figure}
\epsscale{1.0}
\plotone{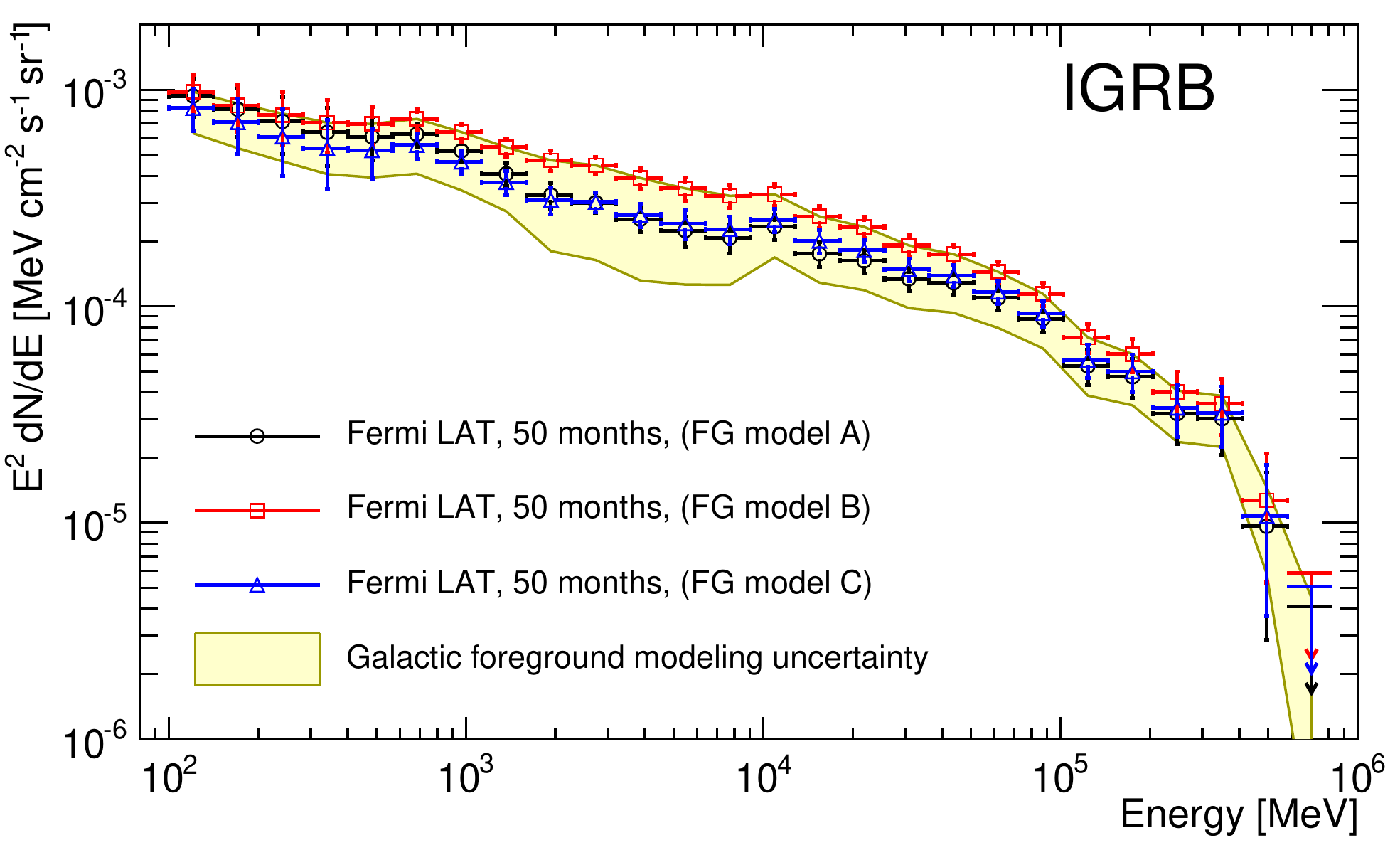}
\caption{Comparison of the derived IGRB intensities for different
  foreground (FG) models. The error bars include the statistical
uncertainty and systematic uncertainties from the effective area
  parametrization, as well as the CR background subtraction (statistical
and systematic uncertainties have been added in quadrature). The shaded
  band indicates the systematic uncertainty arising from uncertainties
in the Galactic foreground: the IGRB intensity
  range spanned by the three benchmark models, the variants described in Section \ref{sec:foregroundSystematics},
  and the normalization uncertainties derived from the high-latitude
  data/model comparison. See Section \ref{sec:systematics} for
  details.}
\label{fig:igrb_spectrum_comparison}
\end{figure}

Including this second study, we find that the normalization of the
IGRB intensity $I_{>100}$ varies by $+15$\%/$-30$\% with
respect to foreground model A depending on our assumptions
about the DGE foreground, while the spectral index varies between 2.26 and 2.34, 
and the cutoff energy between 206 and 374~GeV.
Summarizing the results above, the systematic uncertainty in the IGRB spectrum associated 
with foreground modeling is shown in Figure~\ref{fig:igrb_spectrum_comparison}.

\subsection{Total EGB intensity}


The total EGB as defined in Section \ref{sec:introduction} is a
quantity independent of the instrument and observation time. For the purpose of this 
analysis, we call the sum of the IGRB and the
sky-averaged intensity of high-latitude $|b|>20\deg$ resolved LAT sources the total EGB. 
We note that this definition of the total EGB formally includes a small
fraction of Galactic sources, e.g., millisecond pulsars.
However, the contributions of known high-latitude pulsars extracted from the second LAT pulsar
catalog \citep{Abdo2013:2PC} is less than 5\% of the 
total EGB anywhere in the measured energy range, well below
the level of systematic uncertainty inherent to the IGRB measurement. 
We further check the variation of the total intensity when varying the latitude threshold from 
$|b|>20\deg$ to $|b|>40\deg$ to probe a possible Galactic source contamination. 
We find that the derived intensities are consistent within the
respective uncertainties for each latitude threshold.

\begin{figure}
\epsscale{1.0}
\plotone{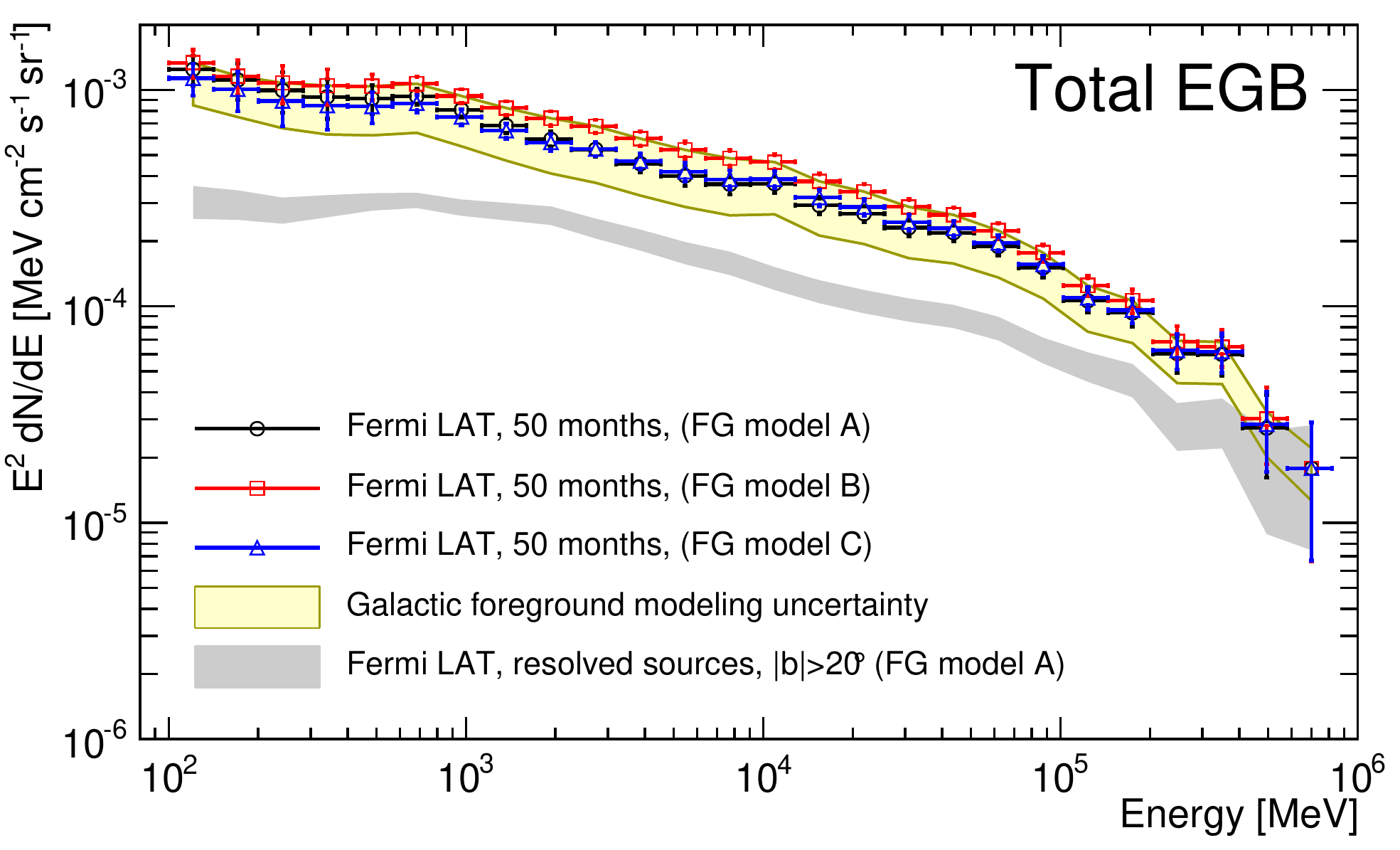} 
\caption{Comparison of the total EGB intensities for different
  foreground models. The total EGB intensity is obtained by summing the IGRB intensity and the 
  cumulative intensity from resolved \textit{Fermi} LAT sources at latitudes $|b|>20\deg$ (gray band).
  See Figure \ref{fig:igrb_spectrum_comparison} for legend. }
\label{fig:egb_spectrum_comparison}
\end{figure}

Figure \ref{fig:egb_spectrum_comparison} compares the total EGB derived for foreground 
models A, B, and C, respectively. Numerical values for the total EGB intensities per energy band are
given in Table \ref{tab:IGRBResults}.
Again, we use a $\chi^2$-regression to
test different functional parametrizations of the spectrum. 
The best-fit parameters for a fit with a power law with an exponential
cutoff as well as $\chi^2$ values for all tested spectral 
models are summarized in Table \ref{tab:TotalFitResults}. For the
total EGB we find similarly as for the IGRB that a power law with an exponential cutoff describes the spectral shape 
significantly better than an unbroken
power law. The cutoff energy is higher for
the total EGB than for the IGRB. As in the case of the IGRB, 
we cannot distinguish an exponential cutoff spectral model from a
broken power law. Results of the spectral fits are summarized
in Table \ref{tab:TotalFitResults}.

Systematic uncertainties in the total EGB spectrum arising from modeling the Galactic foreground
are indicated by the shaded band in Figure \ref{fig:egb_spectrum_comparison}, constructed using the identical methods
described in Section \ref{sec:systematics} for the IGRB.

\clearpage


\clearpage

\clearpage


\newcommand{\tvect}[2]{\ensuremath{\begin{smallmatrix}#1\\#2\end{smallmatrix}}}

\begin{deluxetable}{cccccc}
\tabletypesize{\scriptsize}
\tablecaption{IGRB and total EGB intensities}
\tablewidth{0pt}
\tablehead{
\colhead{Energy range} &  
\colhead{IGRB} & 
\colhead{FG model uncert.} &
\colhead{Total EGB} &
\colhead{FG model uncert.} & 
\colhead{Sources $|b|>20^{\deg}$} \\
\colhead{$[$GeV$]$} & 
\colhead{intensity} &
\colhead{on IGRB} &
\colhead{intensity} &
\colhead{on total EGB} &
\\
&
\colhead{$[$cm$^{-2}$s$^{-1}$sr$^{-1}]$} &
\colhead{$[$cm$^{-2}$s$^{-1}$sr$^{-1}]$} &
\colhead{$[$cm$^{-2}$s$^{-1}$sr$^{-1}]$} &
\colhead{$[$cm$^{-2}$s$^{-1}$sr$^{-1}]$} &
\colhead{$[$cm$^{-2}$s$^{-1}$sr$^{-1}]$} 
}
\startdata

$0.10$ -- $0.14$ & 
$( 2.8 \pm 0.6 ) \times 10^{-6}$ & 
$\tvect{+0.1}{-0.9} \times 10^{-6}$ & 
$( 3.7 \pm 0.6 ) \times 10^{-6}$ & 
$\tvect{+0.3}{-1.2} \times 10^{-6}$ & 
$( 9.0 \pm 1.6 ) \times 10^{-7}$ \\ 
$0.14$ -- $0.20$ & 
$( 1.7 \pm 0.4 ) \times 10^{-6}$ & 
$\tvect{+0.1}{-0.6} \times 10^{-6}$ & 
$( 2.3 \pm 0.4 ) \times 10^{-6}$ & 
$\tvect{+0.1}{-0.8} \times 10^{-6}$ & 
$( 6.2 \pm 1.0 ) \times 10^{-7}$ \\ 
$0.20$ -- $0.28$ & 
$( 1.1 \pm 0.3 ) \times 10^{-6}$ & 
$\tvect{+0.1}{-0.4} \times 10^{-6}$ & 
$( 1.5 \pm 0.3 ) \times 10^{-6}$ & 
$\tvect{+0.1}{-0.5} \times 10^{-6}$ & 
$( 4.1 \pm 0.6 ) \times 10^{-7}$ \\ 
$0.28$ -- $0.40$ & 
$( 6.7 \pm 2.0 ) \times 10^{-7}$ & 
$\tvect{+0.7}{-2.4} \times 10^{-7}$ & 
$( 9.7 \pm 2.0 ) \times 10^{-7}$ & 
$\tvect{+1.2}{-3.2} \times 10^{-7}$ & 
$( 3.0 \pm 0.4 ) \times 10^{-7}$ \\ 
$0.40$ -- $0.57$ & 
$( 4.5 \pm 1.0 ) \times 10^{-7}$ & 
$\tvect{+0.7}{-1.6} \times 10^{-7}$ & 
$( 6.7 \pm 1.0 ) \times 10^{-7}$ & 
$\tvect{+0.9}{-2.2} \times 10^{-7}$ & 
$( 2.2 \pm 0.2 ) \times 10^{-7}$ \\ 
$0.57$ -- $0.80$ & 
$( 3.3 \pm 0.4 ) \times 10^{-7}$ & 
$\tvect{+0.6}{-1.1} \times 10^{-7}$ & 
$( 4.9 \pm 0.4 ) \times 10^{-7}$ & 
$\tvect{+0.7}{-1.6} \times 10^{-7}$ & 
$( 1.6 \pm 0.1 ) \times 10^{-7}$ \\ 
$0.80$ -- $1.1$ & 
$( 1.9 \pm 0.2 ) \times 10^{-7}$ & 
$\tvect{+0.4}{-0.7} \times 10^{-7}$ & 
$( 3.0 \pm 0.2 ) \times 10^{-7}$ & 
$\tvect{+0.5}{-1.0} \times 10^{-7}$ & 
$( 1.1 \pm 0.1 ) \times 10^{-7}$ \\ 
$1.1$ -- $1.6$ & 
$( 1.1 \pm 0.1 ) \times 10^{-7}$ & 
$\tvect{+0.3}{-0.4} \times 10^{-7}$ & 
$( 1.8 \pm 0.1 ) \times 10^{-7}$ & 
$\tvect{+0.4}{-0.6} \times 10^{-7}$ & 
$( 7.1 \pm 0.7 ) \times 10^{-8}$ \\ 
$1.6$ -- $2.3$ & 
$( 6.0 \pm 0.8 ) \times 10^{-8}$ & 
$\pm 2.7 \times 10^{-8}$ & 
$( 1.1 \pm 0.1 ) \times 10^{-7}$ & 
$\pm 0.3 \times 10^{-7}$ & 
$( 4.8 \pm 0.5 ) \times 10^{-8}$ \\ 
$2.3$ -- $3.2$ & 
$( 3.9 \pm 0.4 ) \times 10^{-8}$ & 
$\tvect{+1.9}{-1.8} \times 10^{-8}$ & 
$( 6.9 \pm 0.5 ) \times 10^{-8}$ & 
$\tvect{+1.9}{-2.1} \times 10^{-8}$ & 
$( 3.0 \pm 0.3 ) \times 10^{-8}$ \\ 
$3.2$ -- $4.5$ & 
$( 2.3 \pm 0.3 ) \times 10^{-8}$ & 
$\tvect{+1.3}{-1.1} \times 10^{-8}$ & 
$( 4.2 \pm 0.4 ) \times 10^{-8}$ & 
$\tvect{+1.3}{-1.2} \times 10^{-8}$ & 
$( 1.9 \pm 0.2 ) \times 10^{-8}$ \\ 
$4.5$ -- $6.4$ & 
$( 1.5 \pm 0.2 ) \times 10^{-8}$ & 
$\tvect{+0.8}{-0.6} \times 10^{-8}$ & 
$( 2.6 \pm 0.3 ) \times 10^{-8}$ & 
$\tvect{+0.8}{-0.7} \times 10^{-8}$ & 
$( 1.1 \pm 0.1 ) \times 10^{-8}$ \\ 
$6.4$ -- $9.1$ & 
$( 9.6 \pm 1.5 ) \times 10^{-9}$ & 
$\tvect{+5.4}{-3.8} \times 10^{-9}$ & 
$( 1.7 \pm 0.2 ) \times 10^{-8}$ & 
$\pm 0.5 \times 10^{-8}$ & 
$( 7.3 \pm 0.9 ) \times 10^{-9}$ \\ 
$9.1$ -- $13$ & 
$( 7.6 \pm 1.0 ) \times 10^{-9}$ & 
$\tvect{+3.1}{-2.1} \times 10^{-9}$ & 
$( 1.2 \pm 0.1 ) \times 10^{-8}$ & 
$\pm 0.3 \times 10^{-8}$ & 
$( 4.4 \pm 0.5 ) \times 10^{-9}$ \\ 
$13$ -- $18$ & 
$( 4.0 \pm 0.5 ) \times 10^{-9}$ & 
$\tvect{+2.0}{-1.1} \times 10^{-9}$ & 
$( 6.8 \pm 0.6 ) \times 10^{-9}$ & 
$\tvect{+2.0}{-1.9} \times 10^{-9}$ & 
$( 2.7 \pm 0.3 ) \times 10^{-9}$ \\ 
$18$ -- $26$ & 
$( 2.6 \pm 0.3 ) \times 10^{-9}$ & 
$\tvect{+1.2}{-0.7} \times 10^{-9}$ & 
$( 4.4 \pm 0.4 ) \times 10^{-9}$ & 
$\pm 1.2 \times 10^{-9}$ & 
$( 1.7 \pm 0.2 ) \times 10^{-9}$ \\ 
$26$ -- $36$ & 
$( 1.6 \pm 0.2 ) \times 10^{-9}$ & 
$\tvect{+0.7}{-0.4} \times 10^{-9}$ & 
$( 2.7 \pm 0.2 ) \times 10^{-9}$ & 
$\pm 0.7 \times 10^{-9}$ & 
$( 1.1 \pm 0.1 ) \times 10^{-9}$ \\ 
$36$ -- $51$ & 
$( 1.1 \pm 0.1 ) \times 10^{-9}$ & 
$\tvect{+0.4}{-0.3} \times 10^{-9}$ & 
$( 1.8 \pm 0.2 ) \times 10^{-9}$ & 
$\tvect{+0.4}{-0.5} \times 10^{-9}$ & 
$( 7.3 \pm 0.9 ) \times 10^{-10}$ \\ 
$51$ -- $72$ & 
$( 6.3 \pm 0.8 ) \times 10^{-10}$ & 
$\tvect{+2.0}{-1.7} \times 10^{-10}$ & 
$( 1.1 \pm 0.1 ) \times 10^{-9}$ & 
$\tvect{+0.2}{-0.3} \times 10^{-9}$ & 
$( 4.5 \pm 0.6 ) \times 10^{-10}$ \\ 
$72$ -- $100$ & 
$( 3.6 \pm 0.5 ) \times 10^{-10}$ & 
$\tvect{+1.1}{-1.0} \times 10^{-10}$ & 
$( 6.2 \pm 0.6 ) \times 10^{-10}$ & 
$\tvect{+1.1}{-1.7} \times 10^{-10}$ & 
$( 2.6 \pm 0.3 ) \times 10^{-10}$ \\ 
$100$ -- $140$ & 
$( 1.5 \pm 0.3 ) \times 10^{-10}$ & 
$\tvect{+0.5}{-0.4} \times 10^{-10}$ & 
$( 3.1 \pm 0.4 ) \times 10^{-10}$ & 
$\tvect{+0.5}{-0.9} \times 10^{-10}$ & 
$( 1.5 \pm 0.2 ) \times 10^{-10}$ \\ 
$140$ -- $200$ & 
$( 9.8 \pm 2.0 ) \times 10^{-11}$ & 
$\tvect{+2.7}{-2.6} \times 10^{-11}$ & 
$( 1.9 \pm 0.3 ) \times 10^{-10}$ & 
$\tvect{+0.3}{-0.5} \times 10^{-10}$ & 
$( 9.3 \pm 1.6 ) \times 10^{-11}$ \\ 
$200$ -- $290$ & 
$( 4.7 \tvect{+1.4}{-1.3} ) \times 10^{-11}$ & 
$\tvect{+1.3}{-1.2} \times 10^{-11}$ & 
$( 8.9 \pm 1.7 ) \times 10^{-11}$ & 
$\tvect{+1.3}{-2.4} \times 10^{-11}$ & 
$( 4.1 \pm 1.0 ) \times 10^{-11}$ \\ 
$290$ -- $410$ & 
$( 3.2 \tvect{+1.1}{-1.0} ) \times 10^{-11}$ & 
$\tvect{+0.9}{-0.8} \times 10^{-11}$ & 
$( 6.3 \pm 1.3 ) \times 10^{-11}$ & 
$\tvect{+0.9}{-1.7} \times 10^{-11}$ & 
$( 3.0 \pm 0.8 ) \times 10^{-11}$ \\ 
$410$ -- $580$ & 
$( 7.3 \tvect{+5.7}{-5.1} ) \times 10^{-12}$ & 
$\tvect{+3.8}{-2.9} \times 10^{-12}$ & 
$( 2.1 \tvect{+0.9}{-0.8} ) \times 10^{-11}$ & 
$\tvect{+0.4}{-0.5} \times 10^{-11}$ & 
$( 1.3 \pm 0.6 ) \times 10^{-11}$ \\ 
$580$ -- $820$ & 
$< \, 2.3 \times 10^{-12}$ & 
$$ & 
$( 9.7 \pm 6.0 ) \times 10^{-12}$ & 
$\tvect{+2.3}{-2.8} \times 10^{-12}$ & 
$( 9.0 \pm 5.2 ) \times 10^{-12}$ \\ 

\enddata
\tablecomments{Measured intensities of the IGRB, the total EGB, and the 
identified sources ($|b|>20^{\deg}$) per energy band, when using model A to describe the Galactic foreground.
Uncertainties arising from foreground (FG) modeling are given in
separate columns. Digitized versions of this table and the
corresponding results for foreground models B and C are available in
the online supplementary materials.}
\label{tab:IGRBResults}
\end{deluxetable}

\begin{deluxetable}{lccccccc}
\tabletypesize{\scriptsize}
\tablecaption{Results of the parametric fit of the IGRB}
\tablewidth{0pt}
\tablehead{
\colhead{Foreground} & \colhead{$ I_{100} $} & \colhead{$\gamma$} & 
\colhead{$E_{\mathrm{cut}}$} & 
\colhead{$I_{>100}$} &
\colhead{$\chi^2$/ndof} &\colhead{$\chi^2$/ndof} & 
\colhead{$\chi^2$/ndof} \\
\colhead{model}
& \colhead{$[$MeV$^{-1}$cm$^{-2}$s$^{-1}$sr$^{-1}]$}
&
& \colhead{$[$GeV$]$}
& \colhead{$[$cm$^{-2}$~s$^{-1}$~sr$^{-1}]$}
& \colhead{(PLE\tablenotemark{a})}
& \colhead{(PL\tablenotemark{a})}
& \colhead{(BPL\tablenotemark{a})} 
}
\startdata
Model A & $( 0.95 \pm 0.08 ) \times 10^{-7}$ & $2.32 \pm 0.02$ & $279 \pm 52$ & $ ( 7.2 \pm 0.6 ) \times 10^{-6}$ & $13.9/23$ & $87.5/24$ & $13.5/22$ \\
Model B & $( 1.12 \pm 0.08 ) \times 10^{-7}$ & $2.28 \pm 0.02$ & $206 \pm 31$ & $ ( 8.7 \pm 0.6 ) \times 10^{-6}$ & $7.9/23$ & $151./24$ & $10.6/22$  \\ 
 Model C &  $( 0.78 \pm 0.07 ) \times 10^{-7}$ &  $2.26 \pm 0.02$ &  $233 \pm 41$ &  $ ( 6.2 \pm 0.6 ) \times 10^{-6}$ &  $10.7/23$ &  $106.5/24$ &  $11.3/22$ \\ 
\enddata
\tablenotetext{a}{PLE $=$ power-law plus exponential cutoff ; PL $=$ power-law ; BPL $=$ broken power-law}
\tablecomments{Parameters obtained from a parametric fit of the IGRB
  spectrum. Intensity $I_{100}$, spectral index $\gamma$ and cutoff energy 
$E_{\mathrm{cut}}$ for a fit of the observed spectrum with the function
given in Equation \ref{equ:expcutoff} are shown in columns 2--4. 
The integrated IGRB intensity above 100~MeV, $I_{>100}$, is found in column 5.
A comparison of the $\chi^2$/ndof values
between the fit with the function in Equation \ref{equ:expcutoff}
and alternative spectral models is given in columns
6--8. The $\chi^2$ values include systematic uncertainty.}
\label{tab:FitResults}
\end{deluxetable}

\clearpage

\begin{deluxetable}{lcccc}
\tabletypesize{\scriptsize}
\tablecaption{Impact of foreground model variations on IGRB spectral parameters}
\tablewidth{0pt}
\tablehead{
\colhead{Model variation} & \colhead{$I_{100}$} & 
\colhead{$\gamma$} & \colhead{$E_{\mathrm{cut}}$ } & 
\colhead{$\chi^2$/ndof} \\
& \colhead{$[$MeV$^{-1}$cm$^{-2}$s$^{-1}$sr$^{-1}]$}
&
& \colhead{$[$GeV$]$}
&
}
\startdata
10$\times$ ISRF in Galactic Bulge &  $ (0.96 \pm 0.08 )  \times 10^{-7} $ &  $2.31 \pm 0.02$ &  $273 \pm 50$ & $13.4/23$  \\ 
Random magnetic field strength $3$~$\mu$G  &  $ (0.93 \pm 0.09 )  \times 10^{-7} $ & $2.33 \pm 0.03$ &  $257 \pm 56$ & $14.1/23$  \\ 
Plain diffusion model    &  $ (0.91 \pm 0.08  )  \times 10^{-7} $ &  $ 2.31 \pm 0.02 $ &  $ 264 \pm 52 $ & $ 13.4/23 $  \\ 
Model 38\tablenotemark{a}  (Pulsars trace CR sources, 4kpc CR halo)  &  $ ( 1.06 \pm 0.09 ) \times 10^{-7} $ &  $2.33 \pm 0.02$ &  $367 \pm 75$ & $15.0/23$  \\
Model 62\tablenotemark{a} (Pulsars trace CR sources, 10kpc CR halo)  &  $ ( 0.89 \pm 0.08 ) \times 10^{-7} $ &  $2.31 \pm 0.02$ &  $374 \pm 77$ & $16.7/23$  \\
Model 06\tablenotemark{a,}\tablenotemark{b} (SNR trace CR sources, 4kpc CR halo)  &  $ ( 0.56\pm 0.07 ) \times 10^{-7} $ &  $2.27 \pm0.03 $ &  $399 \pm 92$ & $55/23$ \\
\textit{Fermi} Bubbles template A &  $ ( 1.02 \pm 0.09 ) \times 10^{-7}  $ &  $  2.32 \pm 0.02 $ &  $ 229 \pm 44 $ & $13.0/23$  \\ 
\textit{Fermi} Bubbles template B &  $ ( 1.05 \pm 0.09 ) \times 10^{-7}  $ &  $ 2.31 \pm 0.02 $ &  $ 244 \pm 42 $ & $13.7/23$  \\ 
\hi-to-dust ratio $+10$\% &  $ ( 1.09 \pm 0.09 ) \times 10^{-7} $ &  $2.34 \pm 0.02$ &  $280 \pm 52$ & $12.0/23$  \\ 
\hi-to-dust ratio $-10$\% &  $ ( 0.82 \pm 0.08 ) \times 10^{-7} $ &  $2.30 \pm 0.03$ &  $274 \pm 51$ & $17.5/23$  \\ 
\enddata
\tablenotetext{a}{from \citet{Ackermann2012:Diffuse}}\par
\tablenotetext{b}{Model is not 
well fit by a simple power-law with exponential cutoff spectral hypothesis. A two-component fit is used instead for 
this model for the evaluation of the foreground model systematics. The parameters obtained from this fit are 
$I_{100}^{(0)}=0.69 \times 10^{-7}$~GeV$^{-1}$cm$^{-2}$s$^{-1}$sr$^{-1}$,
$\gamma^{(0)}=$1.74,
$E_{\mathrm{cut}}^{(0)}=0.60$~GeV for the first, and 
$I_{100}^{(1)}=0.16 \times 10^{-7}$~GeV$^{-1}$cm$^{-2}$s$^{-1}$sr$^{-1}$,
$\gamma^{(1)}=2.02$,
$E_{\mathrm{cut}}^{(1)}=183$~GeV for the second component.} 

\tablecomments{Parameters obtained from fits of the IGRB spectrum for
  variants of the DGE foreground model. Specific intensity
$I_{100}$, spectral index $\gamma$, and cutoff energy $E_{\mathrm{cut}}$ 
for a fit of the observed spectrum with the function given in Equation \ref{equ:expcutoff}
are shown in columns 2--4. The  $\chi^2$/ndof values of the fit are
  shown in column 5. }
\label{tab:sysModelVar}
\end{deluxetable}

\begin{deluxetable}{lcccccccc}
\tabletypesize{\scriptsize}
\tablecaption{Results of the parametric fit of the total EGB}
\tablewidth{0pt}
\tablehead{
\colhead{Foreground} & \colhead{$I_{100} $} & \colhead{$\gamma$} & 
\colhead{$E_{\mathrm{cut}}$} & 
\colhead{$I_{>100}$} &
\colhead{$\chi^2$/ndof} &\colhead{$\chi^2$/ndof} & 
\colhead{$\chi^2$/ndof} \\
\colhead{model}
& \colhead{$[$MeV$^{-1}$cm$^{-2}$s$^{-1}$sr$^{-1}]$}
&
& \colhead{$[$GeV$]$}
& \colhead{$[$cm$^{-2}$~s$^{-1}$~sr$^{-1}]$}
& \colhead{(PLE\tablenotemark{a})}
& \colhead{(PL\tablenotemark{a})}
& \colhead{(BPL\tablenotemark{a})} 
}
\startdata
Model A & $ ( 1.48 \pm 0.09 ) \times 10^{-7} $ & $2.31 \pm 0.02$ &  $362 \pm 64$ & $ ( 1.13 \pm 0.07 ) \times 10^{-5} $ & $11.0/23$ & $72.4/24$ & $10.5/22$  \\
Model B &  $ ( 1.66 \pm 0.09 ) \times 10^{-7} $ &  $2.28 \pm 0.01$ & $267 \pm 37$ & $ ( 1.29 \pm 0.07 ) \times 10^{-5} $ & $13.5/23$ & $130./24$ & $11.3/22$  \\ 
Model C &   $ ( 1.28 \pm 0.08 ) \times 10^{-7} $ &   $2.30 \pm 0.02$ &  $366 \pm 71$ &  $ ( 0.98 \pm 0.06 ) \times 10^{-5} $ &   $6.9/23$ &  $91.1/24$ &  $7.7/22$  \\ 
\enddata
\tablenotetext{a}{PLE $=$ power-law plus exponential cutoff ; PL $=$ power-law ; BPL $=$ broken power-law}
\tablecomments{Parameters obtained from fits of the total EGB. 
Intensity $I_{100}$, spectral index $\gamma$ and cutoff energy 
$E_{\mathrm{cut}}$ for a fit of the observed spectrum with the PLE function
given in Equation \ref{equ:expcutoff} are shown in columns 2--4. 
The integrated IGRB intensity above 100~MeV, $I_{>100}$, is found in column 5.
A comparison of the $\chi^2$/ndof values
between the fit with the function in Equation \ref{equ:expcutoff}
and alternative spectral models is given in columns
6--8. The $\chi^2$ values include systematic uncertainties.}
\label{tab:TotalFitResults}
\end{deluxetable}

\clearpage

\section{Discussion and conclusions}


We have refined the measurement of the LAT IGRB intensity relative to
the analysis of \cite{Abdo2010:EGB}, which was based on 10 months of
LAT observations, now using 50 months of accumulated data.
The measurement lower bound has been extended from 200~MeV to 100~MeV, and we report the first 
IGRB measurement with any instrument between 102~GeV and 820~GeV. 
 
The updated LAT IGRB spectrum remains consistent with a featureless power
law between 100~MeV and 100~GeV, and there is now strong evidence for
a high-energy cutoff feature. The spectrum is well described by a power law with an exponential
cutoff over the full analyzed energy range from 100~MeV to
820~GeV. For each of the three benchmark DGE models considered here, the power law index of the IGRB is
$\approx2.3$ and the cutoff energy is $\approx250$~GeV (Table
\ref{tab:FitResults}).

The total EGB is derived by adding resolved high-latitude LAT sources
(taken to be primarily extragalactic) to the measured IGRB
intensity. At an energy of 100~GeV, roughly half of the total EGB
intensity has now been resolved into individual sources by the LAT,
predominantly blazars of the BL Lacertae type. \citep[The demographics of LAT
  sources detected at energies above 10~GeV are discussed
  in][]{Ackermann2013:1FHL}. The relative contribution of resolved
sources becomes even more pronounced at energies exceeding 100 GeV.

\begin{figure}
\epsscale{1.0}
\begin{center}
\plotone{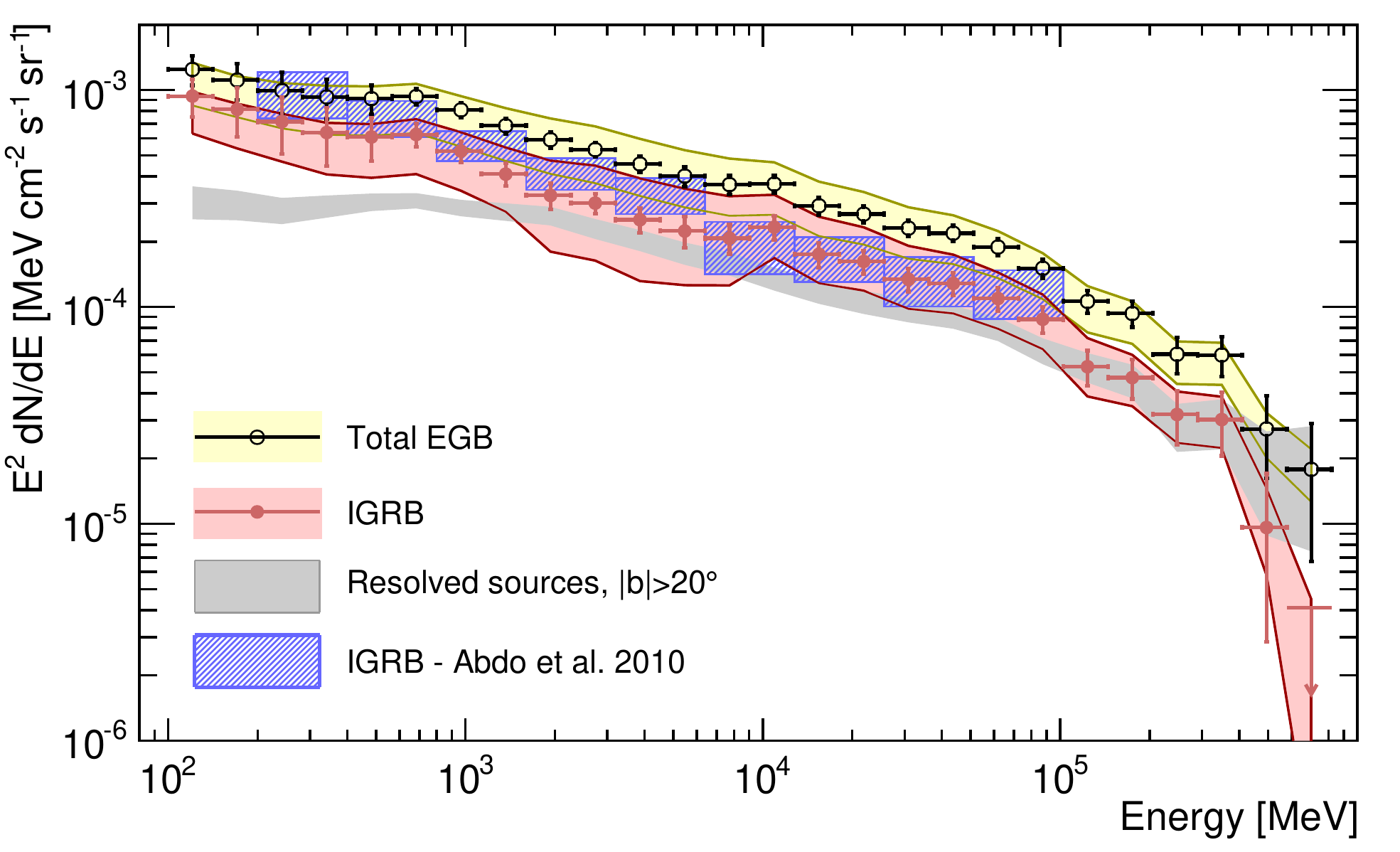} \\
\end{center}
\caption{Comparison of the measured IGRB and total EGB intensities (foreground model A) to 
the first measurement of the IGRB in \citet{Abdo2010:EGB} based on 10 months of LAT data.
The error bars on the LAT measurements include the statistical
uncertainty and systematic uncertainties from the effective area
parametrization, as well as the CR background subtraction. Statistical
and systematic uncertainties have been added in quadrature. The shaded 
bands indicate the systematic uncertainty arising from uncertainties
in the Galactic foreground. 
The total EGB intensity is the sum of the IGRB and the intensity of the 
resolved LAT sources at high Galactic latitudes, $|b|>20\deg$.}
\label{fig:igrb_egb_comparison}
\end{figure}

The intensities of the IGRB and the total EGB are compared to the first LAT measurement of the IGRB in 
\citet{Abdo2010:EGB} in Figure \ref{fig:igrb_egb_comparison}.
The two are compatible within the respective systematic uncertainties. 
Differences can be attributed to the combined effects 
of a more accurate estimate of the CR background at low energies, and changes in the Galactic 
foreground model. Importantly, the model for atmospheric secondaries has been refined to address 
discrepancies between data and simulation. The revised
background rate of misclassified CRs is up to 50\% higher at a few hundred MeV than the older estimates.
This change contributes to a reduced integrated IGRB intensity above 100~MeV of
$7.2\pm0.6\times10^{-6}$~ph~cm$^{-2}$~s$^{-1}$~sr$^{-1}$ in 
comparison to the $1.03\pm0.17\times10^{-5}$~ph~cm$^{-2}$~s$^{-1}$~sr$^{-1}$ reported in \citet{Abdo2010:EGB}.

The intensity resolved into individual sources at latitudes $|b|>20\deg$ did not change substantially
between the two measurements. This is consistent with the findings
reported in \cite{Abdo2010:HighLatitude} and \cite{Ackermann2011:2LAC} that report a 
sky-averaged intensity of sources from Galactic latitudes $|b|>10\deg$ of 
$4.0 \times 10^{-6}$~ph~cm$^{-2}$~s$^{-1}$~sr$^{-1}$ after one year and 
$4.4 \times10^{-6}$~ph~cm$^{-2}$~s$^{-1}$~sr$^{-1}$ after two years of
observations above 100 MeV. This difference corresponds
to only $\approx5$\% of the IGRB intensity.

\begin{figure}
\epsscale{1.0}
\begin{center}
\plotone{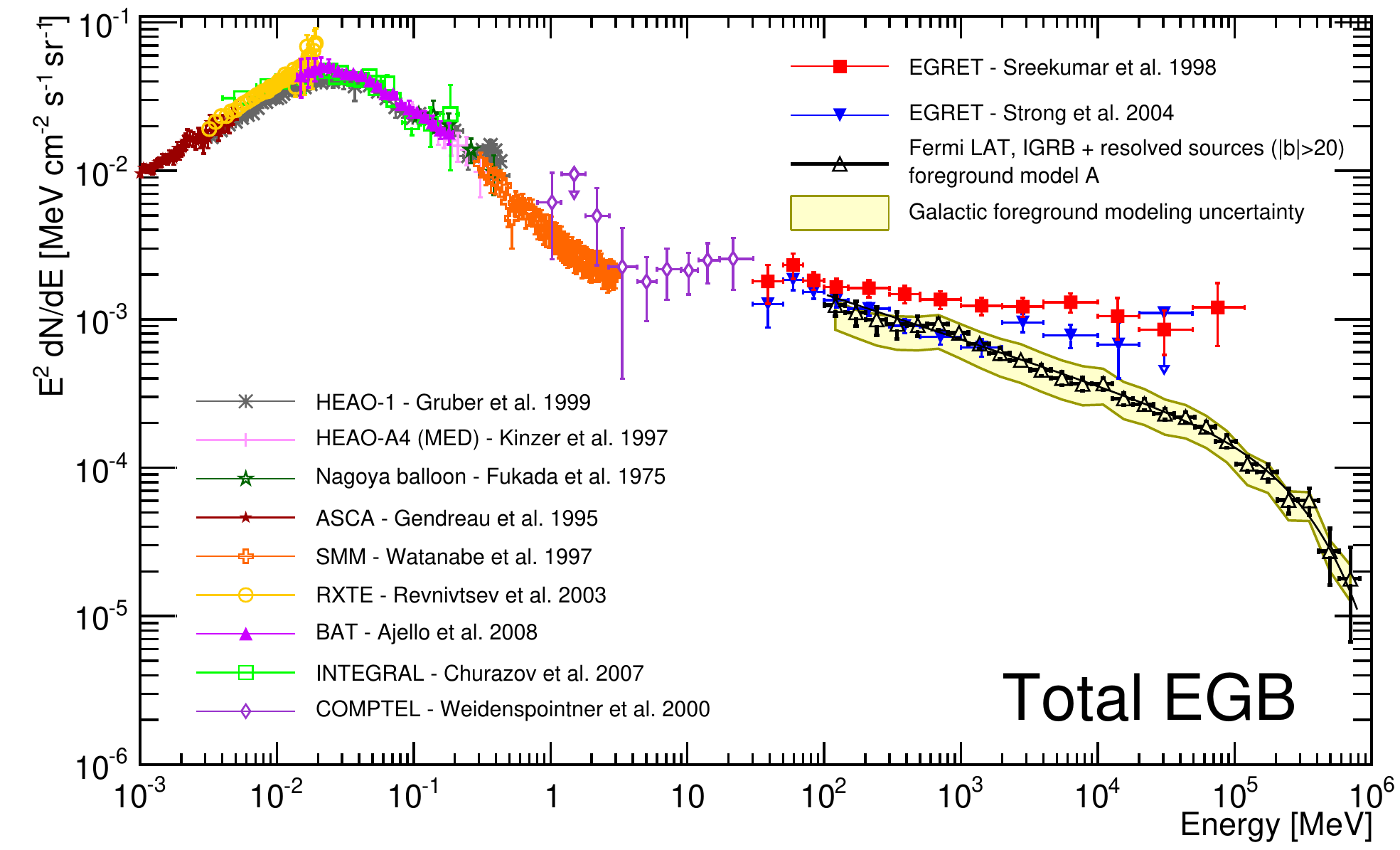} \\
\end{center}
\caption{Comparison of the derived total EGB intensity (foreground model A) to other 
measurements of the X-ray and \gray{} background. 
The error bars on the LAT measurement 
include the statistical uncertainty and systematic uncertainties from the effective area
parametrization, as well as the CR background subtraction. Statistical
and systematic uncertainties have been added in quadrature. The shaded
band indicates the systematic uncertainty arising from uncertainties
in the Galactic foreground. (Note that the EGRET measurements 
shown are measurements of the IGRB. However, EGRET was more than an order
of magnitude less sensitive to resolve individual sources on the sky
than the \textit{Fermi}-LAT.)}
\label{fig:total_comparison}
\end{figure}

Figure \ref{fig:total_comparison} places LAT measurements of the total EGB 
intensity in context with other measurements of the extragalactic
X-ray and \gray{} backgrounds, together spanning nearly nine orders of magnitude in energy between 1~keV and 820~GeV.
\nocite{Gruber:1999}\nocite{Kinzer:1997}\nocite{Fukada:1975}
\nocite{Gendreau:1995}\nocite{Ajello:2008}\nocite{Churazov:2007}\nocite{Weidenspointner:2000}\nocite{Revnivtsev:2003}
\nocite{Strong:2004}\nocite{Watanabe:1997}
There is a good agreement between the total EGB measured by the LAT and
the previous measurement of the IGRB using EGRET data
\citep{Sreekumar:1998,Strong:2004} below 1~GeV. The IGRB measured by the LAT is lower than the EGRET
IGRB measurement, as expected from the greatly superior sensitivity of the LAT to resolve individual sources when compared
to EGRET.

As discussed in Section \ref{sec:introduction}, numerous source
populations as well as truly diffuse processes are expected to
contribute to the EGB intensity. A detailed review of the expected contributions of specific source populations
and diffuse processes is beyond the scope of this work.
Instead, we focus on general constraints that can be applied to extragalactic
\gray{} source populations based on the EGB spectrum, taking into account the effects of EBL attenuation. Other
efforts to statistically characterize the EGB properties considering
the fluctuations of counts in spatial pixels \citep{Malyshev:2011} and
two-point correlation functions \citep{Ackermann2012:Anisotropy} have proven valuable for
constraining the abundance of sources just below the LAT detection
threshold, and similar techniques may be usefully applied to LAT data in the energy range
$>100$~GeV. 


In the interpretation that follows, we use the formalism outlined by
\cite{Murase:2007} and \cite{Inoue:2012} to calculate both the EBL-attenuated
primary signal as well as the electromagnetic cascade emission that would arise from
a variety of generic cosmologically evolving source
populations. We adopt the UV/optical/IR EBL model of \cite{Franceschini:2008} based on observed
galaxy counts, which is found to be consistent with spectral analyses of individually detected \gray{} sources
\citep[e.g.,][]{Ackermann2012:EBL}. The populations are modeled as a collection of sources
sharing the same intrinsic simple power-law
spectral form with a common photon index, $\gamma$, and maximum
energy, $E_{\rm max}$:

\begin{equation}
dN/dE = \begin{cases} N_0 (E/E_0)^{-\gamma}, & E \le E_{\rm max} \\
  0, & E > E_{\rm max} \end{cases}
\end{equation}

\noindent where $E$ is the energy of photons emitted at the source. We
model the evolving comoving volume emissivity (ph
s$^{-1}$ cm$^{-3}$) of source populations without distinguishing between
luminosity or density evolution of the sources. The emissivity
evolution is either (1) parameterized by $j \propto (1+z)^{\beta}$, or
(2) follows the cosmic star formation rate density
\citep{Behroozi:2013}. We consider sources up to
redshifts $z=2$ and $z=3$ in the two cases above,
respectively. Our conclusions do not qualitatively change
  when using a lower maximum redshift of $z=1$. No
attempt is made here to identify sources that could be individually
resolved by the LAT versus those that blur into the unresolved background.

\begin{figure}
\includegraphics[scale=0.42]{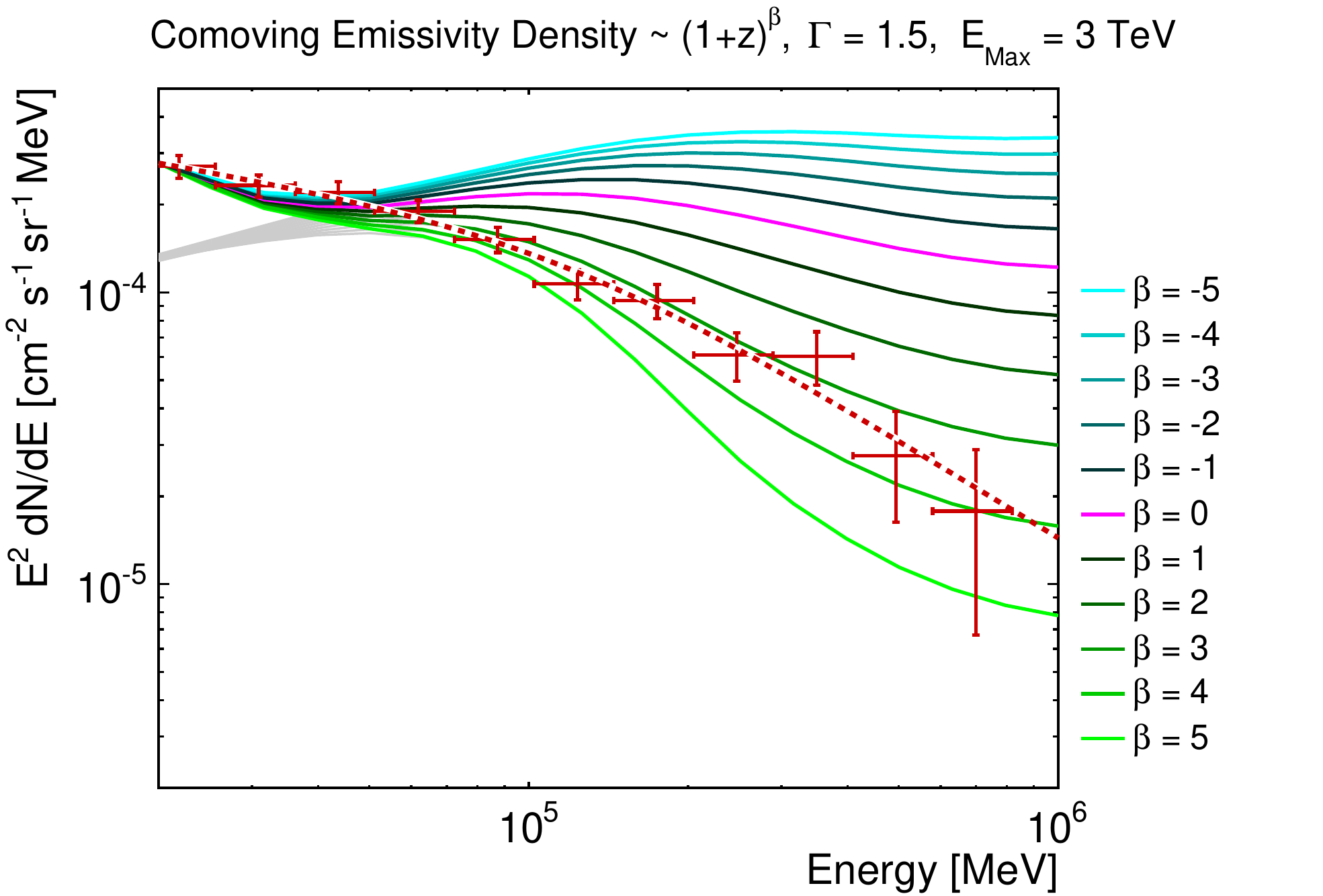}
\includegraphics[scale=0.42]{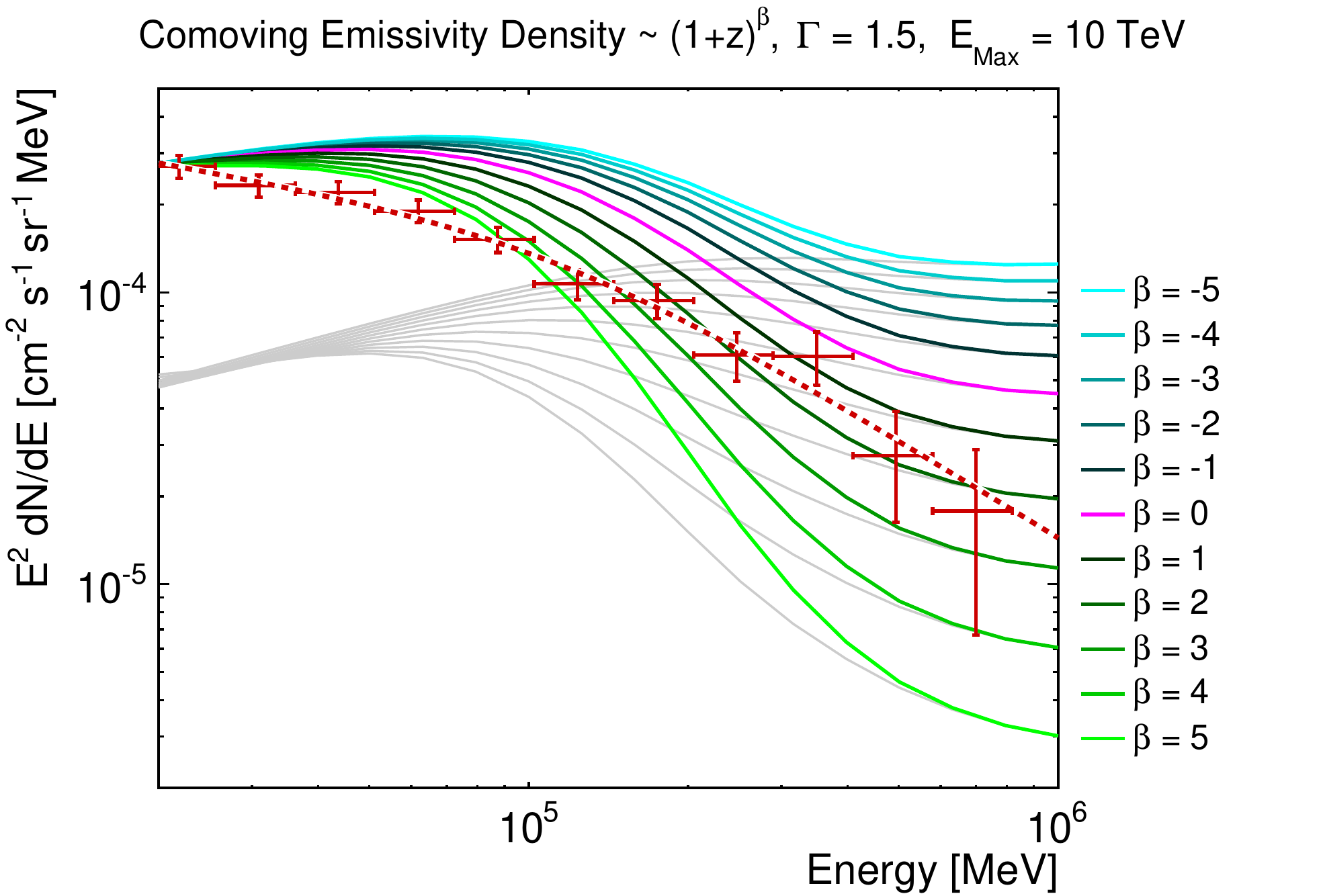}\\
\includegraphics[scale=0.42]{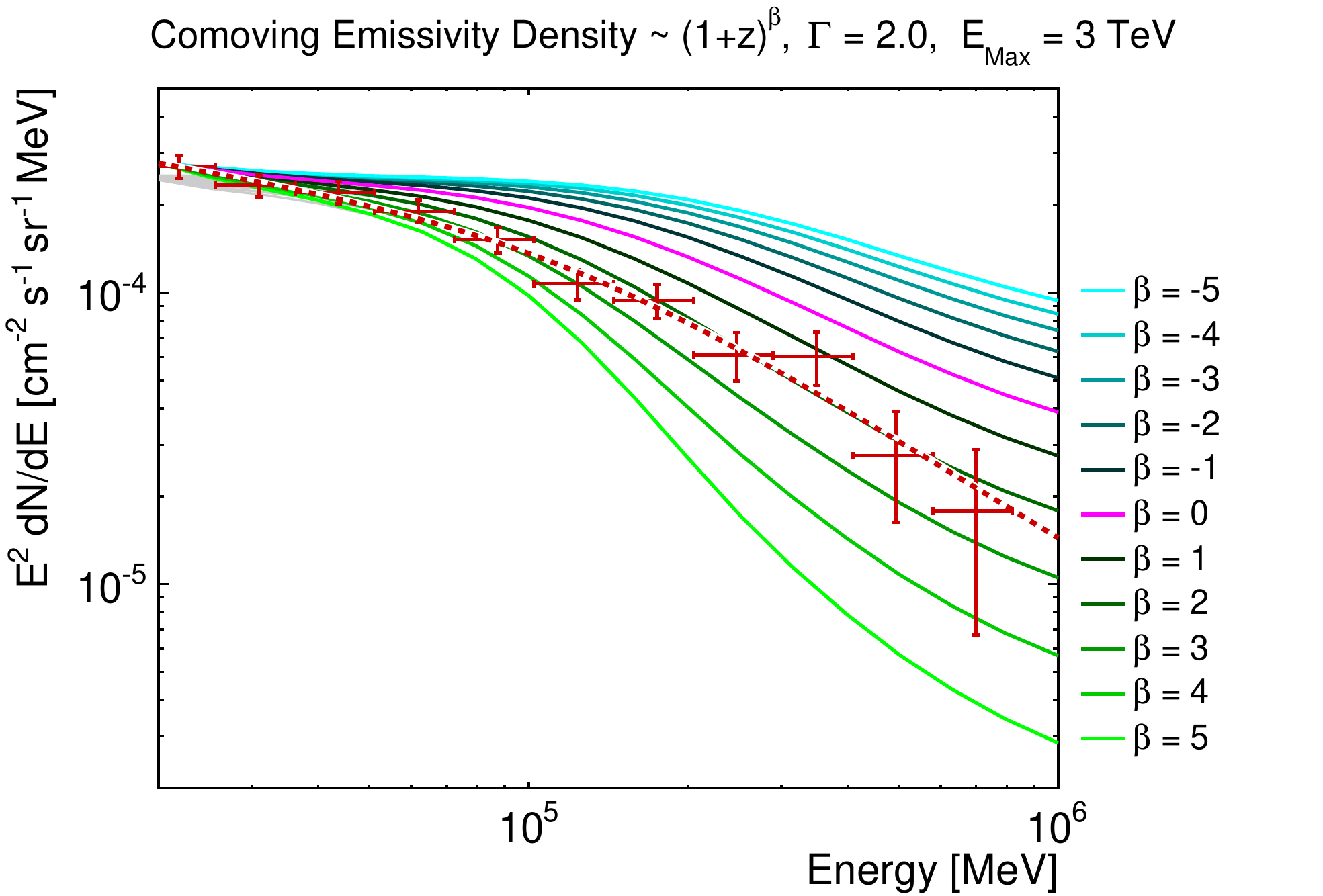}
\includegraphics[scale=0.42]{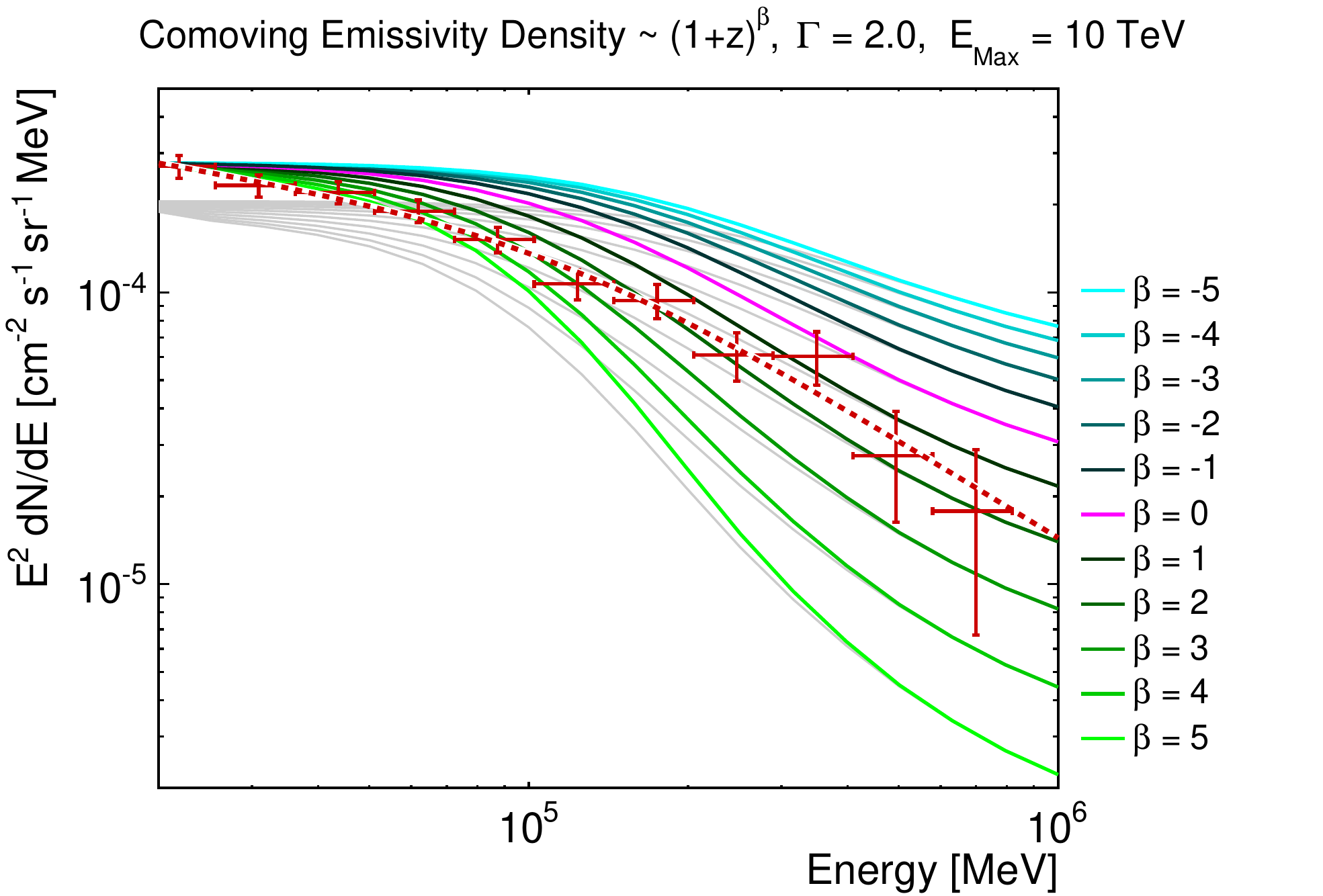}\\
\includegraphics[scale=0.42]{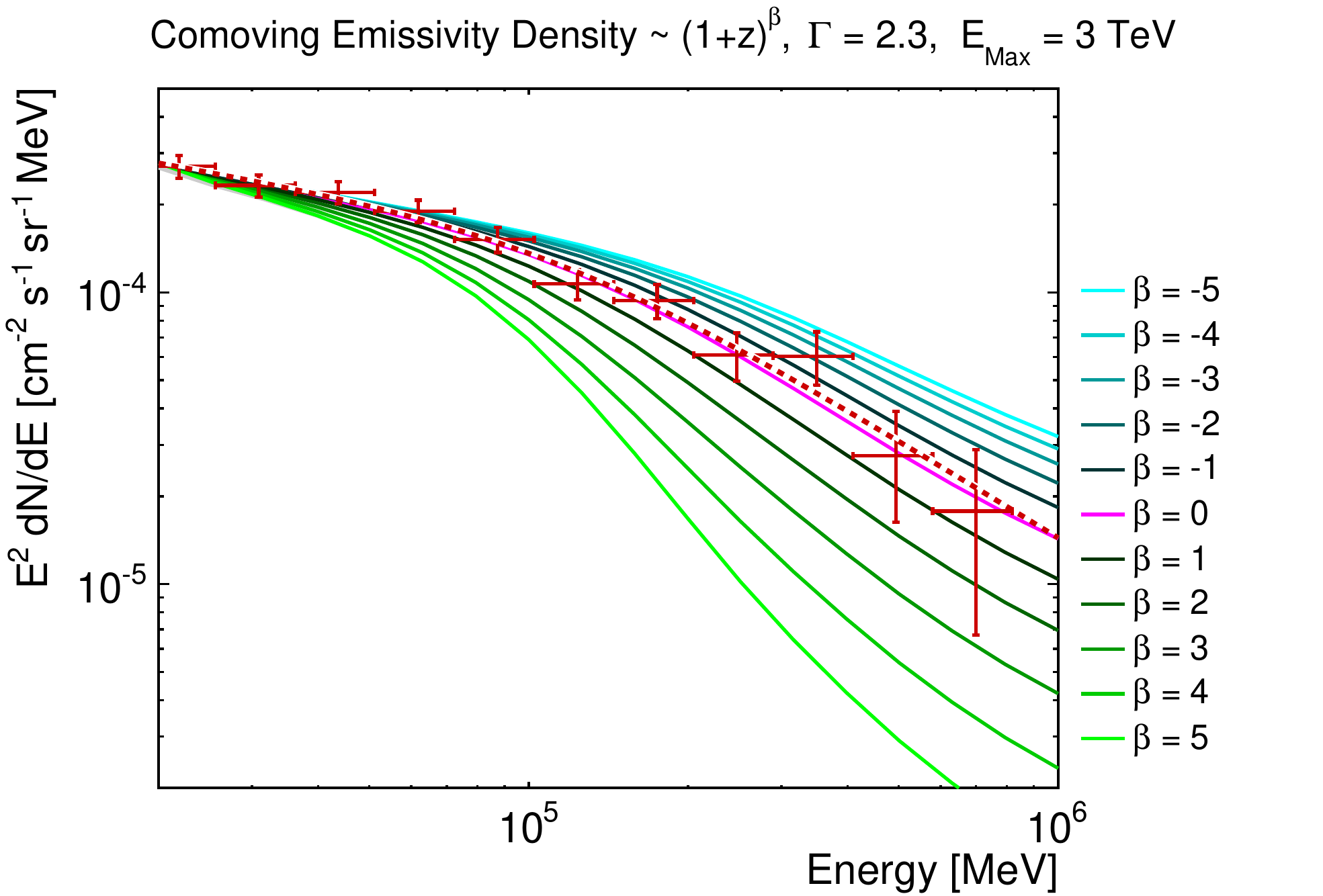}
\includegraphics[scale=0.42]{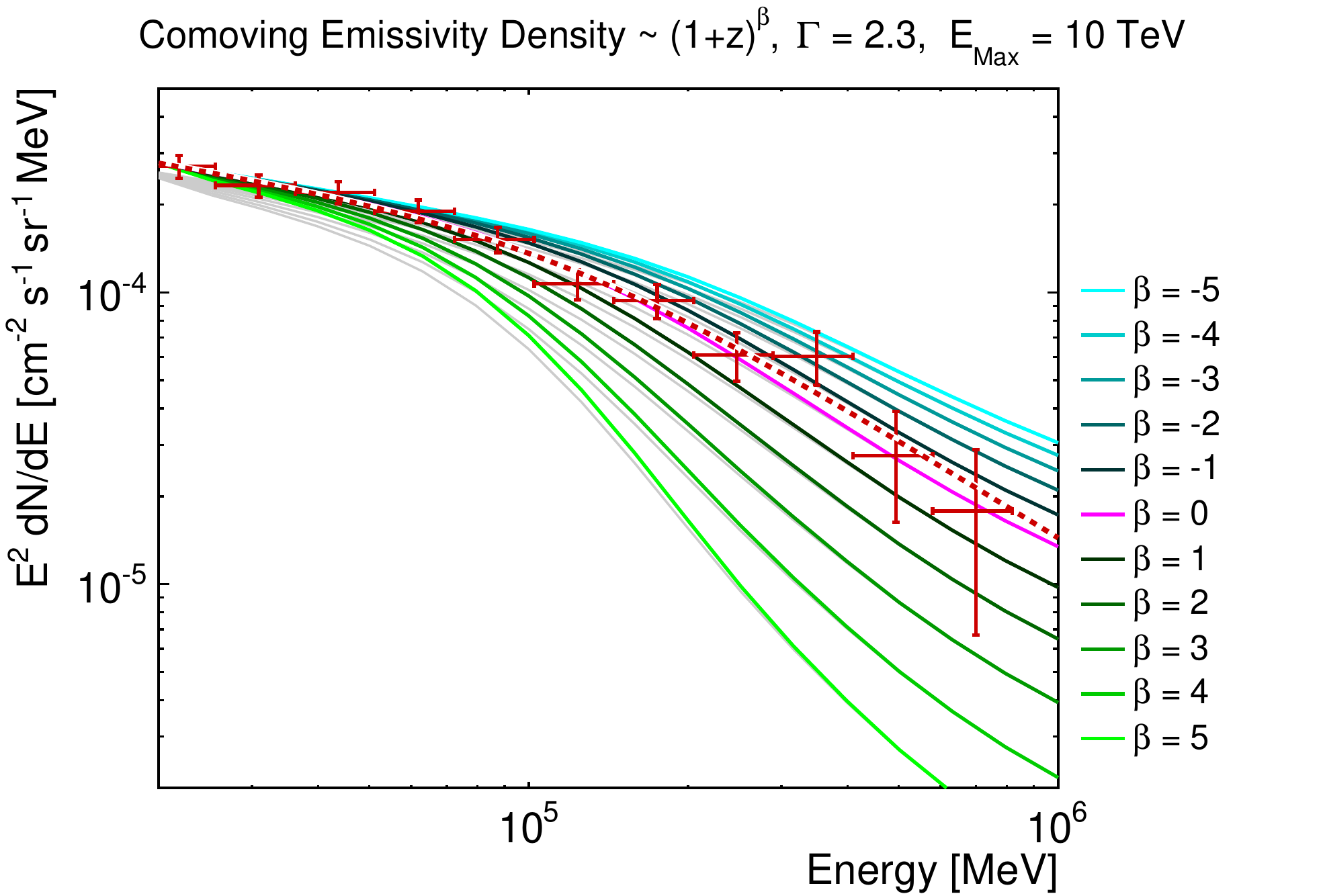}
\caption{EGB contributions of various source populations with
  comoving volume emissivity parameterized by $j \propto (1+z)^{\beta}$ are
  compared to the measured EGB spectrum (foreground model A). Rows are
  differentiated by the assumed photon index of the sources. Left and right
  columns correspond to populations with maximum energies of
  3~TeV and 10~TeV, respectively. Gray curves represent the intensity
  of primary $\gamma$ rays (attenuated by the EBL). Colored curves
  indicate the sum of the primary and cascade components.} 
 \label{fig:egb_sources_parameter}
\end{figure}
\begin{figure}
\includegraphics[scale=0.42]{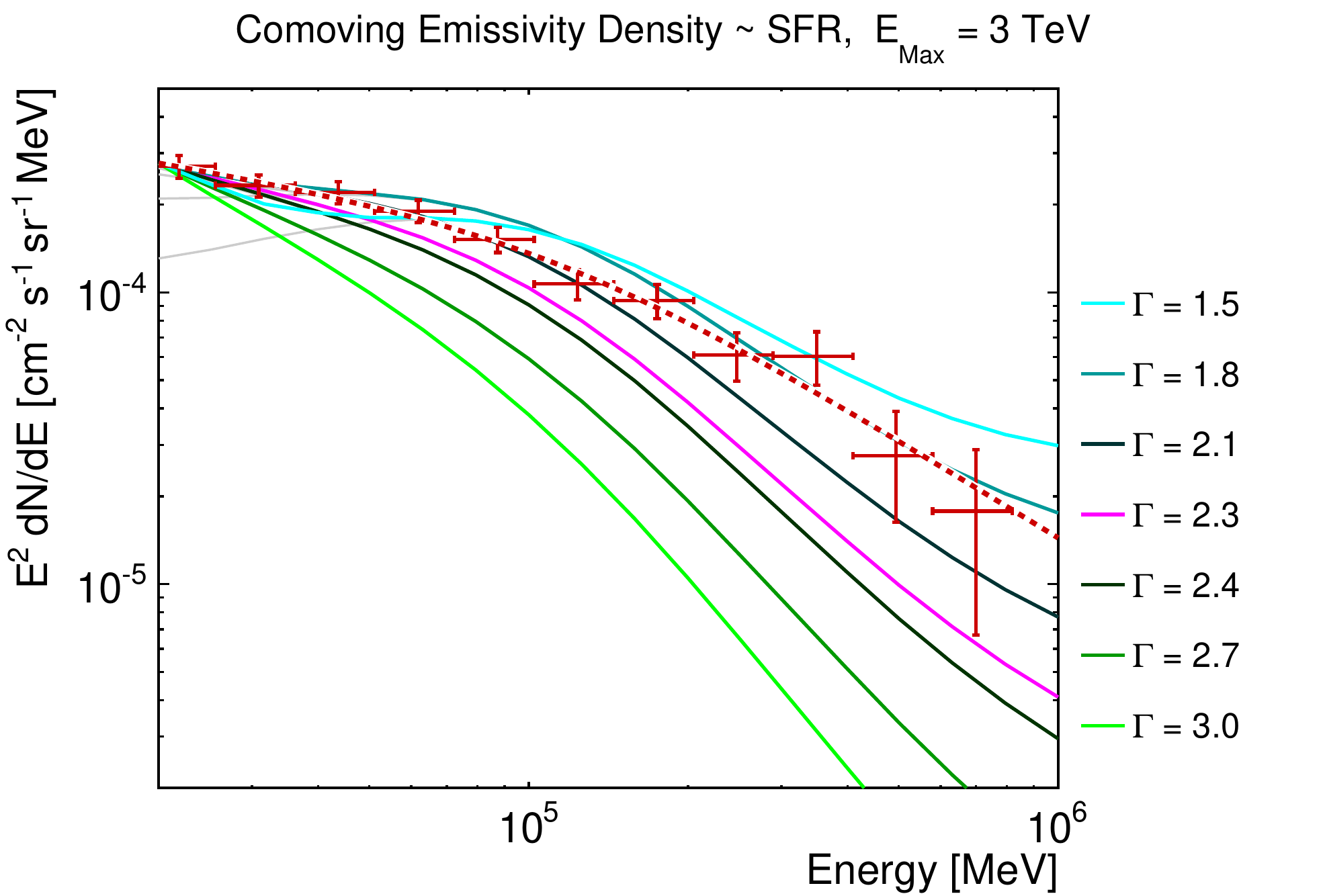}
\includegraphics[scale=0.42]{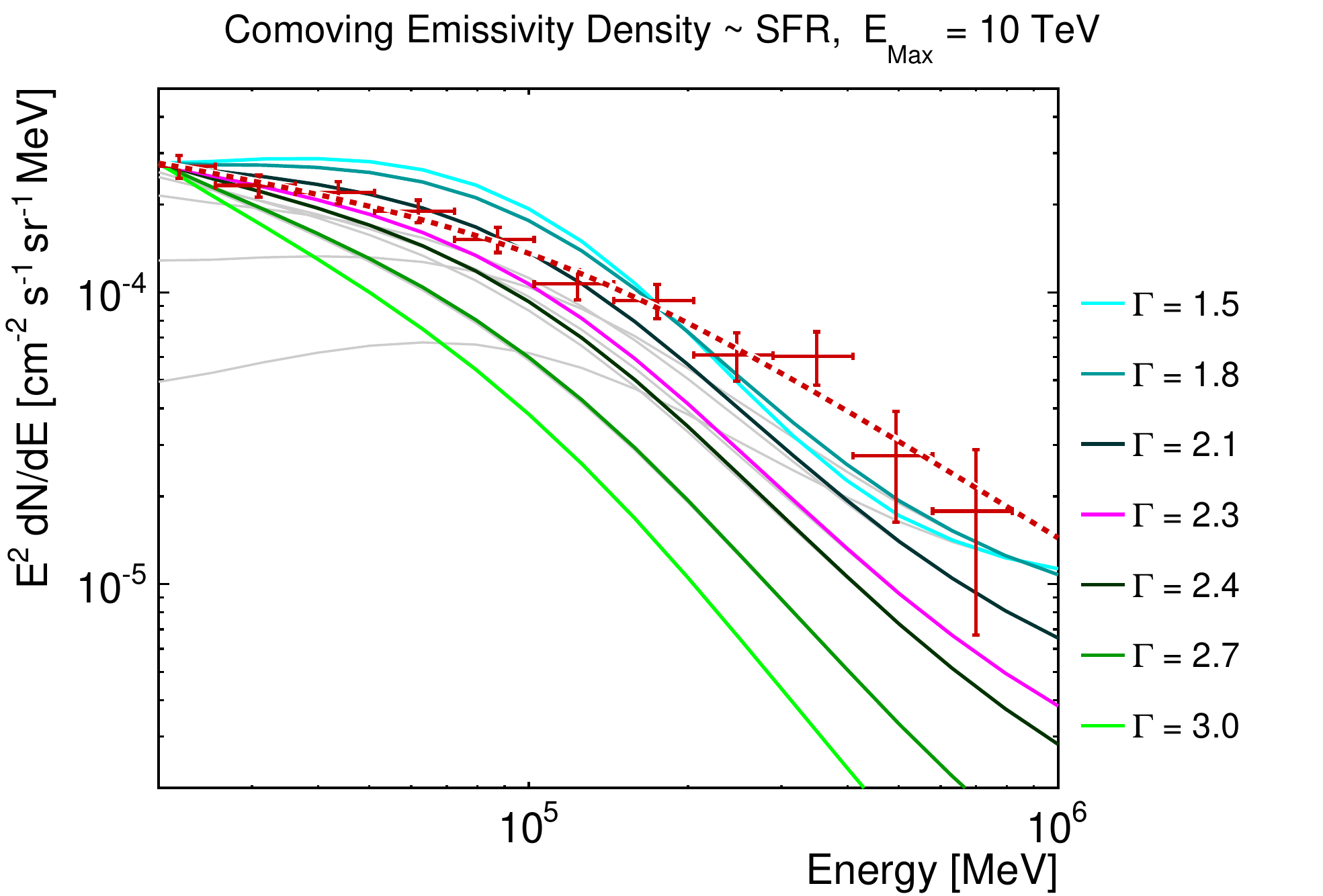}
\caption{EGB contributions of various source population scenarios with
  comoving volume emissivity following the comoving star formation
  rate density are compared to the measured EGB spectrum (foreground
  model A). Each colored curve denotes a source population with a different
  assumed photon index. Left and right columns correspond to populations with maximum energies of
  3~TeV and 10~TeV, respectively. Gray curves represent the intensity
  of primary $\gamma$ rays (attenuated by the EBL). Colored curves
  indicate the sum of the primary and cascade components.
}  
\label{fig:egb_sources_sfr}
\end{figure}

The expected total EGB contributions corresponding to various source population
scenarios are compared to the measured EGB spectrum $>20$~GeV in
Figures \ref{fig:egb_sources_parameter} and \ref{fig:egb_sources_sfr}. In each case, the contribution of the
sources (i.e., sum of primary and cascade components) to the total EGB has been normalized to match that of the measured
total EGB intensity at 20~GeV. Figure \ref{fig:egb_sources_parameter} shows source
populations with parameterized comoving volume emissivities, 
while Figure \ref{fig:egb_sources_sfr} shows populations whose emissivity follows the comoving
star formation rate density. 
In scenarios with photon index $\Gamma=1.5$, the cascade component
can dominate the primary component at energies $<100$~GeV.

Several patterns are apparent in Figures
\ref{fig:egb_sources_parameter} and \ref{fig:egb_sources_sfr}. First,
source populations with negative evolution, especially those with very
hard power-law spectra ($\Gamma<2$) extending to
multi-TeV energies, have difficulty fitting the
high-energy break in the total EGB spectrum and are therefore
unlikely to account for the EGB on their own. On the other hand, source populations whose emissivity
evolves as the cosmic star-forming rate
\citep[corresponding to $\beta\sim3.25$ at low redshifts, e.g.,][]{Hopkins:2006}
also face challenges to match the shape of the total EGB spectrum alone, unless
the spectral properties of those sources are finely tuned. The source
populations which would most readily explain the measured total EGB
spectrum from 100~MeV to nearly 1~TeV are those with photon indices
matching that of the EGB below 100~GeV, namely $\sim2.3$,
and little or no evolution. In fact, the distribution of photon
indices for individual LAT sources detected above 10 GeV is also peaked
near a value of 2.3 \citep{Ackermann2013:1FHL}. These and similar studies
\citep[e.g.,][]{Venters:2010,Inoue:2012,Murase:2012} demonstrate that the power-law
shape of the total EGB spectrum with a single cutoff/break at $\sim250$~GeV
could in principle be explained by a single dominant extragalactic source
population with relatively generic spectral properties and EBL attenuation.
However, more sophisticated modeling efforts taking into account the
specific properties of established extragalactic source classes
are needed to fully understand how sources governed by such diverse
physics produce a nearly featureless EGB spectrum over $\sim4$ decades in energy.



In addition to the implications for source classes comprising the EGB, a
second aspect of this work is a further examination of the
DGE (Section \ref{sec:foreground}). In the effort to accurately
subtract the DGE and thereby isolate the fainter isotropic component, we
considered a wider range of models for CR injection and
propagation in the interstellar medium of the Milky Way than
previously considered, e.g., in \cite{Ackermann2012:Diffuse}. None of the models
tested here simultaneously satisfy constraints from both local CR
measurements and observations of the high latitude \gray{} sky,
particularly in the case of IC emission (see Appendix
\ref{sec:appendixA}). The lack of a clearly preferred DGE model is the largest single source of
systematic uncertainty when measuring the IGRB intensity in the 100~MeV --- 100~GeV energy
range with the LAT. Some of the modifications to commonly used CR injection and diffusion
treatments investigated here may provide interesting future avenues of research.


Further improvements over the Pass 7 event reconstruction and classification are
required to extend EGB measurements with the LAT to both lower and higher
energies. An extension to energies $\sim50$~MeV may help to constrain
the radiative processes contributing to the EGB, e.g., through the
identification of the (redshifted) pionic spectral feature expected
from the interactions of CR nuclei in all galaxies
\citep[e.g.,][]{Stecker:2011,Lacki:2012,Chakraborty:2013}, although no
indication of such a low-energy cutoff is present in the current measurement at
$\sim100$~MeV. An extension to energies $\sim1$~TeV would further clarify the spectra
and evolution of sources that will be studied in detail with
the extragalactic surveys of the High-Altitude Water Cherenkov observatory (HAWC)\footnote{\url{http://www.hawc-observatory.org/}}
and the Cherenkov Telescope Array (CTA)\footnote{\url{http://www.cta-observatory.org/}}. Both
of these spectral extensions to the LAT IGRB measurement may be realized with future Pass 8 analyses
\citep{Atwood2013:Pass8}. Additional insight regarding the sources of the EGB may
come from ongoing studies of the extragalactic background of
high-energy neutrinos \citep{Aartsen:2013}, since the interactions of
very-high and ultra-high energy CRs inevitably create fluxes of both
$\gamma$~rays and neutrinos, the implications of which are discussed by, e.g.,
\cite{Ahlers:2010}, \cite{Berezinsky:2011}, \cite{Wang:2011},
\cite{Gelmini:2012}, and \cite{Murase:2012}.


\clearpage
\acknowledgments

The \fermi{} LAT Collaboration acknowledges generous ongoing support
from a number of agencies and institutes that have supported both the
development and the operation of the LAT as well as scientific data analysis.
These include the National Aeronautics and Space Administration and the
Department of Energy in the United States, the Commissariat \`a l'Energie Atomique
and the Centre National de la Recherche Scientifique / Institut National de Physique
Nucl\'eaire et de Physique des Particules in France, the Agenzia Spaziale Italiana
and the Istituto Nazionale di Fisica Nucleare in Italy, the Ministry of Education,
Culture, Sports, Science and Technology (MEXT), High Energy Accelerator Research
Organization (KEK) and Japan Aerospace Exploration Agency (JAXA) in Japan, and
the K.~A.~Wallenberg Foundation, the Swedish Research Council and the
Swedish National Space Board in Sweden.
Additional support for science analysis during the operations phase is gratefully
acknowledged from the Istituto Nazionale di Astrofisica in Italy and the Centre National d'\'Etudes Spatiales in France.
GALPROP development is partially funded via NASA grant NNX09AC15G.
Some of the results in this paper have been derived using the HEALPix \citep{Gorski:2005} package.





\appendix


\section{Galactic foreground models}
\label{sec:appendixA}

This appendix summarizes the parameters used for modeling the diffuse Galactic foreground emission 
in the benchmark foreground models A, B and C. Please note that a customized 
version of GALPROP is needed to produce the models. The output files of the 
corresponding GALPROP runs are provided in an electronic data repository. 
A link to this repository is distributed in the electronic supplementary material.

\subsection{Foreground model A}
\label{sec:foregroundA}

Foreground model A uses a parametrization of the distribution of 
pulsars in the Galaxy \citep{Lorimer:2006} as the distribution of CR sources (see Figure \ref{fig:sourceDist}) 
where CR electrons and nuclei are injected into the
interstellar medium. The diffusion coefficient for their propagation in the ISM is set to
$D_{xx}=7.0 \times 10^{-28}$~m$^{2}$~s$^{-1}$~$(R/4
\mathrm{~GV})^{0.33}$ where $R$ denotes the rigidity. The CRs are re-accelerated in the ISM with a re-acceleration 
strength (parametrized by the Alfv\'{e}n velocity
$v_A=30$~km~s$^{-1}$) that is constant throughout the Galaxy. The Galactic magnetic field is modeled 
according to \cite{Strong:2011} using a local random field strength of 7.5~$\mu$G.
A CR halo size of 5~kpc is chosen. All nuclei and electrons are injected with an energy spectrum that is 
a broken power law in rigidity. The power-law index of the injection spectrum of protons 
is $1.9$ below 9~GV, $2.45$ between 9~GV and 240~GV, and $2.32$ above 240~GV. For helium, we multiply the injection
spectrum by $R^{0.1}$, and the He/H fraction in the ISG was assumed to be 0.11.
The second break in the energy spectrum and the harder injection spectrum for helium are motivated by 
the results of the measurements of the proton spectrum by the PAMELA
satellite \citep{Adriani:2011a}, and by the CR spectrum inferred from LAT
observations of Earth limb $\gamma$~rays \citep{Ackermann2014:Albedo2}\footnote{We note that the 
AMS-02 collaboration has recently released preliminary data that does not confirm the
hardening of the proton and helium spectra. A test showed that there are  
only negligible effects on the IGRB intensity derived in this work if we use a 
foreground model without a break at 240~GV in the nucleon spectrum.}. 
The \gray{} yield from the interactions of the CR with the ISG was calculated using the 
parametrization of \citet{Kamae:2006}.

The power-law index of the injection spectrum of electrons is $1.5$
below 5~GV, $2.85$ between 5~GV and 25~GV, and $2.32$ above 
25~GV. As in the case of nucleons, the breaks in the spectrum are introduced to obtain a good agreement of the 
model with measurements of the local CR electron spectrum
\citep{Adriani:2011b,Abdo2009:Electrons,Ackermann2012:Positrons}. The break at 5~GV is furthermore motivated by 
the shape of the Galactic synchrotron emission spectrum \citep{Strong:2011}.
Additionally, we assume a high-energy cutoff (implemented as a change in the injection index to $4$ above 1.8~TeV) to 
remain in agreement with the H.E.S.S. measurements of the electron
spectrum up to several TeV \citep{Aharonian:2008}. 

\begin{figure}
\epsscale{.8}
\plotone{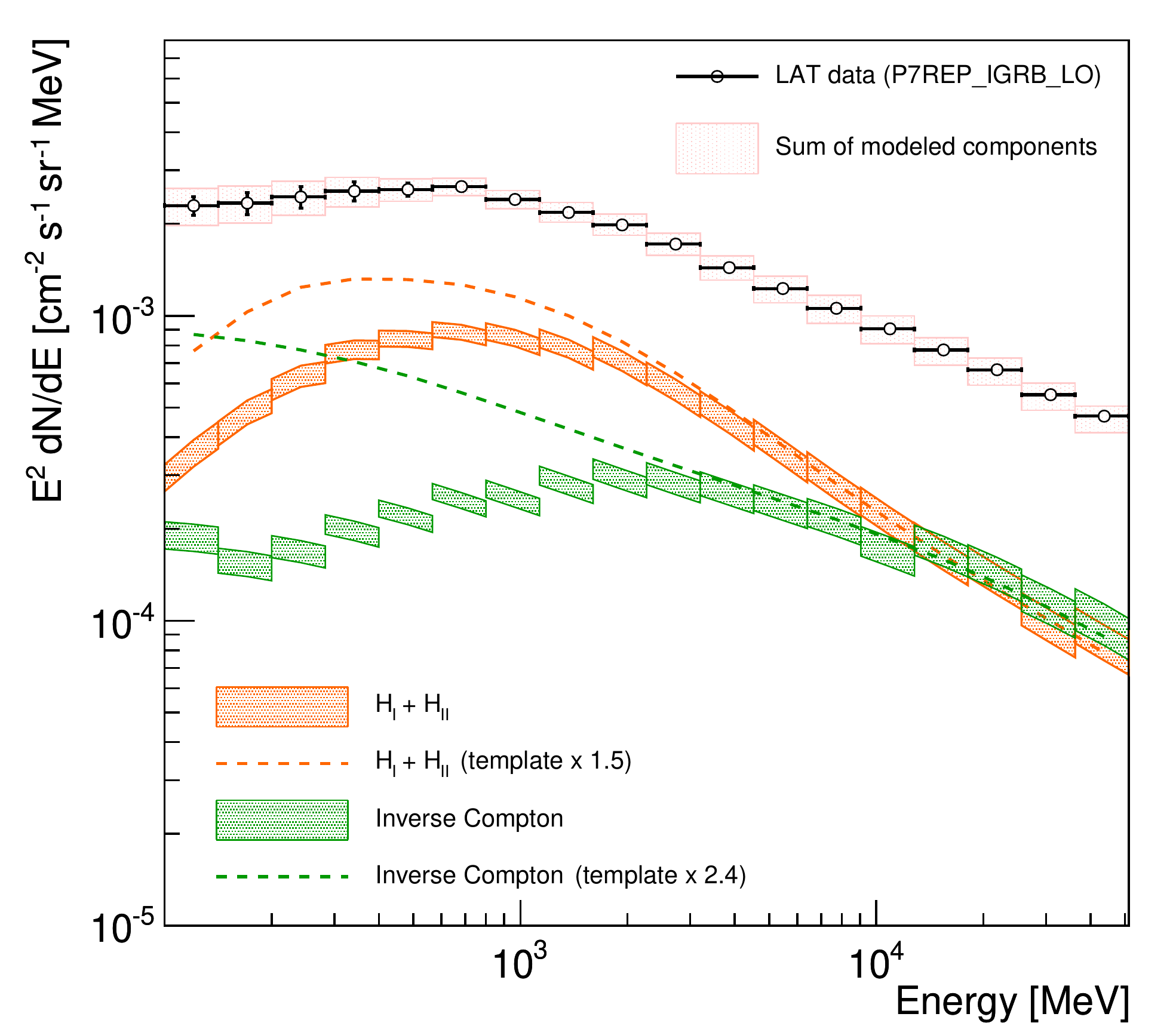}
\caption{Fitted average intensity of the DGE in foreground model A for Galactic 
latitudes $|b|>20\deg$. Contributions from CR interactions with the ISG, and 
contributions from IC are shown separately.  The normalizations of the two components 
are fitted individually in each energy bin. The GALPROP model spectrum that enters the fit is shown as 
dashed lines. The dashed lines are renormalized by the factors indicated in the legend.}
\label{fig:modelGalacticA}
\end{figure}

The doubly broken injection spectra used in foreground model A
improve the model/fit agreement in our maximum likelihood fit
of the \gray{} data.
However, there is only negligible effect on the derived IGRB spectrum.
Figure \ref{fig:modelGalacticA} shows a comparison between
the expected \gray{} spectra (from foreground model A) and 
the spectra obtained by the maximum likelihood procedure when fitting
the model templates to the \gray{} data (see also Section 
\ref{sec:analysis}). For the purpose of this comparison, we extend the
upper bound of the energy range of the low-energy fit 
from 13~GeV to 51~GeV.
The $\gamma$ rays arising from CR interactions with ISG and originating from the 
IC process are shown separately. The input model spectra are
renormalized in the figure to allow for a better comparison of the predicted and 
the fitted spectral shapes. The renormalization factors are determined in the 
energy band between 6.4~GeV and 51~GeV (cf. Section \ref{sec:analysis}).
The numerical values of the renormalization factors for the \hi{}~+~\hii{} and inverse Compton 
templates of our benchmark models are displayed in 
Figures \ref{fig:modelGalacticA}, \ref{fig:modelGalacticB}, and \ref{fig:modelGalacticC}.

The predicted and observed spectra of $\gamma$ rays from CR interactions with ISG agree well at 
energies above a few GeV, besides a moderate renormalization factor.
 A harder spectrum is seen at energies below a few GeV in the fit compared 
to the model. In this energy range, the local interstellar spectrum of
CRs is difficult to measure due to the effects of solar modulation. 

The IC emission model over-predicts the fitted IC
emission at low energies by a factor of up to $\sim4$, while
at high energies it under-predicts the emission by a factor of~2.4. 
Again, the spectral shapes of the model and the fit to LAT data show good
agreement at energies above a few GeV, so the (rescaled) model predictions from foreground 
model A can be used for the high-energy analysis, where the foreground model is fixed.

\subsection{Foreground model B}
\label{sec:foregroundB}

A significantly better agreement between the predicted and fitted IC
emission is found for foreground model B, which includes an
additional population of electron-only sources located near the Galactic center. 
In this model, the bulk of the IC emission arises from the electrons
injected near the Galactic center, and the sources of CR nuclei do
not produce a significant fraction of the local CR electron flux. The
spatial distribution that is used for this additional electron-only 
source population is shown in Figure \ref{fig:sourceDist}. 
We further assume that the Galactic center sources inject electrons following a
power-law spectrum with index $1$ below 20~GV and $2.05$
above 20~GV. We also assume the same high-energy cutoff as in foreground model A to 
maintain agreement with H.E.S.S. measurements
\citep{Aharonian:2008}. The additional electron source population must
be located close to the Galactic center in order to produce bright
enough IC emission to match \gray{} observations without
over-predicting the local CR electron spectrum at high energies
(assuming that the model diffusion parameters are unchanged).

\begin{figure}
\epsscale{.8}
\plotone{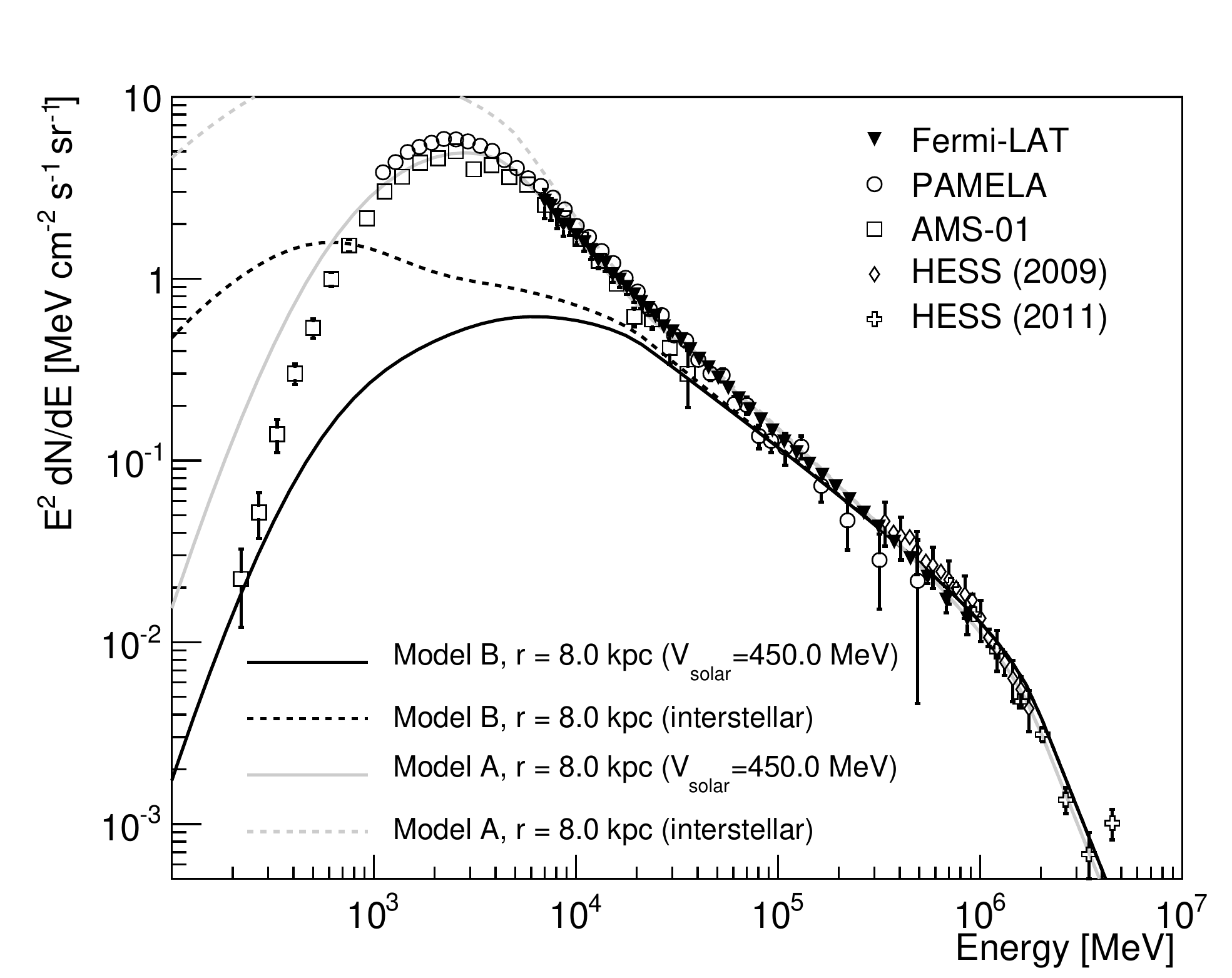}
\caption{Predicted local CR electron spectrum of foreground models A and B in comparison to measurements of the spectrum.
LAT data are taken from \citet{Ackermann2010:electrons7GeV}, PAMELA data from \citet{Adriani:2011b}, AMS-01 data
from \citet{Aguilar:2002}, H.E.S.S. data from \citet{Aharonian:2009} and \citet{Egberts:2011}.}

\label{fig:especSourcepop}
\end{figure}

Figure \ref{fig:especSourcepop} shows the electron spectrum produced
by model B in comparison to measurements, and the electron spectrum predicted by model A. 
There is a clear deficit of electrons below 20~GV in this model. However, these electrons could be easily
supplied by a local source or source population of very
soft electrons without a large impact on the total amount of IC
emission. Natural candidates would be very old supernova remnants (SNRs) in the 
solar neighborhood, e.g., Loop~I where we see high-energy \gray{} emission from the shell in LAT data.
It is, however, beyond the scope of this paper to speculate about and
address the nature of such local soft electron sources. We ignore this potential local contribution for the IGRB analysis. 

\begin{figure}
\epsscale{.8}
\plotone{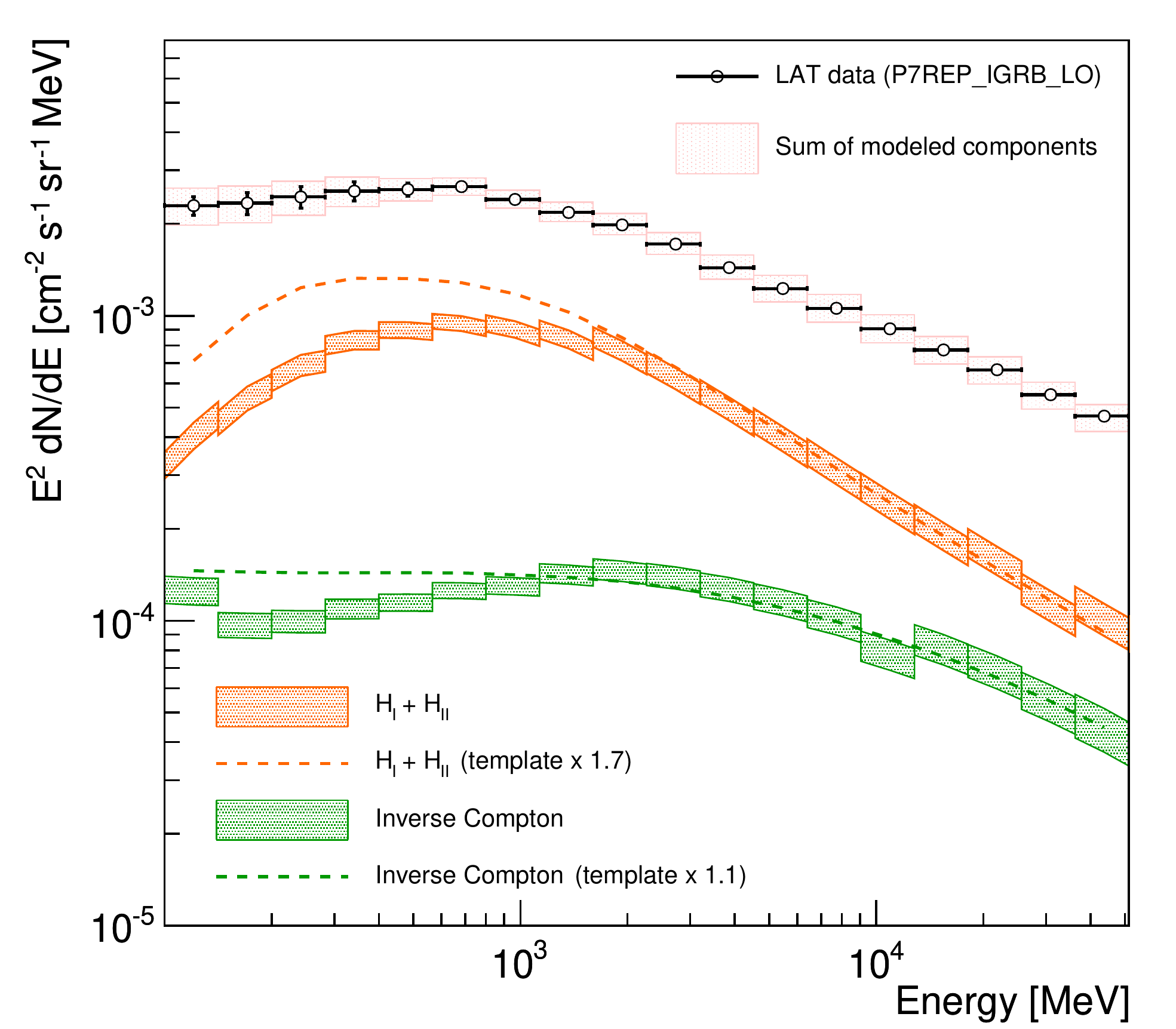}
\caption{Fitted average intensity of the DGE in foreground model
  B. See Figure \ref{fig:modelGalacticA} for a description.}
\label{fig:modelGalacticB}
\end{figure}

The same propagation and injection parameters for nuclei are used in
foreground model B as in model A. Modeled 
and fitted spectra are compared in Figure \ref{fig:modelGalacticB}. 
The renormalization factor for $\gamma$ rays from the 
interactions of CRs with ISG increases from 1.5 in foreground model A
to 1.7 in foreground model B. This is a large change,
and further investigations should be undertaken to understand if the
model B value is still within the bounds of our current uncertainties concerning the total column
density of the high-latitude gas, the \gray{} emissivity of the ISG,
and the gradient in the CR spectrum. For the purpose
of this IGRB analysis, we accept the renormalized spectrum as a valid
fit of the Galactic foreground. It can be further seen from Figure
\ref{fig:modelGalacticB} that the IC spectrum is now well described by the model both in normalization
and in shape besides a small discrepancy at the lowest energies.
 
\subsection{Foreground model C}
\label{sec:foregroundC}

Foreground model C represents a class of models in which
the CR diffusion and re-acceleration vary significantly throughout the Galaxy.
Diffusion and re-acceleration are parametrized within the transport
equation implemented in GALPROP via the spatial diffusion coefficient
$D_{xx}$ and the Alfv\'{e}n velocity $v_A$. We use a simple model to vary the diffusion
coefficient and Alfv\'{e}n velocity by connecting their values to the
strength of the regular and random Galactic magnetic fields.
Following the approximation of \cite{Strong:2007}, the diffusion coefficient is set to

\begin{equation}
D_{xx} (R,r,z) = D_{xx}^{0} \left( \frac{\delta B(r,z)}{\delta B_{\odot}} \cdot \frac{B_{\odot}}{B(r,z)} \right)^{-2} 
        \left( \frac{R}{R_0} \cdot \frac{B_{\odot}}{B(r,z)} \right)^{0.33}
\end{equation}


\noindent where $B$ is the strength of the regular field and $\delta B$
corresponds to the strength of the random field at radius~$r$ from the Galactic center and height~$z$ above the Galactic plane. 
$B_{\odot}$ and $\delta B_{\odot}$ denote their values at the position of the solar system. $R_{0}=4$~GV is the reference
rigidity, and $D_{xx}^{0}$ is the local diffusion coefficient at reference 
rigidity. The diffusion coefficient is constrained to $D_{xx}(R_0,r,z) \le 10^{30}$~cm$^{2}$~s$^{-1}$ at
reference rigidity. This constraint ensures that the mean free path of the CRs stays below the kpc scale 
for particles up to tens of TV in rigidity\footnote{We tested that the specific choice of this upper bound on the 
diffusion coefficient is irrelevant by increasing the maximum value for $D_{xx}(R_0,r,z)$ by one order of magnitude. The effects
on the predicted \gray{} emission were negligible.}.

\begin{figure}
\epsscale{1.1}
\plottwo{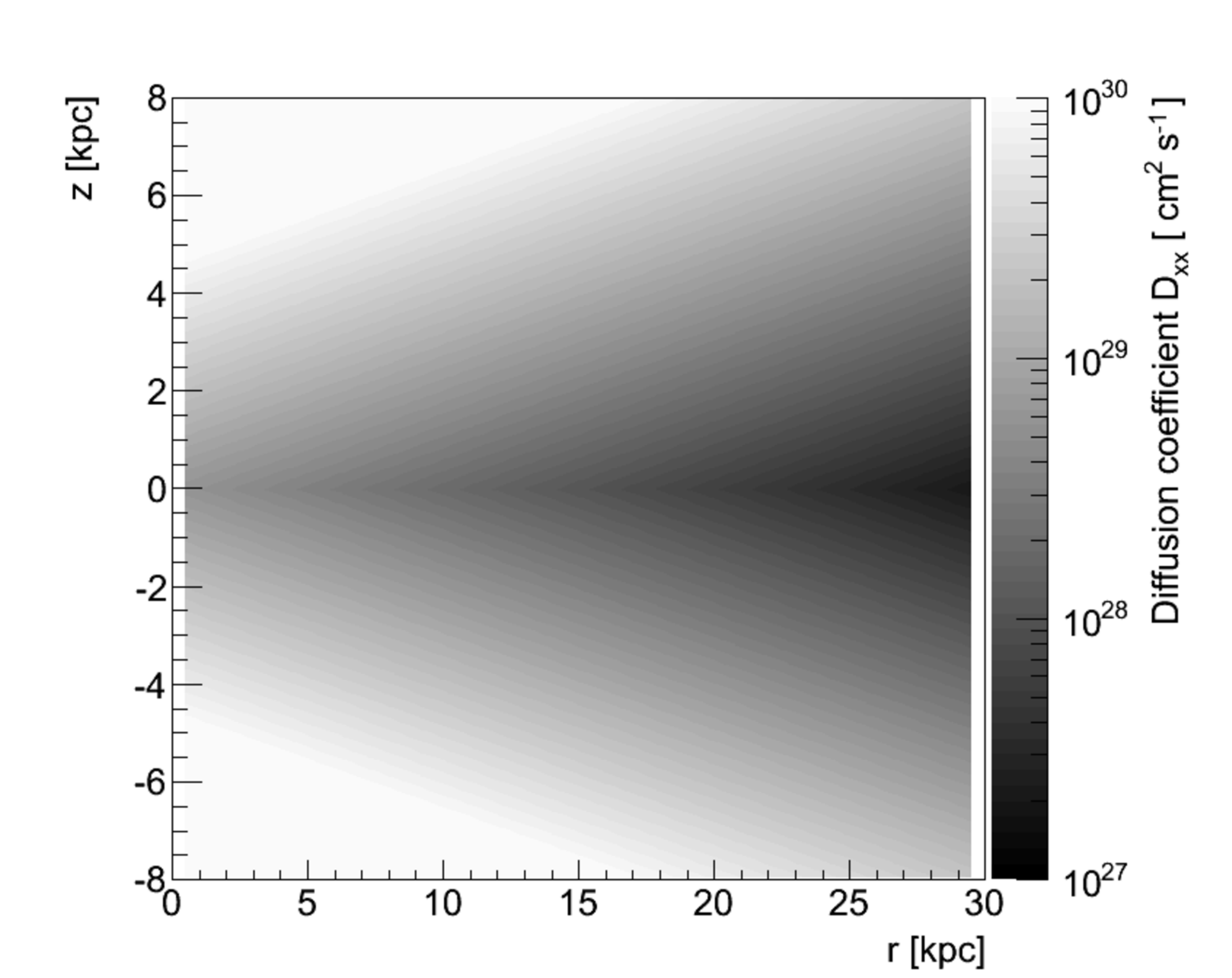}{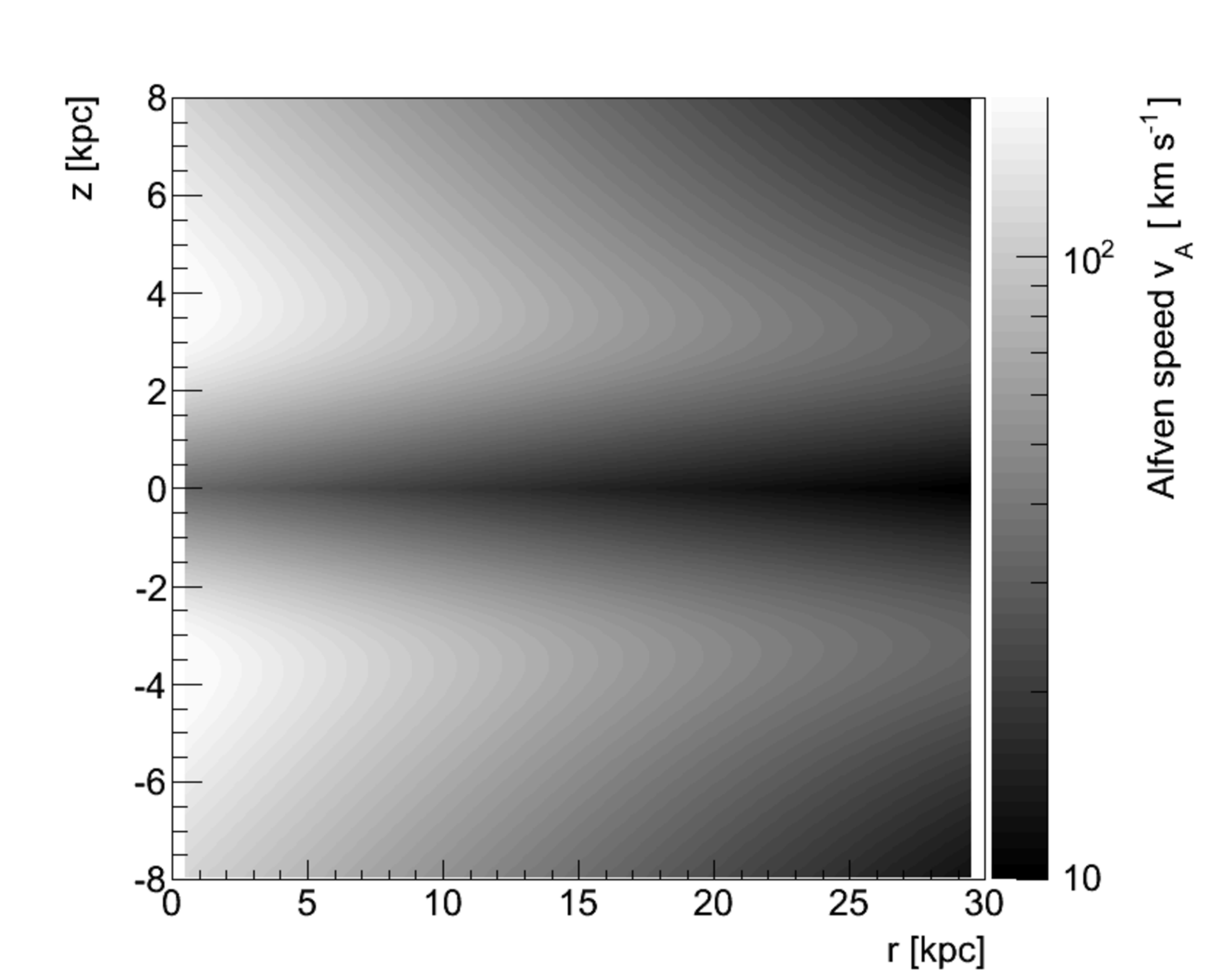}
\caption{Diffusion coefficient ({\it left}) and Alfv\'{e}n velocity ({\it right}) used in foreground model C. Both are
 functions of distance from the Galactic center ($r$) and height above the Galactic plane ($z$).}
\label{fig:diffandvalfven}
\end{figure}

We parametrize the Alfv\'{e}n velocity as $v_A \propto B_{\mathrm{tot}}/\sqrt{\rho}$, where $B_{\mathrm{tot}}$ is the 
total magnetic field strength and $\rho$ the density of ions 
in the interstellar medium. For the ion density, the same model is
used as in the rest of this work \citep{Gaensler:2008}.
Simple models are assumed for the random and the
regular magnetic field components with exponential scale heights 
and scale lengths. The regular magnetic field strength is assumed to be 
4~$\mu$G at the position of the Sun, with a scale length of 11~kpc
and a scale height of 4~kpc. The random field strength is assumed to be 
4~$\mu$G, constant in the Galactic plane with a scale height of 2~kpc. 
The scale heights are in good agreement with scale heights found from equilibrium conditions
 \citep{Kalberla:1998}. The field strengths are in qualitative agreement with recent 
 studies of the Galactic synchrotron emission in \citet{Orlando:2013},
that take radio polarization into account.
The extent of the CR halo for this model is set to 8~kpc; we note that constraints on the
halo size found in earlier studies based on the \BeTen/\BeNine{} 
ratio apply only to models with a static diffusion coefficient.
Figure \ref{fig:diffandvalfven} shows the diffusion coefficient and
the Alfv\'{e}n velocity as a function of the Galactocentric radius~$r$ 
and the height above the Galactic plane~$z$.

\begin{figure}
\epsscale{.8}
\plotone{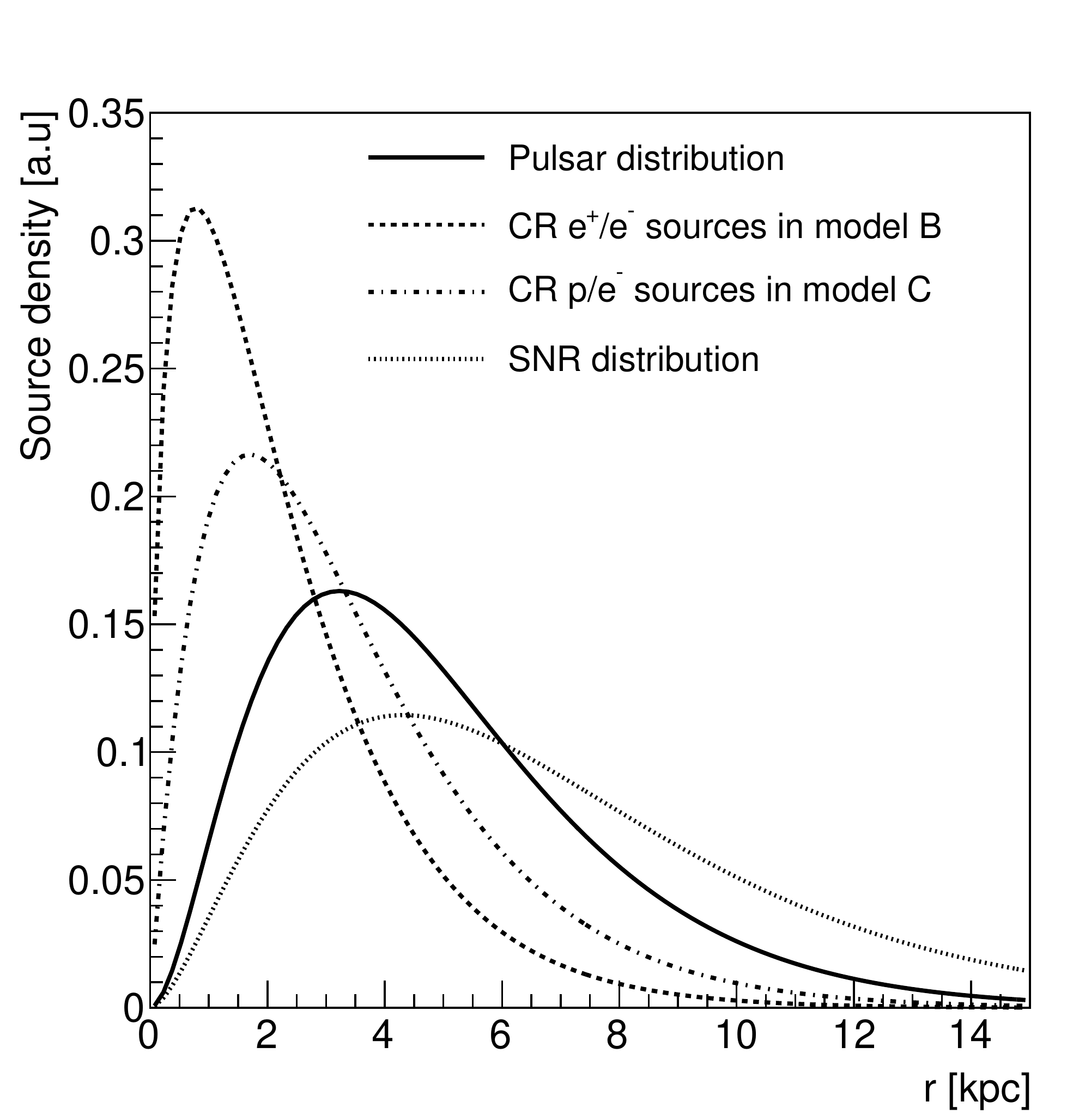}
\caption{Parametrizations of the radial CR source distribution used in
  this work. The pulsar distribution is taken from \citet{Lorimer:2006}, the 
SNR distribution from \citet{Case:1998}. The curves in the figure are normalized to 
unit integrated source density, the actual normalizations used in the models are derived
from comparisons of predicted and measured local CR proton and electron intensities. } 

\label{fig:sourceDist}
\end{figure}

A customized version of the GALPROP code (see above) is used that
allows the modeling of propagation scenarios in which the spatial and momentum 
diffusion are functions of radial distance from the Galactic center
and height above the Galactic plane. As for foreground models
A and B, GALPROP is used in its 2D mode that solves the transport equation on a 2D spatial grid
in Galactocentric radius and height $(r, z)$ around the Galactic center.
The CR source distribution assumed in model C is more peaked toward the Galactic center than the pulsar distribution 
in model A (see Figure \ref{fig:sourceDist}). The high-energy
injection spectra for CR electrons and protons are the same as for model A, power laws in rigidity with an index of $2.32$. 
However, an injection spectrum with an index of $1.9$ below 13~GV, and $2.40$ between 13~GV and 240~GV is used for the 
CR protons in model C, slightly different from the spectrum used in model A. For the 
CR electrons the injection spectral index is $1.5$ below 4.5~GV, and $2.70$ between 4.5~GV and 25~GV.
These modifications in the injection spectrum improve the agreement of the local CR spectra predicted by model C with
measurements.

\begin{figure}
\epsscale{.8}
\plotone{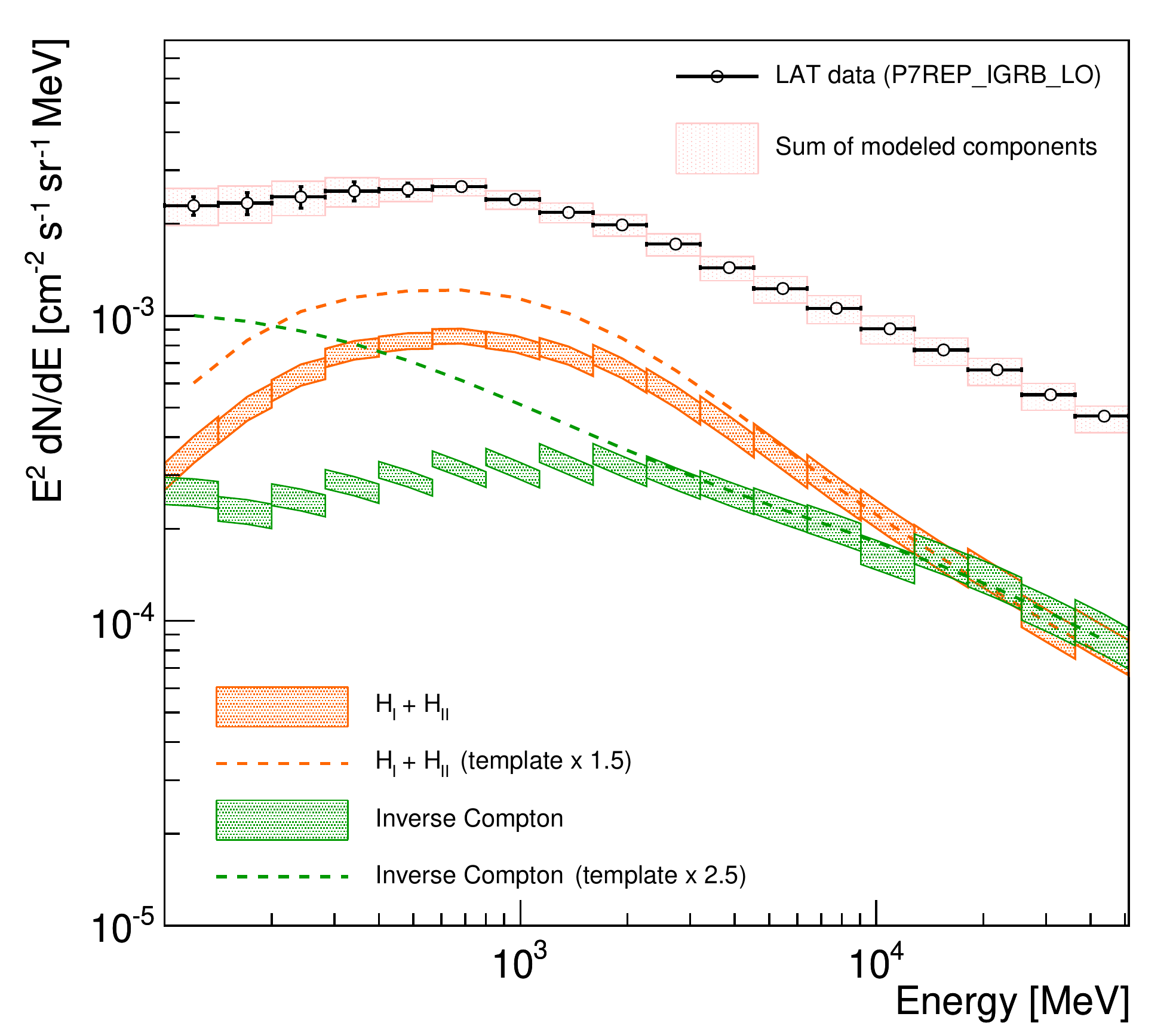}
\caption{Fitted average intensity of the DGE in foreground model
  C. See Figure \ref{fig:modelGalacticA} for a description.}
\label{fig:modelGalacticC}
\end{figure}

Modeled and fitted \gray{} spectra are compared in Figure \ref{fig:modelGalacticC}. 
The renormalization factor for \gray{s} from the interactions of CRs 
with the ISG is~1.5 as for model A. The total intensity of the
IC emission is similarly under-predicted by a factor of~2.5 at energies above a few GeV, while it is over-predicted at low energies. 
However, an interesting aspect of this model is that it predicts a flatter CR gradient
than models A and B, and therefore predicts a higher \gray{}
emissivity in the outer Galaxy. It was found in two other studies
\citep{Abdo2010:2ndQuadrant,Ackermann2011:3rdQuadrant} that the
emissivity derived from LAT observations is indeed higher in the 
outer Galaxy than predicted by diffuse emission models of class A. 
We do not discuss the CR gradient of model C further here as it is not relevant for the IGRB analysis.
\clearpage


\clearpage

\section{Residual maps}
\label{sec:appendixB}

\begin{figure}
\begin{center}
\epsscale{.9}
\plotone{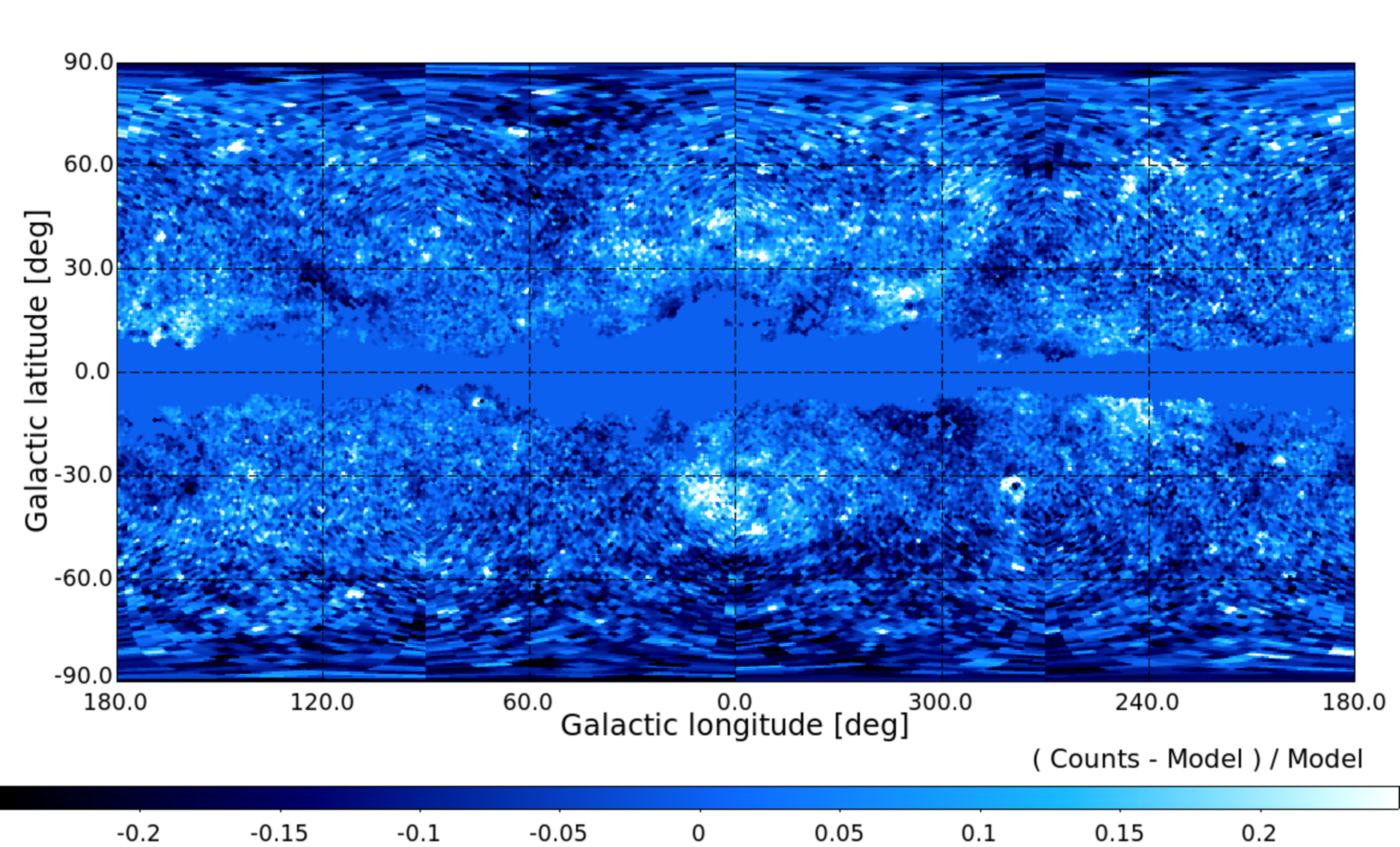}
\plotone{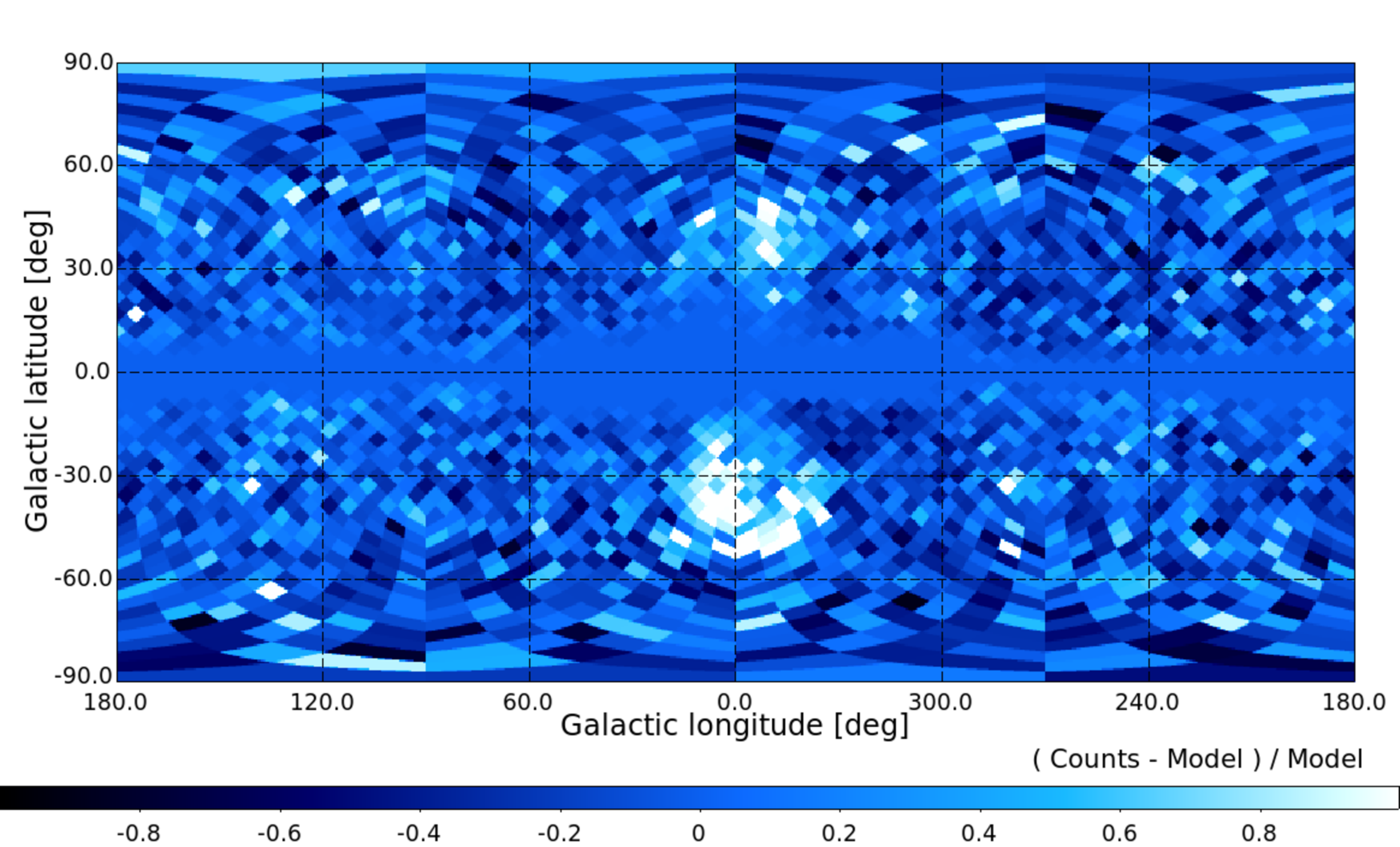}
\end{center}
\caption{Residual maps for foreground model A used in this
  analysis. The fractional difference in counts between the actual data and the fitted
model is shown in the figures. {\it Upper:} All counts above 100~MeV are included in the map. The
pixel size is 0.8~deg$^2$ (HEALPix order 6). {\it Lower:} 
Only counts above 13~GeV are included in the map. The pixel size has been increased to 13~deg$^2$
(HEALPix order 4) to account for the reduced count statistics at higher energies.}
\label{fig:resmapA}
\end{figure}

Figure \ref{fig:resmapA} shows residual maps of the relative
deviations in the number of expected and observed counts 
when using foreground model A for the DGE. The
first map shows the residual for all counts above 100~MeV,
the second map shows the residual for counts above 13~GeV. Multiple
structures are visible in the former map while the latter
is dominated by the \textit{Fermi} Bubbles.

\begin{figure}
\begin{center}
\epsscale{.9}
\plotone{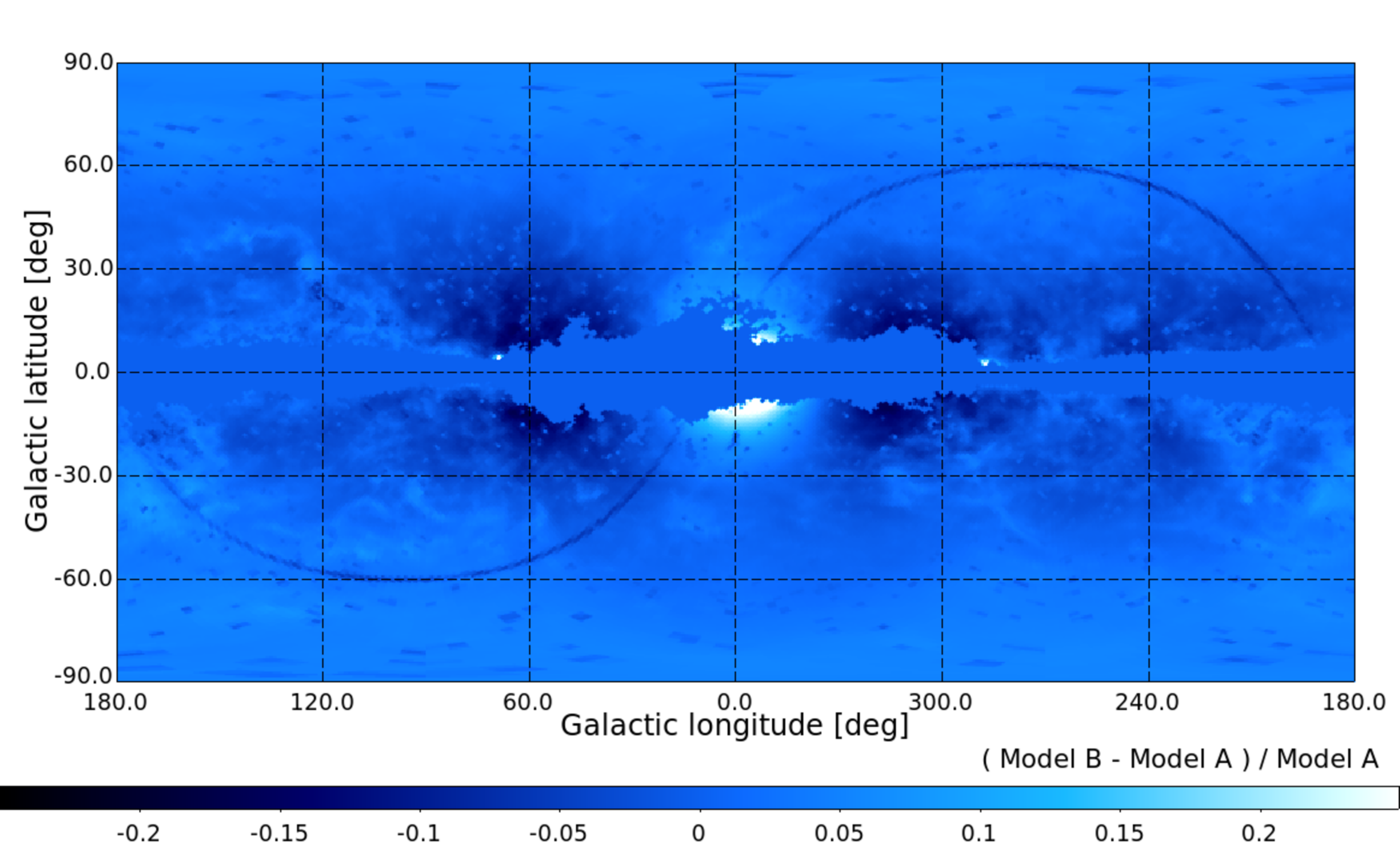}
\plotone{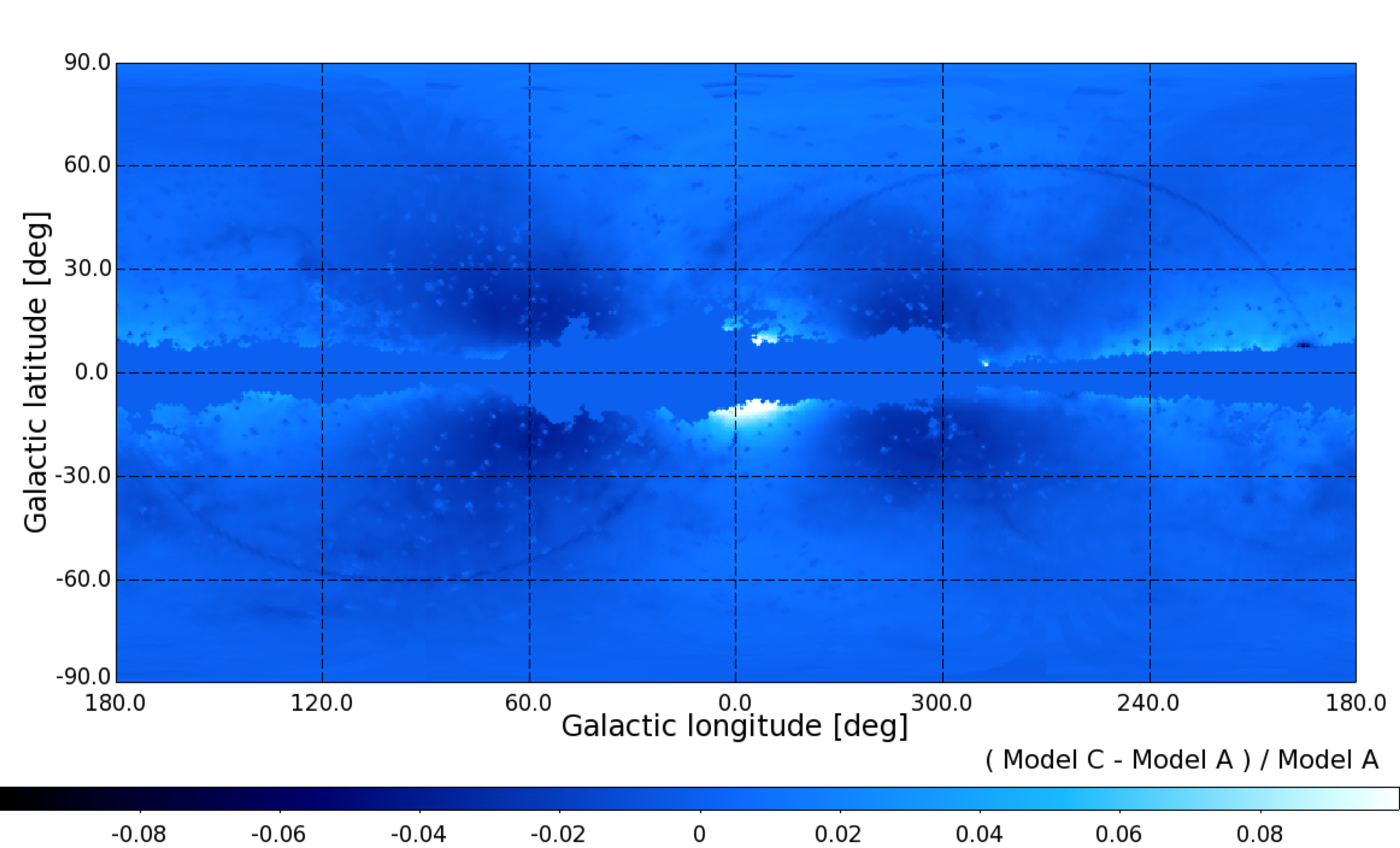}
\end{center}
\caption{Differences in foreground models A, B, and C. The fractional difference in the number of
predicted counts above 3~GeV between the models using the alternative Galactic foregrounds, B and C and the 
model using Galactic foreground A are displayed in the maps. The pixel size is 0.8~deg$^2$ (HEALPix order 6).
{\it Upper:} Fractional difference for model B. {\it Lower:} Fractional difference for model C.}
\label{fig:resmapBC}
\end{figure}

Figure \ref{fig:resmapBC} visualizes the difference in the predicted
\gray{} emission when using Galactic foreground models 
B or C instead of model A. These differences are more prominent at
energies above a few GeV, where the IC emission contributes a larger
fraction of the total \gray{} emission than at few hundreds of
MeV. Therefore, the relative deviation in predicted
counts above 3~GeV is shown in the maps when using foreground models B and C in the fit
with respect to using foreground model A.

For foreground model B, a higher \gray{} intensity is predicted close
to the Galactic center region arising from the IC emission 
of electrons that originate from the additional source population we
introduced in this model. For foreground model C, regions with higher intensity can be observed towards 
the outer Galaxy, reflecting the more efficient transport of CRs into the outer Galaxy
by the modified propagation scheme used in model C.
 
\clearpage

\clearpage

\section{Criteria for masking regions in the likelihood fit}
\label{sec:appendixC}

Approximately 90\% of the ISG is atomic (\hi{}), ionized (\hii{}), or
molecular hydrogen gas (\htwo{}). The remainder is mostly helium,
and is generally assumed to be mixed uniformly with the hydrogen.
The column density and distribution of atomic gas in the Galaxy can be estimated from surveys of the 21~cm
hyperfine-structure transition line of the hydrogen atom \citep{Kalberla:2005}. 
The distribution of molecular hydrogen gas can be inferred indirectly from surveys of the 2.6~mm J(1$\rightarrow$0) 
transition of the CO molecule by assuming a proportionality (usually
called the \Xco{} factor) between the intensity of this line
integrated over frequency ($W_{\rm CO}$) and the column density of \htwo{} gas
\citep{Dame:2001}. 
In both cases, the velocity component of the gas parallel to the line
of sight is measured via the doppler shift of the transition line.
When combined with a model of the Galaxy rotation curve, the observed velocity can be
converted into a measurement of Galactocentric radius, allowing for a determination of
the gas density distribution along the line of sight.
In this work, we use the gas
distribution model of \citep{Ackermann2012:Diffuse}, where the
total gas column density is distributed in 17 Galactocentric rings
spanning the radial range from 0 to 50~kpc. Due to the small scale heights of the 
gas (a few tens of pc for \htwo{} and a few hundreds of pc for \hi{}),
most of the gas outside of our local Galactic neighborhood will
appear concentrated around the Galactic plane. Even in our local
neighborhood, most of the \htwo{} gas is concentrated in isolated
clouds at low Galactic latitudes.

We use this fact to exclude regions of the sky from our likelihood fit
that have a significant column density of \htwo{} gas
($W_{\rm CO} > 2.5$~K~km~s$^{-1}$) along the respective line of sight as
well as lines of sight with a significant column density of \hi{} gas
($N_{\rm HI} > 5 \times 10^{20}$~cm$^{-2}$)
located beyond our local solar neighborhood (8~kpc~$<$~r~$<$10~kpc)
according to the gas distribution model in \citet{Ackermann2012:Diffuse}.
Independent of the gas column densities found, Galactic latitudes $|b|<5\deg$ are also excluded.
The exclusion of such regions simplifies the likelihood analysis
considerably. Specifically, models of the DGE in the remaining
mostly high-latitude parts of the sky do not depend on assumptions about the
CO-to-\htwo{} conversion factor \Xco, or on how
accurately we model variations of the CR density throughout the Galaxy
leading to variations of the \hi{} emissivity for
gas at larger distances. The excluded regions cover a total of 17\% of
the sky.

\bibliographystyle{apj}
\bibliography{extragalactic_long}

\clearpage

\end{document}